\documentclass[11pt,a4paper]{article}
\pdfoutput=1
\usepackage{jheppub}

\usepackage[top=4.5cm, bottom=0.cm, outer=-0.25cm, inner=4.75cm]{geometry}
\usepackage{placeins}
\usepackage{latexsym}
\usepackage[bbgreekl]{mathbbol}
\usepackage{mathtools}
\usepackage{amsmath,amssymb,amsthm,amsbsy,amsfonts,mathrsfs}
\usepackage{hyperref}
\usepackage{multirow}
\usepackage{slashed}
\usepackage{float}
\usepackage{esint}
\usepackage{epsfig}
\usepackage{lscape}
\usepackage{titlesec}
\usepackage{enumitem}   

\usepackage{graphicx}
\usepackage[utf8]{inputenc}
\usepackage{subfigure}
\usepackage{nicefrac}
\usepackage{braket}
\usepackage{rotating}
\usepackage{afterpage}
\usepackage{bbm}
\usepackage{color,xcolor}
\usepackage{cancel}
\usepackage{stackrel}
\usepackage[framemethod=tikz]{mdframed} 
\usepackage{empheq}
\usepackage[many]{tcolorbox}
\usepackage{simplewick}
\usepackage{pgf}
\usepackage{tikz}
\usetikzlibrary{shapes.geometric}
\usetikzlibrary{decorations.markings}
\usetikzlibrary{positioning}
\usetikzlibrary{fit}
\usepackage{tkz-berge}
\usepackage{rubikcube}
\usetikzlibrary{patterns}
\usepackage{tikz-3dplot}

\tikzstyle{vertex}=[circle, draw, inner sep=0pt, minimum size=6pt]
\newcommand{\vertex}{\node[vertex]}
\tikzstyle{checkpauli}=[rectangle,draw, minimum size=6pt]
\newcommand{\checkpauli}{\node[checkpauli]}
\tikzstyle{diamondnode}=[diamond,draw, minimum size=6pt]

\newcommand{\Depth}{4}
\newcommand{\Height}{4}
\newcommand{\Width}{4}

\newcommand{\pmark}[1]{\begin{tikzpicture}[overlay,remember picture]\node(#1)at (-1em,.7ex){};\end{tikzpicture}}
\newcommand{\smark}[1]{\begin{tikzpicture}[overlay,remember picture]\draw(#1)--(0,.7ex);\end{tikzpicture}}
\newcommand{\vpmark}[1]{\begin{tikzpicture}[overlay,remember picture]\node(#1)at (-.3em,5ex){};\end{tikzpicture}}
\newcommand{\vsmark}[1]{\begin{tikzpicture}[overlay,remember picture]\draw(#1)--(-.3em,-1ex);\end{tikzpicture}}

\newcommand{\AxisRotator}[1][rotate=0]{%
    \tikz [x=0.25cm,y=0.60cm,line width=.2ex,-stealth,#1] \draw (0,0) arc (-150:150:1 and 1);
}

\makeatletter

\pgfkeys{/pgf/.cd,
    ellipse ratio/.code={\pgfkeyssetvalue{/pgf/ellipse ratio}{#1}},
    ellipse ratio/.initial=1
}

\pgfdeclareshape{newellipse}
{
  \inheritsavedanchors[from=ellipse]
  \inheritanchorborder[from=ellipse]
  \savedanchor\radius{%
    \pgf@y=.5\ht\pgfnodeparttextbox%
    \advance\pgf@y by.5\dp\pgfnodeparttextbox%
    \pgfmathsetlength\pgf@yb{\pgfkeysvalueof{/pgf/inner ysep}}%
    \advance\pgf@y by\pgf@yb%
    \pgf@x=.5\wd\pgfnodeparttextbox%
    \pgfmathsetlength\pgf@xb{\pgfkeysvalueof{/pgf/inner xsep}}%
    \advance\pgf@x by\pgf@xb%
    \pgfkeysgetvalue{/pgf/ellipse ratio}{\ratioscale}
    \pgfmathsetmacro\widthfactor{sqrt(\ratioscale^2+1)/\ratioscale}
    \pgfmathsetmacro\heightfactor{sqrt(\ratioscale^2+1)}
    \pgf@x=\widthfactor\pgf@x%
    \pgf@y=\heightfactor\pgf@y%
    \pgfmathsetlength\pgf@yc{\pgfkeysvalueof{/pgf/minimum height}}%
    \ifdim\pgf@y<.5\pgf@yc%
      \pgf@y=.5\pgf@yc%
    \fi%
    \pgfmathsetlength\pgf@xc{\pgfkeysvalueof{/pgf/minimum width}}%
    \ifdim\pgf@x<.5\pgf@xc%
      \pgf@x=.5\pgf@xc%
    \fi%
    \pgfmathsetlength{\pgf@xb}{\pgfkeysvalueof{/pgf/outer xsep}}%
    \pgfmathsetlength{\pgf@yb}{\pgfkeysvalueof{/pgf/outer ysep}}%
    \advance\pgf@x by\pgf@xb%
    \advance\pgf@y by\pgf@yb%
  }

  \inheritanchor[from=ellipse]{center}
  \inheritanchor[from=ellipse]{mid}
  \inheritanchor[from=ellipse]{base}
  \inheritanchor[from=ellipse]{north}
  \inheritanchor[from=ellipse]{south}
  \inheritanchor[from=ellipse]{west}
  \inheritanchor[from=ellipse]{mid west}
  \inheritanchor[from=ellipse]{base west}
  \inheritanchor[from=ellipse]{north west}
  \inheritanchor[from=ellipse]{south west}
  \inheritanchor[from=ellipse]{east}
  \inheritanchor[from=ellipse]{mid east}
  \inheritanchor[from=ellipse]{base east}
  \inheritanchor[from=ellipse]{north east}
  \inheritanchor[from=ellipse]{south east}

  \inheritbackgroundpath[from=ellipse]
}
\makeatother

\newmdenv[%
middlelinecolor=blue!20!,
middlelinewidth=1pt,
backgroundcolor=blue!10!,
roundcorner=10pt
]{identity}

\newmdenv[%
middlelinecolor=red!20!white,
middlelinewidth=1pt,
backgroundcolor=red!10!white,
roundcorner=10pt,
subtitlebelowline=true,
frametitle={Calculation},
frametitlefont={\normalfont\bfseries\sffamily\color{red!40!white}},
]{calculation}

\def\){\right)}
\def\({\left( }
\def\]{\right] }
\def\[{\left[ }

\def\NO{\nonumber}

\newcommand{\be}{\begin{equation}}
\newcommand{\ee}{\end{equation}}

\def\bea{\begin{eqnarray}}
\def\eea{\end{eqnarray}}

\def\bal#1\eal{\begin{align}#1\end{align}}

\def\bald{\begin{aligned}}
\def\eald{\end{aligned}}

\def\bsub{\begin{subequations}}
\def\esub{\end{subequations}}

\def\beqx{\begin{displaymath}}
\def\eeqx{\end{displaymath}}

\newcommand{\bmat}{\left(\begin{array}}
\newcommand{\emat}{\end{array}\right)}

\def\a{\alpha}

\def\n{\nu}

\def\x{\xi}
\def\z{\zeta}

\def\G{\Gamma}

\def\X{\Xi}

\def\cx{{\cal X}}

\def\bo{{\raise-.3ex\hbox{\large$\Box$}}}               
\def\face{{\raise.2ex\hbox{$\displaystyle \bigodot$}\mskip-2.2mu \llap {$\ddot
        \smile$}}}                                   
\def\>{\rangle}                                      
\def\<{\langle}                                      

\def\leftrightarrowfill{$\mathsurround=0pt \mathord\leftarrow \mkern-6mu
        \cleaders\hbox{$\mkern-2mu \mathord- \mkern-2mu$}\hfill
        \mkern-6mu \mathord\rightarrow$}        
\def\dvec#1{\vbox{\ialign{##\crcr
        \leftrightarrowfill\crcr\noalign{\kern-1pt\nointerlineskip}
        $\hfil\displaystyle{#1}\hfil$\crcr}}}           

\def\-{\hphantom{-}}

\def\bbx#1\ebx{\begin{empheq}[box={\tcbhighmath[colframe=blue!20!white,colback=blue!10!white]}]{align} #1 \end{empheq}}

\title{Holography, cellulations and error correcting codes}

\author[a,b]{Marika Taylor}
\author[a,c]{and Charles Woodward}

\affiliation[a]{STAG Research Centre, Highfield, University of Southampton, SO17 1BJ Southampton, UK}  
\affiliation[b]{School of Mathematical Sciences, Highfield, University of Southampton, SO17 1BJ Southampton, UK}
\affiliation[c]{School of Physics and Astronomy, Highfield, University of Southampton, SO17 1BJ Southampton, UK}

\emailAdd{M.M.Taylor@soton.ac.uk}
\emailAdd{C.A.Woodward@soton.ac.uk}

\abstract{Quantum error correction codes associated with the hyperbolic plane have been explored extensively in the context of the AdS$_3$/CFT$_2$ correspondence. In this paper we initiate a systematic study of codes associated with holographic geometries in higher dimensions, relating cellulations of the spatial sections of the geometries to stabiliser codes. We construct analogues of the HaPPY code for three-dimensional hyperbolic space (AdS$_4$), using both absolutely maximally entangled (AME) and non-AME codes. These codes are based on uniform regular tessellations of hyperbolic space but we note that AME codes that preserve the discrete symmetry of the polytope of the tessellation do not exist above two dimensions. We also explore different constructions of stabiliser codes for hyperbolic spaces in which the logical information is associated with the boundary and discuss their potential interpretation. We explain how our codes could be applied to interesting classes of holographic dualities based on gravity-scalar theories (such as JT gravity) through toroidal reductions of hyperbolic spaces.}


\begin{document}  
	\maketitle



\section{Introduction}
\label{sec:intro}

In recent years there has been increasing interest in relations between quantum information and holography. A key connection is the relationship between the hyperbolic plane, viewed as the spatial slice of AdS$_3$, and quantum error correction codes. AdS/CFT is usually discussed in continuum language but the quantum code picture of holography relies on discretisations of the hyperbolic plane with which simple quantum error correcting codes can be associated. 

The best known class of such codes is the HaPPY codes \cite{Pastawski2015HolographicCorrespondence}. These are based on regular uniform tessellations of the hyperbolic plane, with a corresponding graph state respecting the maximal discrete symmetry group of the tessellation. The codes are related to perfect tensors and absolutely maximally entangled (AME) states \cite{PastawskiCodeGeometries}, properties which immediately give rise to Ryu-Takayanagi behaviour \cite{Ryu2006AspectsEntropy} for the entanglement i.e. the entanglement of a boundary region scales with the length of the associated discrete geodesic through the bulk. 

While perfect tensors have elegant properties that facilitate the analysis of the code properties, the properties of AME/perfect tensor codes do not reflect the expected behaviour of the dual conformal field theory. For example, correlation functions in a conformal field theory have power law fall-off but the correlation functions induced by AME states do not admit such behaviour. However, there are various ways to adapt the HaPPY construction to give rise to the expected two-dimensional CFT behaviour, ranging from using random tensors to almost perfect tensors \cite{Bhattacharyya:2016hbx,Bhattacharyya:2016zyk,Hayden:2016cfa,Qi:2018shh}. Thus for many purposes the HaPPY construction is viewed as a useful toy model for physical behaviour. 

\bigskip

Despite the considerable study of codes, there are still many important conceptual questions about quantum error correction codes for holographic spacetimes. Most of the code constructions are based around two-dimensional spatial slices of static (or stationary) three-dimensional geometries. Dynamics has been explored in \cite{Gesteau2020TheDynamics,Niermann2021HolographicSpacetime,Kohler2018ToyHamiltonians} but is primarily restricted to evolutions that respect constant curvature. The code constructions are associated with discretisations of the spacetime, and the relation to continuum geometry is not well understood. (See however \cite{Verstraete:2010ft,Haegeman:2011uy} for discussions of taking the continuum limit in the context of tensor networks and entanglement.) 

While AdS$_3$ gravity is often used as a toy model for holography, it misses important features of generic holographic dualities. Three-dimensional gravity is not dynamical and negative Einstein curvature implies constant negative Riemann curvature. In higher-dimensional gravity dynamics would change the Riemann curvature and any discrete approach/code mapping should incorporate this feature. Holographic dualities involve additional fields (scalars, gauge fields, fermions) and the quantum error correction codes should be able to capture these additional degrees of freedom. There have been some previous attempts to explore these issues. For example, by generalising the tensors to represent a Bacon-Shor code, one can include gauge fields in quantum error correcting codes (in two spatial dimensions) \cite{Cao:2020ksw,Cao:2021ibt}. The construction of codes in dimensions higher than two has been considered in \cite{Kohler2018ToyHamiltonians}; this approach uses so-called perturbation gadgets to construct codes associated with discretisations of higher dimensional spaces. A recent review of holographic quantum error correction can be found in \cite{JahnHolographicReview}. 

\bigskip

The main goal of this work is to initiate the systematic construction of codes associated with AdS spacetimes in general dimensions. As we discuss in section \ref{sec:tess}, AdS codes are relevant not only in the context of the standard conformal AdS/CFT dualities. Reduction of AdS spacetimes on tori gives rise to scalar/gravity theories for which the holographic duals are theories with dimensionally running couplings. A specific example of considerable current interest is the reduction of AdS$_3$ on a circle to give JT gravity, for which the details of the holographic dictionary were studied in \cite{Cvetic:2016eiv,Taylor:2017dly}. However, the relation between these bulk scalar/gravity theories and dual quantum field theories with generalized conformal structure holds much more broadly \cite{Kanitscheider2008PrecisionBranes} and it was shown in \cite{Kanitscheider:2009as} that all such dualities could be understood in terms of toroidal reductions of AdS (with the torus not necessarily having an integer dimension). 

In section \ref{sec:tess} we discuss how one would carry out a toroidal reduction of a hyperbolic tessellation, illustrating our discussions with the example of the hyperbolic plane reduced on a circle, which is relevant for the AdS$_3$/JT gravity case. This discussion shows how one would obtain codes associated with the lower-dimensional scalar/gravity theories from higher dimensional hyperbolic codes. 

\bigskip

In sections \ref{sec:qecc} and \ref{sec:happy3d} we develop the construction of codes associated with regular uniform tessellations of hyperbolic geometries in three and higher dimensions. The reason for beginning with a HaPPY type approach based on regular uniform tessellations is the desirable properties captured by such codes e.g. Ryu-Takayanagi entanglement, pushing behaviour, greedy algorithm. While one would anticipate the need to adapt the construction to obtain CFT entanglement and correlation function behaviour, the two-dimensional HaPPY code is a useful starting point for more realistic constructions. 

One of the key aspects of the HaPPY construction is the relationship between perfect tensors and absolutely maximally entangled states and we review this in detail in section \ref{sec:qecc}. For higher dimensional hyperbolic space a number of subtleties are encountered in using regular uniform tessellations and perfect tensors to construct codes. Firstly, the number of regular uniform tessellations decreases with dimension e.g. there are only four such tessellations for three dimensional hyperbolic space. This contrasts with the infinite number of such tessellations in two dimensions. 

Secondly, one cannot assign qudits to the polytope of the tessellation in a way that the assignment both preserves the discrete symmetry of the polytope and corresponds to a perfect tensor/absolutely maximally entangled state. The prototype HaPPY code is based on pentagons and the physical qubits associated to each side of the pentagon are equivalent to each other in the corresponding graph state. Standard classifications of absolutely maximally entangled (AME) states \cite{HelwigAMESecret,Helwig2013AbsolutelyStates,Huber2016AbsolutelyExist,Goyeneche2015AbsolutelyMatrices} imply that one cannot, for example, have the qudits associated with each face of a three-dimensional polytope of a regular hyperbolic tessellation being equivalent in the corresponding graph state. 

This might sound surprising but in fact many hyperbolic plane tessellations are also incompatible with simultaneously preserving the discrete symmetry of the polygon cell and corresponding to an AME state. The new issue in higher dimensions is that the number of regular uniform tessellations is sparse, and all the polytopes corresponding to these tessellations are not compatible with maximal discrete symmetry and AME states. In section \ref{sec:happy3d} we present two alternative ways of addressing this issue. The first is to relax the requirement of AME/perfect tensor; we construct codes that respect the discrete symmetry of each polytope cell of the tessellation but are not AME. The second approach is to work with AME codes, but now the assignment of the qudits to polytope cell faces does not respect the discrete symmetry of the polytope. The latter implies that one has to be careful with the concatenation of cells and we give an example of how this can be done consistently. There are many future directions to develop these codes further, which we discuss in section \ref{discussion}. 

\bigskip

The underlying principle of the HaPPY code is that logical qubit information is encoded in each cell of the tessellation with this information pushed to the boundary via physical qubits. HaPPY and related codes have natural interpretations in terms of spacetime reconstruction: one can ask questions about which parts of the boundary are necessary for reconstruction of a given region in the bulk. 

While most holographic constructions of hyperbolic codes are based on the principle of encoding logical information into each cell, this is not the only possible way to construct codes for hyperbolic spaces. It is known that one can map cellulations of spaces into other types of quantum error correction codes, namely CSS codes. As we discuss in section \ref{css-tes} the logical information in such codes is associated both with the asymptotic boundary of the space as well as with internal boundaries/defects. 

CSS codes associated with the hyperbolic plane have been constructed in earlier literature. In section \ref{css-tes} we review such constructions and explain how CSS codes associated with higher dimensional hyperbolic spaces can be constructed. We note that these codes could potentially be used in two distinct ways. Firstly, if the code is associated with the entire (regulated) hyperbolic space, then the logical information at the boundary would be interpretable in the dual conformal field theory. One could think of these codes as in some sense reversing the logic of HaPPY: in these codes, one can ask questions about how logical information in the boundary is encoded through a physical qubit network in the bulk. 

The second potential use of the CSS code construction is as a way to implement local holography. Suppose one considers a region of a hyperbolic space with a boundary at finite distance, and associates a CSS code to a tessellation of this region with logical information being encoded at the boundary of the region. Each region is thus locally holographic, with logical information associated with the boundary. One could then envisage concatenating codes associated with neighbouring regions, pushing the logical information all the way out to the asymptotic boundary. This approach could be a natural starting point for incorporating dynamics, with the curvature and encoding in each region potentially evolving with time.  

\bigskip

The plan of this paper is as follows. In section \ref{sec:tess} we discuss tessellations and cellulations for geometries that arise in holographic correspondences. In section \ref{sec:qecc} we explore in detail the construction of quantum error correction codes associated with two-dimensional tessellations, focusing particularly on AME states and the concatenation of codes between cells.  In section \ref{sec:happy3d} we construct codes associated with hyperbolic geometries in three and higher dimensions, giving examples of both AME and non-AME codes. In section \ref{css-tes} we discuss alternative constructions of codes associated with hyperbolic tessellations, CSS codes in which the logical qubits are encoded through global properties of the tessellation. We conclude and explore directions for future research in section \ref{discussion}. In the appendices \ref{key-codes} and \ref{Pauli} we provide summaries of key properties of stabiliser codes.

\section{Holographic geometries, polytopes and tessellations}
\label{sec:tess}

Throughout this paper we will be exploring cellulations of holographic geometries and associated codes using several classes of representative examples. Our examples are primarily based on maximally symmetric geometries as these are both the best understood holographic geometries and furthermore their cellulations are well studied by mathematicians. 

Our main example of holographic geometry is perhaps unsurprisingly Anti-de Sitter for which spatial slices are hyperboloids, as this is the basis for the best understood holographic correspondence. Earlier literature on holographic codes has been primarily based around $AdS_3$ whose spatial slices are hyperbolic planes. In this section we will discuss cellulations of hyperbolic spaces in general dimensions as a first step towards generalising $AdS_3$ constructions to higher dimensions. 

\bigskip

We will also be interested in holographic geometries with compact directions, focussing on two distinct representative classes of examples. The first class of examples is products of AdS with spheres which of course arise frequently in holographic dualities. For example, AdS$_3$ spacetimes often occur as products with two and three dimensional spheres: 
\be
ds^2 = \frac{1}{z^2} \left ( dz^2 + dx^2 - dt^2 \right ) + {\cal R}^2 d \Omega_n^2 \label{kkn}
\ee
where $d \Omega_n^2$ is the metric on $S^n$ with $n =2$ or $3$. Here we represent the anti-de Sitter spacetime in Poincar\'{e} coordinates to contrast against the case discussed below but the spacetime is regular and global coordinates would cover the entire spacetime. The simplest prototype for compact additional dimensions of this type would be AdS$_3 \times $ S$^1$, for which we can express the metric as 
\be
ds^2 = \frac{1}{z^2} \left ( dz^2 + dx^2 - dt^2 \right ) + {\cal R}^2 d y^2 \label{kk}
\ee
where the anti-de Sitter space has unit radius and ${\cal R}$ is the radius of the circle direction $y$. We will discuss tessellation of this prototypical example below.  

\bigskip

Our second class of examples relates to toroidal compactifications of higher dimensional AdS spacetimes.  These examples are interesting as from the lower dimensional perspective they give rise to holographic dualities with running couplings: reducing the CFT on a torus, the dual field theory is from the lower dimensional perspective a theory with generalised conformal structure i.e. conformal invariance broken only by a single running coupling. 

Such dualities are amongst the simplest prototypes for non-conformal gauge/gravity dualities and arise in the context of dualities associated with D-branes and fundamental strings in \cite{Itzhaki:1998dd}. The detailed holographic dictionary for non-conformal branes and the associated generalized conformal structure was studied in \cite{Kanitscheider2008PrecisionBranes}. It was shown in \cite{Kanitscheider:2009as} that generic dualities with generalized conformal structure could be interpreted in terms of compactifications of AdS on tori, with the dimension of the torus not necessarily being integral.

As an illustrative case, we can express the metric for AdS$_{(\sigma + 3)}$ in the form
\be
ds^2 = \frac{1}{z^2} \left ( dz^2 + dx^2 - dt^2 + d { \bf y}  \cdot d {\bf y}_{\sigma} \right ) \label{non-compact}
\ee
where ${\bf y}$ denotes $\sigma$ coordinates. When the $y$ coordinates are not periodic the metric simply describes the Poincar\'e patch of anti-de Sitter space in $(\sigma + 3)$ dimensions, with the Poincar\'{e} horizon being at $z \rightarrow \infty$. 

If the coordinates ${\bf y}$ are periodically identified, and thus parameterise a torus $T^{\sigma}$, the metric above has a conical singularity as $z \rightarrow \infty$; one cannot view the metric above as covering part of a regular manifold. Nonetheless this situation is of physical interest: it corresponds to the dual CFT being compactified on a torus $T^{\sigma}$ and the conical singularity can be cloaked by a horizon at finite temperature. 

In this setup we can relate negative curvature geometries in $(\sigma + 3)$ dimensions to solutions of three dimensional Einstein-scalar gravity as follows. Pure gravity solutions in $(\sigma + 3)$ dimensions satisfy the equations of motion following from the action
\be
S = L \int d^{\sigma + 3} x \sqrt{- G} \left ( R (G) + (\sigma +1)(\sigma + 2) \right )
\ee
where 
\be
L = \frac{l^{\sigma +1}}{ 16 \pi G_{\sigma + 3}}
\ee
with $l$ the AdS radius and $G_{\sigma + 3}$ the Newton constant.  

Now consider a diagonal reduction ansatz for the metric $G$: 
\be
ds^2 = ds_3^2 + e^{\frac{2 \phi}{\sigma}} d {\bf y} \cdot d {\bf y}_{\sigma}
\ee
The three dimensional metric and scalar field $\phi$ satisfy the equations of motion following from the reduced action
\be
S = L V_{\sigma} \int d^3 x \sqrt{ -g} e^{\phi} \left ( R(g) +  \left ( \frac{\sigma - 1}{\sigma} \right ) (\partial \phi)^2 + (\sigma + 1) (\sigma + 2 ) \right ),
\ee
where $V_{\sigma}$ is the volume of the compactified directions ${\bf y}$. One particular solution of these reduced equations is
\be
ds^2 = \frac{1}{z^2} \left ( dz^2 + dx^2 - dt^2 \right ) \qquad
e^{\phi} = \frac{1}{ z^{\sigma}}, \label{run}
\ee
i.e. the reduction of the higher dimensional AdS solution. The key conceptual difference relative to \eqref{kk} is that the radii of the circle directions scale with the AdS radius, rather than being fixed. 

\bigskip

The above example relates to three dimensional Einstein scalar theories. However, the general picture holds in generic dimensions providing a generic class of non-conformal holographic dualities. A particularly interesting case of considerable recent interest is the circle reduction of AdS$_3$ itself i.e. in the metric 
\be
ds^2 = \frac{1}{z^2} \left ( dz^2 + dx^2 - dt^2  \right ) \label{non-c2t}
\ee
the $x$ direction is taken to be periodic with radius $R_x$. Reduction of AdS$_3$ gravity on a circle results in a two-dimensional theory that is equivalent to JT gravity \cite{Kanitscheider2008PrecisionBranes,Cvetic:2016eiv}, which is again dual to a (one-dimensional) theory with a dimensionally running coupling. Such two-dimensional backgrounds have been discussed extensively in the context of SYK dualities following the well-known work of \cite{Maldacena:2016hyu,Maldacena:2016upp}. In later sections we will discuss tessellations for these backgrounds and how these may relate to quantum codes. 



\subsection{Tessellations of hyperbolic spaces}  \label{tes-gen}

In this section will review how classes of holographic geometries can be tessellated and cellulated by polytopes, beginning with the very familiar case of two dimensional tessellations. We start with the definition of a tessellation for a two-dimensional (Riemannian) manifold. A tessellation consists of a covering of the manifold ${\cal M}$ by a set of polygons $\{ P_j \}$, each of which is associated with a distance preserving function $\phi_j : P_{j} \rightarrow {\cal M}$. This implies that any two points on a polygon which are associated with $(a,b)$ satisfy
\be
D_{P_j} (a,b) = d_{\cal M} (\phi_j(a), \phi_j (b))
\ee
where $D_{P_j}$ and $d_{\cal M}$ are distances in the polygon and manifold, respectively. In two dimensions tessellations are commonly referred to as tilings. 

\bigskip

Turning to holographic applications, in the context of AdS$_3$, the spatial section is the hyperbolic plane $H_2$, tessellations of which are very well studied within the mathematics literature. From the perspective of physical applications it is often natural to focus on tessellations that respect discrete symmetry groups that are subgroups of the continuous symmetry group of hyperbolic space. The most symmetric tessellations are based on regular uniform tilings. 

 It is a well known result in mathematics, following the famous work of Coxeter \cite{Coxeter}, that the hyperbolic plane admits an infinite number of regular tilings. {\it Regular} tilings can be characterised by the Schl\"{a}fli pair $\{ p,q \}$, where $p$ is the $p$-gonal regular polygon and $q$ denotes the number of line segments associated with each vertex of the polygon (or equivalently the number of p-gons at each vertex). Every positive integer pair such that
\be
\frac{1}{p} + \frac{1}{q} < \frac{1}{2} 
\ee
gives a hyperbolic tiling. The specific tilings used in the context of the original HaPPY code \cite{Pastawski2015HolographicCorrespondence} 
are $\{ 5, 4 \}$ (and dual, see below), which is an example of tiling which is both regular and uniform.

\bigskip

A {\it uniform} tiling is a tiling that has regular polygons as faces and is vertex transitive so there is an isometry mapping any vertex onto any other. Uniform tilings can be described by their vertex configuration, a notation for representing the sequence of faces around the vertex. The vertex configuration gives the number of sides of faces going around the vertex. For example $a. b. c $ denotes a vertex that has three faces around it, faces with sides $a$, $b$ and $c$. With this notation the regular pentagonal tiling is denoted $5.5.5.5$. 

Uniform tilings may be regular, if also face and edge transitive, but can also be quasi regular (edge transitive but not face transitive) or semi-regular (if neither edge nor face transitive). In the context of holography it is most natural to work with uniform regular tilings, as one would expect transitivity of faces, edges and vertices. 
All uniform tilings generate {\it dual} uniform tilings; in the context of two-dimensional tessellations, duality relates the vertices of one tessellation to the edges of the dual. For the uniform regular tilings, each tiling $ \{ p,q \}$ has a dual tiling $\{ q,p \}$. For example, the dual of the pentagonal tiling 
with four pentagons meeting at each vertex, $\{ 5, 4 \}$, is a square tiling with five squares meeting at each vertex $ \{ 4, 5 \}$. 

\bigskip

The main focus of this paper is generalising codes based on tessellations to spatial dimensions higher than two. To describe higher dimensional tessellations we first need to define the relevant polytopes. A $d$-dimensional (Euclidean) polytope $P$ is a compact subset of $d$ dimensional Euclidean space bounded by a finite number $k$ of $(d-1)$ dimensional hyperplanes; compactness implies that $k > d$. A $d$ dimensional polytope has facets which are $(d-1)$ dimensional polytopes; these facets themselves have facets that are $(d-2)$ dimensional polytopes and so on. The set of all the $i$ dimensional sub-polytopes are referred to as $i$ cells. 

A tessellation of a $d$-dimensional Riemannian manifold ${\cal M}$ formally consists of a set of $d$-dimensional polytopes $\{ P_j \}$ embedded via isometries $\phi_j : P_j \rightarrow {\cal M}$. The images of the $i$ cells under the mapping are also $i$ cells. Tessellations for which the isometric maps are  replaced by injective homeomorphisms are called cellulations; a cellulation does not respect distances (and therefore ${\cal M}$ need not necessarily admit a metric) but a cellulation does respect the topology of the manifold. 

\bigskip


Given this general definition of tessellations, let us now consider higher dimensional AdS spacetimes whose spatial sections are hyperbolic. We begin with AdS$_4$, with spatial section $H^3$. Again, from a physical perspective, it is natural to look first at tessellations that respect discrete subgroups of the continuous symmetry group, i.e. regular and uniform tessellations. These are well studied within the mathematics literature. 

Three-dimensional hyperbolic space may be tessellated with regular polytopes that are regular polyhedra and are characterised by Schl\"{a}fli symbols $\{ p, q \}$. Here $p$ refers to the face type of the regular polyhedron while $q$ is the vertex figure, which is the polygon obtained by connecting vertices that are one edge away from a given vertex. For a regular polyhedron the vertex figure is always a regular polygon. An example is $\{ 5, 3 \}$, the regular dodecahedron with pentagonal faces and three edges around each vertex. Hyperbolic geometry is obtained, as in the two dimensional case, when
\be
\frac{1}{p} + \frac{1}{q} < \frac{1}{2}. 
\ee
Topologically a regular two-dimensional tessellation may be viewed as a polyhedron such that the angular defect at the vertex is zero and this is why the Schl\"{a}fli $\{ p, q \}$ symbols arise in both contexts. 

\bigskip

The polytope clearly does not uniquely define how the three-dimensional hyperbolic space is tessellated. For regular tessellations (also called honeycombs), the tessellation is defined by the polytope $\{ p, q \}$ together with the edge figure $\{ r \}$, i.e. the number of polyhedra around each edge. This information may be summarised in the Schl\"{a}fli symbol $\{ p, q, r \}$. 

A key difference relative to two dimensions is that the number of regular uniform tessellations is finite. Indeed, the only four regular compact three-dimensional hyperbolic honeycombs are shown in Table \ref{ads4:table}. 

\begin{table}[h!]
\begin{center}
 \begin{tabular}{||c c c c||} 
 \hline
 Name & Schl\"{a}fli symbol & Polytope & Edge figure \\ [0.5ex] 
 \hline\hline
 Icosahedral honeycomb & $\{ 3,5,3 \}$ & Icosahedron $\{ 3, 5 \}$ & $\{ 3 \}$  \\ 
 \hline
 Order 5 cubic honeycomb & $\{ 4, 3, 5 \}$ & Cube $ \{ 4,3 \}$  & $\{ 5 \}$ \\
 \hline
 Order 4 dodecahedral honeycomb & $ \{ 5,3,4 \}$ & Dodecahedron $\{ 5,3 \}$ & $ \{ 4 \}$  \\
 \hline
 Order 5 dodecahedral honeycomb & $ \{ 5,3,5 \}$ & Dodecahedron $ \{ 5,3 \}$ & $\{ 5 \}$ \\ [1ex] 
 \hline
\end{tabular}
\end{center}
\caption{The four regular compact hyperbolic honeycombs of $H^3$.}
\label{ads4:table}
\end{table}

Under duality operations, the cells and vertices are interchanged, and the faces and edges. We illustrate the corresponding Schl\"{a}fli symbols for these dual honeycombs in Table \ref{ads4dual:table}.

\begin{table}[h!]
\begin{center}
 \begin{tabular}{||c c c||} 
 \hline
 Name & Schl\"{a}fli symbol & Dual \\ [0.5ex] 
 \hline\hline
 Icosahedral honeycomb & $\{ 3,5,3 \}$ & $\{ 3, 5, 3 \}$  \\ 
 \hline
 Order 5 cubic honeycomb & $\{ 4, 3, 5 \}$  & $\{ 5,3,4 \}$ \\
 \hline
 Order 4 dodecahedral honeycomb & $ \{ 5,3,4 \}$ &  $ \{ 4, 3 ,5  \}$  \\
 \hline
 Order 5 dodecahedral honeycomb & $ \{ 5,3,5 \}$ &  $\{ 5, 3, 5 \}$ \\ [1ex] 
 \hline
\end{tabular}
\end{center}
\caption{Dual honeycombs of $H^3$.}
\label{ads4dual:table}
\end{table}

Hyperbolic honeycombs in higher dimensions can similarly be classified \cite{Coxeter}. For tessellations of spatial slices of $AdS_5$, the regular compact honeycombs of $H^4$ are
\begin{equation}
\{ 3,3,3,5 \} \qquad \{4,3,3,5\} \qquad \{5,3,3,5 \} \qquad \{5,3,3,4\} \qquad \{5,3,3,3 \}. \label{regular4d}
\end{equation} 
Analogously to the 3d honeycombs, the Schl\"afli symbol $\{p,q,r,s\}$ captures the four-dimensional polytope $\{ p,q,r \}$ and via $s$ the structure of the honeycomb i.e. how the polytopes fit together. The relevant 4d regular polytopes are $\{3,3,3\}$ (5-cell with tetrahedron 3d projection); $\{4,3,3\}$ (tesseract, with cubic 3d projection) and $\{5,3,3\}$ (120-cell, with tricontahedron 3d projection). (Note that the canonical honeycomb for $R^4$ is $\{4,3,3,4\}$.) For five dimensional hyperbolic space $H^5$ there is only one regular honeycomb ($\{3,4,3,3,3\}$). Interestingly, for hyperbolic spaces in dimensions greater than five there are no regular compact honeycombs.

\subsubsection{Tessellations of AdS$_3 \times$ S$^n$} \label{AdSS}

In this section we will consider possible tessellations for holographic geometries involving compact spheres, using the example of \eqref{kkn} to illustrate the discussions. The spatial sections of AdS$_3 \times $ S$^n$ are H$^2 \times$ S$^n$. Consider first the prototype case of $n=1$. Clearly since the space is a direct product we can take direct products of tessellations for each component. For $S^1$ the relevant polytopes are closed line segments; these are regular and are represented by the Schl\"afli symbol $\{ \}$. Uniform regular tessellations of H$^2 \times $ S$^1$ can thus be characterised as $\{ p , q \} \times \{ \}$. A special case of this is where the circle is covered by one single segment. 

Now consider the case of $n=2$, i.e. the two sphere. Uniform tessellations of the sphere are well documented and in particular the regular uniform tessellations include $\{ p, q \}$ where
\be
\frac{1}{p} + \frac{1}{q} > \frac{1}{2}.
\ee
There are a finite number of such regular uniform tessellations with $p > 2$, and the total number of polygons required to cover the sphere is in each case finite.  One can also consider tessellations of the form $\{2 , m \}$ as this manifestly satisfies the requirement above for any value of $m$. Here the sphere is divided into $m$ equivalent segments; for such tessellations one can take $ m \rightarrow  \infty$ i.e. there are an infinite number of possible tessellations.  

Having discussed the possible regular tessellations, let us now turn to what would be the most natural approach in the context of holographic codes. 
In holographic dualities of the type \eqref{kkn} the compact space plays a qualitatively different role to the non-compact hyperbolic space: the radius of the compact space is fixed and does not depend on the hyperbolic radial coordinate i.e. the renormalization group scale. As we review later, the basic idea of the encoding map of a code is to represent interior regions of the hyperbolic space by logical qubits: the tessellation captures the increase in the number of qubits required as one approaches the conformal boundary. 

As the radius of the compact space is fixed, it is not clear tessellating the compact space non-trivially would be the natural choice for the encoding map approach to holography, rather than working with the trivial tessellation which is the compact space itself (tessellations are $\{ p, q \} \times S^2$). The latter manifestly preserves the full symmetry group of the compact space, while any non-trivial tessellation breaks the symmetry group to a discrete subgroup. Thus, for holographic geometries in which the compact part has a fixed radius, tessellations and associated codes are perhaps most naturally constructed by treating the code qubits to transform in appropriate representations of the compact space symmetry group. We will however use non-trivial tessellations of spheres in section \ref{css-tes}, in the context of the topologically spherical regulated 
boundary of hyperbolic space itself.

\subsubsection{Toroidal compactification of hyperbolic spaces} \label{sec:tor}

Now let us turn to toroidal compactifications of hyperbolic spaces. In this set up the compact direction has a radius that depends on the holographic scale; toroidal identifications have fixed points and thus the holographic geometries are not regular manifolds. Mathematics literature focuses on tessellations and cellulations of regular manifolds and such irregular toroidal compactifications do not fall within the usual classifications. Nevertheless, as discussed above, such geometries arise rather generically in holography and it is interesting to explore how one could relate these to quantum codes. 

We will explore two qualitatively different ways to cellulate a toroidal compactification which are distinguished by whether they preserve a discrete subgroup of the toroidal symmetry. Let us illustrate this discussion using the example given in \eqref{non-compact}. One can manifestly cellulate the space locally with cells of the type 
$\{ p, q \} \times T^{\sigma}$. The torus itself can also be non-trivially cellulated by breaking each circle direction into segments. In both approaches the cellulation will break down as $z \rightarrow \infty$, reflecting the conical singularity. This can be addressed by excising this region i.e. considering a cellulation with a boundary near to $z \rightarrow \infty$.  

\begin{figure}[h]
\begin{center}
\includegraphics[scale=0.4]{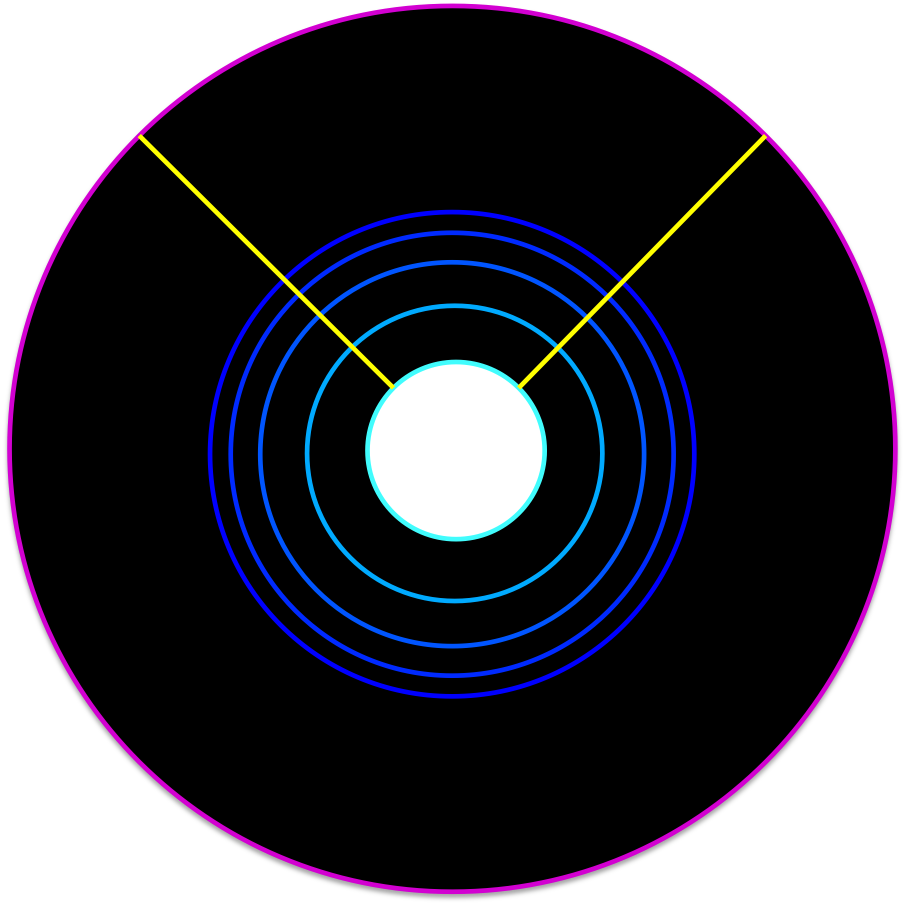}
\end{center}
\caption{\label{fig:cc} Compactified hyperbolic plane, tiled by annuli of equal area. The yellow lines are identified to compactify the plane, creating a conical singularity which is shown as excised (white region).}
\end{figure}

This type of cellulation is similar to those discussed in the previous subsection. However, in the situation described in the previous subsection the scale of the compact geometry is fixed. For toroidally compactified hyperbolic spaces the scale of the torus is not constant but varies with $z$. To capture this feature the volumes of the toroidal components of the cells $T^{\sigma}$ cannot be constant but should increase as cells approach the boundary. The constant negative curvature therefore has to be built in by hand through the volumes of the cells. 

The simplest example to visualise is the case of the hyperbolic plane itself, with the $x$ direction periodic; this is relevant for the relationship between AdS$_3$ reduced on a circle to give JT gravity. The associated tiling is shown in Figure~\ref{fig:cc}. Each annulus is of equal proper area, and accordingly the width of the annulus decreases along the radial direction. One can implement any radius of the $x$ direction by identifications, but these result in a conical singularity that needs to be excised from the tiling. The identifications are shown in yellow, and the excised region is shown in white. 

\bigskip

A second approach to cellulating a toroidal compactification involves the use of local regular tessellations such as those described in section 
\ref{tes-gen} with appropriate identifications. In this case the cells are of constant proper volume, with the negative curvature captured by the structure of the cellulation. Again one will need to excise a hole in the centre of the space, reflecting the singularity as $z \rightarrow \infty$. 

The simplest class of examples relate to circular reduction of the hyperbolic plane. Consider a generic hyperbolic $\{ p, q \}$ tessellation, with a regular $p$-gon at the centre. Such a tessellation has a discrete rotational symmetry $Z_p$ associated with the symmetry group of the $p$-gon. We can construct from this a cellulation of a negative curvature space with conical singularity, by removing the central $p$-gon. One then identifies two of the vertices of the missing pentagon, which will necessarily be related by a certain $Z_p$ transformation. One then continues to remove pentagons and identify vertices related by the same $Z_p$ transformation. 

\begin{figure}[h]
\begin{center}
\includegraphics[scale=0.3]{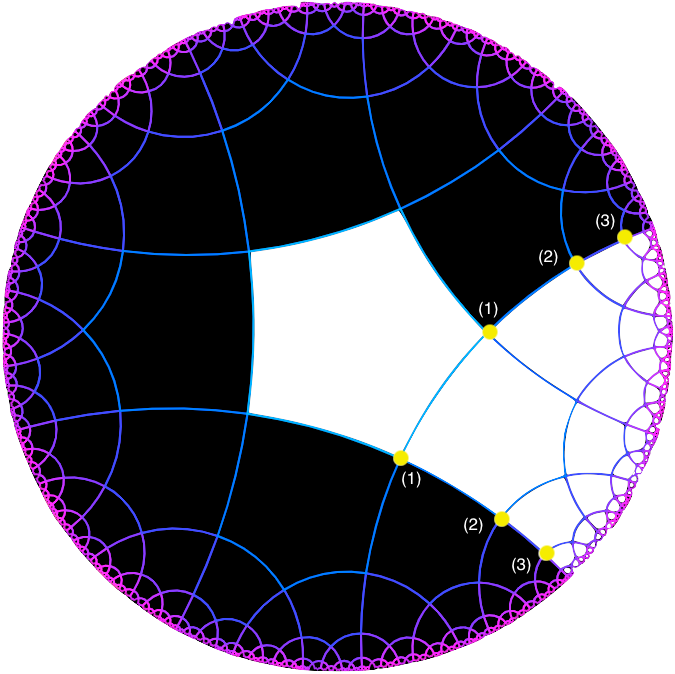}
\end{center}
\caption{\label{fig:cone} $\{5,4 \}$ tessellation, with central pentagon removed. Numbered vertices, and the edges that link them, which are all related by $Z_5$ transformations, are identified.}
\end{figure}

An example based on a $\{5,4 \}$ tessellation is shown in Figure~\ref{fig:cone}. The tessellation preserves a discrete $Z_5$ symmetry group. The numbered vertices and associated edges are related by a $Z_5$ transformation and are identified. This identification is clearly only possible if the central pentagon is removed. The resulting tiling describes a space that is locally hyperbolic but which has an excised conical singularity.  By suitable choice of $\{ p,q \}$ and discrete symmetry group, one can realise different radii of the $x$ direction. 

The generalisation to higher dimensions would work similarly. Tilings based on regular polytopes have discrete symmetry groups, which are subgroups of the continuous rotation group of the hyperbolic space. An analogous process of removing the central cell and identifying faces/edges etc that are related by a discrete symmetry transformation will result in a tessellation of a space in which one direction is compact and there is an excised conical singularity. 

\bigskip

The two approaches have complementary advantages. The first approach preserves the continuous symmetry group of the toroidal directions and therefore compactification and restriction to zero modes on the compact space are straightforward. However, the tiling is not based on a regular tessellation and therefore does not directly inherit standard geometric properties associated with tessellations. 

The second approach breaks the continuous symmetry group of the toroidal directions but is obtained from regular tessellations via quotienting and therefore inherits properties from the original tessellation. However, as the symmetry in the compact directions is broken, one cannot straightforwardly describe the tiling using only the lower-dimensional perspective. Dynamically, one often focuses on the zero mode sector of a toroidal reduction, but in this setup it would not be straightforward to identify and retain only zero modes.  We will comment on the features of codes associated with both classes of tilings in our conclusions.

\subsection{Cellulations and graphs}

In the final part of this section we review how cellulations of manifolds can be described by graphs known as Hasse diagrams. These graphs are useful in relating cellulations to certain classes of quantum error correcting codes, as we discuss in section \ref{css-tes}.

In a tessellation the polytopes are embedded via isometries as described in section \ref{tes-gen}. If we are primarily interested in embeddings in which the cells do not intersect and the overlap between the cells is invariant, embedding via isometries may be an unnecessarily strong condition. The term cellulation refers to embedding of polytopes via (injective) homeomorphisms. In a cellulation the topology of the manifold is respected but the manifold does not necessarily have a metric i.e. a notion of distance. 

Hasse diagrams are multipartite graphs that show how cells within a cellulation are connected. Each node in the diagram represents a cell in the cellulation. Two nodes can be connected to each other only if
\begin{enumerate}
\item The cell corresponding to one node is contained in the cell corresponding to the other.
\item In addition, the difference of dimensionalities of these cells is one. 
\end{enumerate}
Each level in the Hasse diagram thus corresponds to the set of all $i$-dimensional subpolytopes ($i$ cells) and the total number of levels for a D-dimensional cellulation is $D$: the polytope itself is usually not included as part of the Hasse diagram, so the levels range from $0$ (vertices) to $(D-1)$. 

\begin{figure}
\begin{center}
\begin{tikzpicture}
	\node[draw] at (-1.5,1) (u1) {$f_1$}; 
	\node[draw] at (-0.5,1) (y1) {$f_2$};
	\node[draw] at (0.5,1) (w1) {$f_3$}; 
	\node[draw] at (1.5,1) (x1) {$f_4$};
		\node[draw] at (-2.5,0) (a2) {$e_1$}; 
	\node[draw] at (-1.5,0) (b2) {$e_2$};	
		\node[draw] at (-0.5,0) (c2) {$e_3$}; 
	\node[draw] at (0.5,0) (d2) {$e_4$};
	\node[draw] at (1.5,0) (e2) {$e_5$}; 
	\node[draw] at (2.5,0) (f2) {$e_6$};	
	
	\node[draw] at (-1.5,-1) (u3) {$v_1$}; 
	\node[draw] at (-0.5,-1) (y3) {$v_2$};
	\node[draw] at (0.5,-1) (w3) {$v_3$}; 
	\node[draw] at (1.5,-1) (x3) {$v_4$};

	\path
		(u1) edge (a2)
		(u1) edge (b2)
		(u1) edge (c2)
		(y1) edge (a2)
		(y1) edge (d2)
		(y1) edge (e2)
		(w1) edge (b2)
		(w1) edge (e2)
		(w1) edge (f2)
		(x1) edge (c2)
		(x1) edge (d2)
		(x1) edge (f2)	
		(u3) edge (a2)
		(u3) edge (b2)
		(u3) edge (e2)
		 
		(y3) edge (a2)
		(y3) edge (c2)
		(y3) edge (d2)
		(w3) edge (b2)
		(w3) edge (c2)
		(w3) edge (f2)
		(x3) edge (d2)
		(x3) edge (e2)
		(x3) edge (f2)

	;
\end{tikzpicture}
\end{center}
\caption{\label{fig:hasse} Hasse diagram for the tetrahedron.}
\end{figure}
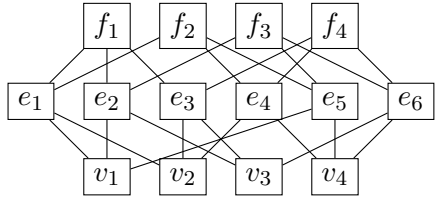

A simple example of a Hasse diagram is a tetrahedron, shown in Figure~\ref{fig:hasse}. Suppose that the faces of the tetrahedron are labelled as $(f_1,f_2,f_3,f_4)$ and its six edges are labelled as $(e_1,e_2,e_3,e_4,e_5,e_6)$, with the four vertices being $(v_1,v_2,v_3,v_4)$. The Hasse diagram shows the connections between faces, edges and vertices. Note that this diagram omits the single $3$-cell itself i.e. the tetrahedron. 

\bigskip

To connect with quantum codes, we will need to use the following property of cellulations:
\begin{itemize}
\item If $c_{i+1}$ and $c_{i-1}$ are nodes at levels $(i+1)$ and $(i-1)$, respectively, then the number of cells at level $i$ that are connected to both $c_{i+1}$ and $c_{i-1}$ is either zero or two. 
\end{itemize}
The significance of this property is that, as reviewed in \cite{Breuckmann2018PhDCode}, it relates to the existence of a map between three consecutive levels of a cellulation and a CSS (Calderbank, Shor and Steane) code:
\begin{itemize}
\item Any subgraph of a Hasse diagram consisting of three consecutive levels defines the Tanner graph of a CSS code. 
\end{itemize}
The key properties of CSS stabiliser codes are summarised in Appendix A, and we will explain Tanner graphs further in section \ref{css-tes}. 

\section{Connecting tessellations with error correcting codes}
\label{sec:qecc}

In this section we will discuss how tessellations of hyperbolic spaces can be related to quantum error correcting codes, explaining the approach of HaPPY \cite{Pastawski2015HolographicCorrespondence}, focusing on aspects of the construction that are key in generalising to higher dimensions. Throughout this section we will reference various well known properties of quantum error correcting codes which are for convenience summarised in appendices. To generalise the construction to higher dimensions we will in particular need to explore in detail the properties of absolutely maximally entangled states and how these relate to the tessellation used. 


\subsection{Stabiliser construction of HaPPY codes in two spatial dimensions} \label{stabconhappy}

Perhaps the best known example of a holographic code is the HaPPY code proposed in \cite{Pastawski2015HolographicCorrespondence}. This code is based on a regular $\{ p,q \}$ tessellation of hyperbolic space viewed as a spatial section of $AdS_{3}$. The basic premiss of the code is the following. Each of the polygons is associated with a certain {\it perfect tensor} structure. Perfect tensors have an even number of indices and symmetry properties that relate to the entanglement structure of the associated quantum state/code. The main example used in \cite{Pastawski2015HolographicCorrespondence} is a $[[5, 1, 3]]$ stabilizer code, encoding one logical qubit into five physical qubits with code distance three. 

\begin{figure}[h]
\begin{center}
\includegraphics [scale =0.3] {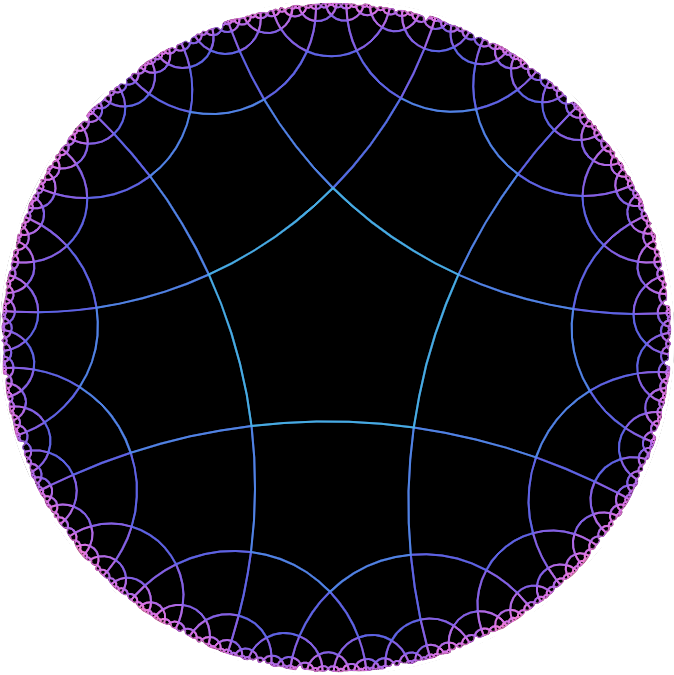}
\end{center}
\caption{\label{fig:54} $\{5,4\}$ Tessellation of hyperbolic plane. Associated with each pentagon there is one logical qubit and five physical qubits, each of which is associated with an edge.}
\end{figure}


Before we describe the HaPPY construction in detail, let us summarise the main geometric features. The realisation is based on a $\{ 5, 4 \}$ tessellation, shown in Figure~\ref{fig:54}. For every pentagon there is a ``logical'' qubit associated with the entire pentagon while edges connecting vertices are associated with five ``physical" qubits. The conformal boundary of the spatial section is regulated and the edges intersecting the regulated boundary are associated with the physical qubits in a discretisation of the boundary theory. 

\begin{figure}[h]
\begin{center}
\includegraphics [scale =0.35] {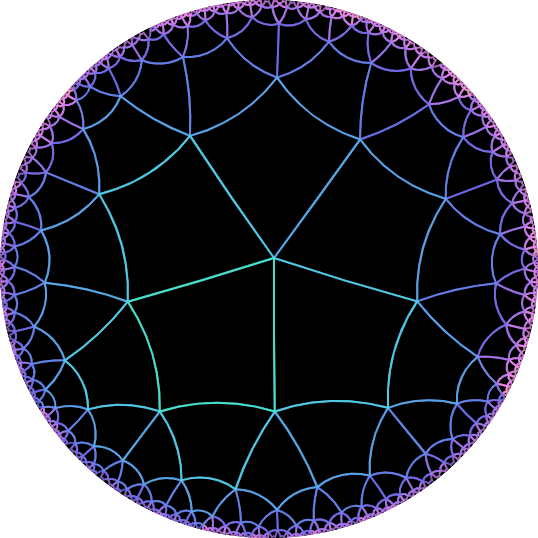}
\end{center}
\caption{\label{fig:45} $\{4,5\}$ Tessellation of hyperbolic plane. In the HaPPY code each node represents the uncontracted logical qubit leg of the perfect tensor, and each edge represents the contraction of physical qubit tensor legs.}
\end{figure}

There is a rank six perfect tensor describing the code and this is shown in the dual $\{ 4,5 \}$ tessellation, shown in Figure~\ref{fig:45}. Each node in the $\{ 4,5 \}$ tessellation is associated with an uncontracted leg of the tensor (the logical qubit). The legs of the tensor associated with the physical qubits are contracted against the corresponding legs in the neighbouring pentagon; in the $\{4,5 \}$ tessellation these are the edges linking nodes. 

\bigskip

\begin{figure}[h]
\begin{center}
\begin{tikzpicture}[mystyle/.style={draw,shape=circle,fill=black, inner sep=0pt, minimum size=4pt}]
\def\ngon{5}
\node[draw, regular polygon,regular polygon sides=\ngon,minimum size=3cm] (p) {};
\foreach\x in {1,...,\ngon}{
    \node[mystyle=\x] (p\x) at (p.side \x){};
};
\node[style={draw,shape=circle,fill=red, inner sep=0pt, minimum size=4pt}=p.center] at (p.center) {}; 
\end{tikzpicture}
\caption{\label{pent}{A single pentagon constructed using the HaPPY approach associated with a single logical qubit (that one diagrammatically represents by placing in it's centre) and five physical qubits (one associated with each edge).}}
\end{center}
\end{figure}
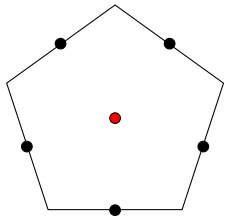

We next explore in detail the construction presented in \cite{Pastawski2015HolographicCorrespondence} from first principles; the details of this construction are required in developing higher dimensional generalisations in the following section. 
We consider fixed time slices of the bulk space AdS$_{3}$ corresponding to the two-dimensional hyperbolic plane H$^{2}$. The models presented are then realisations of regular, uniform hyperbolic tessellations that discretise H$^{2}$ in a maximally symmetric fashion. The HaPPY construction utilises the pentagonal tiling of two-dimensional hyperbolic space, given by Schl\"{a}fi symbol $\{5,4\}$. For each pentagon, one places a single physical qubit on each edge. A final qubit is placed in the centre of the pentagon which can later be shown to have the interpretation of a logical qubit, see Figure~\ref{pent}. 

\bigskip

More generally, for each such system of this type, one associates a graph $G = (V,E)$, where $V = \{1,\dots,n\}$ is a finite set of vertices and $E$ are the corresponding edges such that $n=5$ for the pentagon. The vertices of this graph are chosen to be the qubits in the system. In order to preserve maximal symmetry, the graph is chosen such that neighbouring qubits on the pentagon are connected via an edge. The central qubit of the pentagon is chosen to be connected to every other qubit in the pentagon. Hence our graph looks as shown in Figure~\ref{maxsym}. 

\begin{figure}[h]
\centering
\begin{minipage}[b]{0.4\textwidth}
\centering
\begin{tikzpicture}[mystyle/.style={draw,shape=circle,fill=black, inner sep=0pt, minimum size=4pt}]
\def\ngon{5}
\node[draw, regular polygon,regular polygon sides=\ngon,minimum size=3cm] (p) {};
\foreach\x in {1,...,\ngon}{
    \node[mystyle=\x] (p\x) at (p.side \x){};
};
\foreach\x in {1,...,\ngon}{\draw (p.side \x) -- (p.center) {};
};
\draw (p.side 1) -- (p.side 2) {};
\draw (p.side 2) -- (p.side 3) {};
\draw (p.side 3) -- (p.side 4) {};
\draw (p.side 4) -- (p.side 5) {};
\draw (p.side 5) -- (p.side 1) {};
\node[style={draw,shape=circle,fill=black, inner sep=0pt, minimum size=4pt}=p.center] at (p.center) {};
\end{tikzpicture}
\caption{\label{maxsym} {The maximally symmetric construction of graph $G$ for a single pentagon.}}
\end{minipage}
\hfill
\begin{minipage}[b]{0.4\textwidth}
\centering
\begin{tikzpicture}[mystyle/.style={draw,shape=circle,fill=black, inner sep=0pt, minimum size=4pt}]
\def\ngon{5}
\node[draw, regular polygon,regular polygon sides=\ngon,minimum size=3cm] (p) {};
\foreach\x in {1,...,\ngon}{
    \node[mystyle=\x] (p\x) at (p.corner \x){};
};
\foreach\x in {1,...,\ngon}{\draw (p.corner \x) -- (p.center) {};
};
\node[style={draw,shape=circle,fill=black, inner sep=0pt, minimum size=4pt}=p.center] at (p.center) {}; 

\end{tikzpicture}
\caption{The resulting graph (state) to which $G$ stabilisers are assigned.}
\label{AMEpentagon}
\end{minipage}
\end{figure}

\subsection{Graph States and AME states}
\label{GSAME}

In this section we discuss the relation between graphs and absolutely maximally entangled states. Given a graph $G = (V,E)$ representing the system, one can define a graph state $\ket{G}$. One will first construct this in terms of the controlled-Z gate operator $CZ_{ij}$, defined in \ref{contZ}, before noticing it can be extended to have an interpretation in the stabiliser formalism. For any graph $G$ consisting of $n$ vertices (system of $n$ qudits), the corresponding graph state $\ket{G} \in \mathcal{H}^{\otimes n}$ is defined by
\be
\ket{G} := \prod_{i>j} CZ_{ij}^{A_{ij}} \ket{+}^{\otimes n}
\ee 
where $A_{ij} \in \mathbbm{Z}_p$ are called weights and form the $n \times n$ adjacency matrix $A \in \mathbbm{Z}_p^{n \times n}$. Here, $p$ is prime and the Hilbert space $\mathcal{H} \cong \mathbbm{C}_p$. The adjacency matrix encapsulates all the relevant information about the connectivity of the graph, such that a weight zero operator represents no edge joining the vertices $i$ and $j$ and weight one operators represent a single edge connecting them and so on. The initial prepared state $\ket{+}^{\otimes n}$ represents joint $+1$ eigenstate of $X_i$ arising from the $X$-eigenbasis
\be
\ket{+}=F^\dagger \ket{0} = \frac{1}{\sqrt{D}} \sum_{l=0}^{D-1} \omega^{-0l} \ket{l},
\ee
where $F$ is the Fourier gate. For example in the most simple case, where one considers qubits; $\ket{+}=(\ket{0} + \ket{1})/\sqrt{2}$, the Fourier gate reduces to the Hadamard gate
\be
H = \frac{1}{\sqrt{2}} \begin{bmatrix} 1 & 1 \\ 1 & -1 \end{bmatrix}
\ee
and the controlled-Z operator is $CZ=\ket{00}\bra{00} + \ket{01}\bra{01} + \ket{10}\bra{10} - \ket{11}\bra{11}$.

\bigskip
 
Since graph states are just a special class of stabiliser states, an alternative but equivalent way of defining a graph state is possible using the generators of the stabiliser group. Any $n$-qudit graph state may be represented as $n$ operators,
\be
\label{graphstab}
g_i = X_i \prod_{j=1}^n Z_j^{A_{ij}},
\ee
which provide the minimal set of generators of the stabiliser group. The graph state can then be thought of as the common $+1$ eigenspace of these generators;
\be
g_i \ket{G} = \ket{G}, \; \text{for} \; i=1,\dots,n.
\ee
One should note that for the case of qubits, graphs can only have edges of weight 0 or 1. This is because if one imposes a weight 2 edge, $Z^2 = Z^0 = \mathbbm{1}$ and is the same as a weight zero operator. When extending to qudits of a more general dimension $D$, one now has $Z^D = \mathbbm{1}$ and so one can have weights $A_{ij}$ from 0 to $D-1$. This allows for graphs to have multiple edges between vertices as indicated in Figures \ref{singleline} and \ref{doubleline}. The only constraints on each adjacency matrix element $A_{ij}$ are that these weights are symmetric (i.e. $A_{ij} = A_{ji}$) and that there are no weights connecting a vertex to itself $A_{ii}$ or in other words, there are no loops present in $G$.

\begin{figure}[h]
\centering
\begin{minipage}[b]{0.4\textwidth}
\centering
\begin{tikzpicture}[mystyle/.style={draw,shape=circle,fill=black}]
  \node[mystyle] at (0,0) (i) {};
  \node[mystyle] at (3,0) (j) {};
  \draw (i) -- (j);
  \node at (-0.5,0) (ii) {i};
   \node at (3.5,0) (jj) {j};
\end{tikzpicture}
\caption{A single $Z$ operator between 2 nodes corresponds to a single edge in the graph state.}
\label{singleline}
\end{minipage}
\hfill
\begin{minipage}[b]{0.4\textwidth}
\centering
\begin{tikzpicture}[mystyle/.style={draw,shape=circle,fill=black}]
  \node[mystyle] at (0,0) (i) {};
  \node[mystyle] at (3,0) (j) {};
  \draw[transform canvas={yshift=-1.5pt}] (i) -- (j);
  \draw[transform canvas={yshift=1.5pt}] (j) -- (i);
  \node at (-0.5,0) (ii) {i};
   \node at (3.5,0) (jj) {j};
\end{tikzpicture}
\caption{$Z^2$ operator between two nodes. For qubits $Z^2 = \mathbbm{1}$ this is the same as the two nodes being disconnected.}
\label{doubleline}
\end{minipage}
\end{figure}

\bigskip

The generalised Clifford group $\mathcal{C}_n$ is the (group-theoretic) normaliser of the Pauli group $\mathcal{P}_n$. In other words it is the group of unitary operators $U$ which map the Pauli group onto itself: $U \mathcal{P}_n U^\dagger = \mathcal{P}_n$. Further, the local Clifford group $\mathcal{C}_n^l \subseteq \mathcal{C}_n$ is the $n$-fold tensor product of the Clifford group of order one ($\mathcal{C}=\mathcal{C}_1$). When considering the Pauli group acting on qubits, the Clifford group is simply generated by the Hadamard gate $H$, the phase gate $P$ and the CNOT gate $U_{CNOT}$. One also notes that under conjugation, a unitary operator $U$ that fixes the stabiliser group $S$ of a quantum error correcting code is an encoded operation. Hence, $S' = USU^\dagger$ implies that $\ket{c'} = U\ket{c}$ is a codeword stabilised by every stabiliser element in $S'$, where $\ket{c}$ is stabilised by every element in $S$.

\bigskip

As proven in \cite{Bahramgiri2006}, two stabiliser states with generator matrices $A$ and $B$ are equivalent under the action of the local Clifford group if and only if there exist invertible matrices $U$ and $Y$ such that $B=UAY$ and where $Y$ can be represented as
\be
Y = \begin{pmatrix} E & F \\ E' & F' \end{pmatrix}
\ee
where 
\be
E = \text{diag}(e_1,\dots, e_n), \quad F =  \text{diag}(f_1,\dots, f_n),
\ee
\be
E' = \text{diag}(e'_1,\dots, e'_n), \quad F' =  \text{diag}(f'_1,\dots, f'_n),
\ee
and $e_i f'_i - f_i e'_i =1 \; \forall i$. Another important result of \cite{Bahramgiri2006} is that every stabiliser state is equivalent under the action of the local Clifford group to a graph state. This means that one only needs to consider graph states when considering entanglement properties of stabiliser states since for any stabiliser state there will exist an equivalent graph state that shares the same entanglement properties.

\bigskip

Since graph states are a subclass of the stabiliser states, then these arguments trivially hold for all graph states. Diagrammatically, one can claim that two graph states are equivalent under local Clifford transformations if and only if one there exists a sequence consisting of the following operations on a vertex $v$, such that one can obtain one graph state from the other;
\begin{enumerate}[label=(\roman*)]
\item One multiplies the weight of each edge connected to vertex $v$ by $b \in \mathbb{Z}_p$ and $b \neq 0$.
\item One transforms the elements of the adjacency matrix as $A_{jk} \rightarrow A_{jk} + a A_{vj} A_{vk}$ where $a\in \mathbb{Z}_p$ and $j\neq k$.
\end{enumerate}

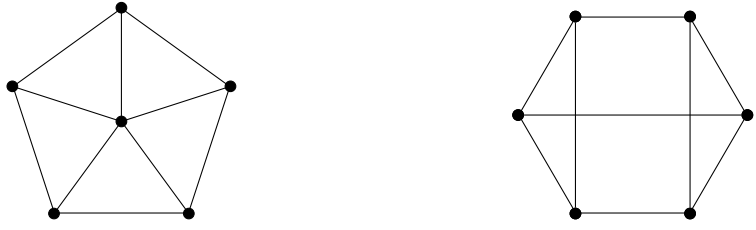
\begin{figure}[h]
\centering
\begin{minipage}[b]{0.4\textwidth}
\centering
\begin{tikzpicture}[mystyle/.style={draw,shape=circle,fill=black, inner sep=0pt, minimum size=4pt}]
\def\ngon{5}
\node[draw, regular polygon,regular polygon sides=\ngon,minimum size=3cm] (p) {};
\foreach\x in {1,...,\ngon}{
    \node[mystyle=\x] (p\x) at (p.corner \x){};
};
\foreach\x in {1,...,\ngon}{\draw (p.corner \x) -- (p.center) {};
};
\node[style={draw,shape=circle,fill=black, inner sep=0pt, minimum size=4pt}=p.center] at (p.center) {}; 

\end{tikzpicture}
\end{minipage}
\begin{minipage}[b]{0.4\textwidth}
\centering
\begin{tikzpicture}[mystyle/.style={draw,shape=circle,fill=black, inner sep=0pt, minimum size=4pt}]
\def\ngon{6}
\node[draw, regular polygon,regular polygon sides=\ngon,minimum size=3cm] (p) {};
\foreach\x in {1,...,\ngon}{
    \node[mystyle=\x] (p\x) at (p.corner \x){};
};
\foreach\x in {1,...,\ngon}{
    \node[mystyle=\x,label={[font=\small,text=blue]$$}]  (p\x) at (p.corner \x) (corner\x) {};
};
\draw (corner1) -- (corner5);
\draw (corner2) -- (corner4);
\draw (corner3) -- (corner6);

\end{tikzpicture}
\end{minipage}
\caption{Two possible graph states that are locally Clifford equivalent to AME(6,2). The two are related by a local complementation.}
\label{complement}
\end{figure}

When considering qubits, operation (i) is always just the identity. For $a=1$, the second operation (ii) is known as local complementation and has been extensively studied \cite{Hein2003Multi-partyStates, Nest2003}. An example of a local complementation of a graph state is shown in Figure \ref{complement}. Here the previous graph state, locally Clifford equivalent to AME(6,2) is considered (left graph in Figure \ref{complement}) and after applying a local complementation, one obtains a new graph state (right graph in Figure \ref{complement}). This new graph state is then still locally Clifford equivalent to AME(6,2) as can be checked using its stabiliser generators. 

\bigskip

Key features of the $\{5,4\}$ tessellation relate to the graph state given in Figure \ref{AMEpentagon} being an absolutely maximally entangled state represented by the notation AME$(6,2)$. For those unfamiliar with absolutely maximally entangled (AME) states, a brief review is detailed in section \ref{AMEstates} or, for more detailed discussions, see the expansive literature \cite{HelwigAMESecret, Helwig2013AbsolutelyApplications, Helwig2013AbsolutelyStates, Huber2016AbsolutelyExist, Goyeneche2015AbsolutelyMatrices}. There are a number of known AME states that have been shown to be locally Clifford equivalent to graph states.  A mechanism was developed in \cite{Helwig2013AbsolutelyStates} for determining bipartite entanglement in graph states and hence being able to determine whether a graph state corresponds to an AME state. 

\subsection{Absolutely Maximally Entangled States}
\label{AMEstates}

In this section we define an absolutely maximally entangled (AME) state through several equivalent definitions and we summarise a method for testing whether a graph state is AME. Our discussion provides a concise summary of the key results in \cite{Helwig2013AbsolutelyApplications, Helwig2013AbsolutelyStates} and more details can be found in these works.

\bigskip

\textbf{Definition:}
\textit{An absolutely maximally entangled state $\ket{\Phi} \in  \mathcal{H}$ is a pure state consisting of $N$ qudits of local dimension $D$, with the total Hilbert space $\mathcal{H} \cong (\mathbb{C}^D)^{\otimes N}$. Therefore, $\ket{\Phi} \in \mathcal{H}_1 \otimes \mathcal{H}_2 \otimes \cdots \otimes \mathcal{H}_N$ where $\mathcal{H}_i \cong \mathbb{C}^D$, such that it satisfies the following equivalent properties.}

\begin{enumerate}
\item \textit{For any possible bipartition of $P = \{1,\dots,N \}$ into disjoint sets $A$ and $B$ such that $A\cup B = P$, $\ket{\Phi}$ is maximally entangled. That is, $\ket{\Phi}$ can be expressed as 
\begin{equation}
\ket{\Phi} = \frac{1}{\sqrt{D^m}} \sum_{k \in \mathbbm{Z}^m_D} \ket{k_1}_{B_1} \cdots \ket{k_m}_{B_m} \ket{\phi(k)}_A,
\end{equation}
where one has assumed $m = |B| \leq |A| = N - m$ without loss of generality and $\braket{\phi(k)|\phi(k')} = \delta_{k k'}$.}

\item \textit{Every possible subset of parties $A \subset P$ with $|A| = \lfloor \frac{N}{2} \rfloor$ gives rise to a reduced density matrix that is totally mixed; $\rho_A = D^{-\lfloor \frac{N}{2} \rfloor} \mathbbm{1}_{D^{\lfloor\frac{N}{2} \rfloor}}$.}
\item \textit{Every possible subset of parties $A \subset P$ with $|A| \leq \frac{N}{2}$ gives rise to a reduced density matrix that is totally mixed.}
\item \textit{Every possible subset of parties $A \subset P$ with $|A| = \lfloor \frac{N}{2} \rfloor$ gives rise to a von Neumann entropy that is maximal, $S(A) = \lfloor \frac{N}{2} \rfloor \log D$.}
\item \textit{Every possible subset of parties $A \subset P$ with $|A| \leq \frac{N}{2}$ gives rise to a von Neumann entropy that is maximal, $S(A) = |A| \log D$.}

\textit{One represents absolutely maximally entangled states defined in this way using the notation AME($N,D$).}

\end{enumerate}

The method used to distinguish whether a graph state is AME can be explained in the following way. This method was presented in \cite{Helwig2013AbsolutelyStates}; for a more detailed discussion, we refer the reader to the original paper. In order to proceed,  we need to make use of the following definitions:

\bigskip

\textbf{Definition:} \textit{Consider the graph state $\ket{G} \in \mathcal{H}^{\otimes n}$ shared between a set of parties $P$. Then for some subset $K \subset P$, the state represented by the graph $G$ such that all vertices in $K$, and all edges that are connected to the parties in $K$, are removed, define the truncated graph state $\ket{G^{\backslash K}}$.}

\bigskip

\textbf{Definition:} \textit{The i-th row of the $n \times n$ adjacency matrix $A \in \mathbb{Z}_p^{n \times n}$ can be denoted $A_i = (A_{i1}\dots A_{in})$. One defines the quantity $A_i\backslash K$ to be the row vector $A_i$ with entries $\{A_{ik_1}\dots A_{ik_m}\}$ removed such that the elements of the subset $K = \{k_1, k_2, \dots k_m\}$ are between $1$ and $n$.}

\bigskip

Now suppose we have a graph state $\ket{G}$ with adjacency matrix $A$ and the party subsets $K= \{k_1, k_2, \dots k_m\}$ with $m = \lfloor \frac{n}{2} \rfloor$. Representing $A_{k_i} \backslash K$ as the $k_i$-th row of the adjacency matrix with elements $\{A_{k_i k_1} \dots A_{k_i k_m} \}$ removed, we can show the graph state is absolutely maximally entangled if and only if the vector quantity $A_{k_i}\backslash K$ are linearly independent in $\mathbb{Z}_p^{n-m}$. 

\bigskip

We now turn to the AME(6,2) state and use its corresponding graph state, given in Figure \ref{AMEpentagon}, to show that this state is absolutely maximally entangled. This serves as a concrete example of how one can check whether a graph state is AME and develop understanding of the method involved. 

The first question to ask is which bipartitions must be considered. Bipartitions for which one set has just a single node are trivial and will clearly provide linearly independent $A_{k_i} \backslash K$. Hence the only bipartitions that need to be considered are all possible combinations of sets of two nodes (corresponding to four nodes in the other set in the bipartition) and sets of three nodes (corresponding to another three nodes in the other set in the bipartition). One must consider every possible combination of nodes for each of these cases; however using the symmetries in the geometry it is obvious that many of these cases are analogous to each other. For brevity, we only give the example of a single one of these bipartitions visually here to outline the method; similar graphs may be drawn for all other possible bipartitions with $m=\lfloor \frac{n}{2} \rfloor$.

\bigskip

\begin{figure}
\centering
\begin{tikzpicture}[mystyle/.style={draw,shape=circle,fill=black, inner sep=0pt, minimum size=4pt},scale=1.5, transform shape,  every node/.style={draw},every newellipse node/.style={inner sep=0pt}]
\def\ngon{5}
\node[draw, regular polygon,regular polygon sides=\ngon,minimum size=3cm] (p) {};
\foreach\x in {1,...,\ngon}{
    \node[mystyle=\x, label={[font=\small,text=blue]$\x$}] (p\x) at (p.corner \x) (corner\x){};
};
\foreach\x in {1,...,\ngon}{\draw (p.corner \x) -- (p.center) {};
};
\node[style={draw,shape=circle,fill=black, inner sep=0pt, minimum size=4pt}, label={[font=\small,text=blue]$6$}] at (p.center) (center) {}; 
 \path (corner1) -- (center) -- (corner2);
 \node[fit=(corner1)(center)(corner2), red, newellipse, xscale=1.1,yscale=1.1, label={[font=\small,text=red, below, label distance = -5]$K$}] {};

\end{tikzpicture}
\caption{Bipartition of AME(6,2) into $K=\{1,2,6\}$ and $L=\{3,4,5\}$.}
\label{62isAME}
\end{figure}
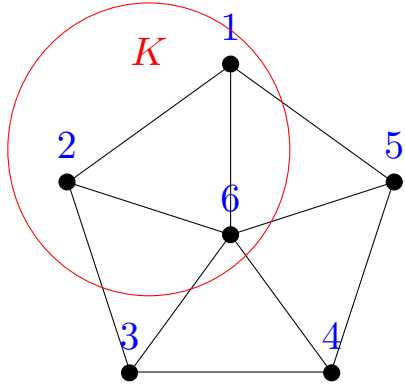

Consider the graph state given in Figure \ref{AMEpentagon}. Here we choose to bipartition the sets into $K=\{1,2,6\}$ and $L=\{3,4,5\}$, where $K$ has been explicitly drawn in Figure \ref{62isAME}. Calculating the relevant vectors from the edges connecting nodes in $K$ and those in $L$
\begin{align}
&A_1 \backslash\{1,2,6\} = (0,0,1) \\
&A_2  \backslash\{1,2,6\} = (1,0,0) \\
&A_6  \backslash\{1,2,6\} = (1,1,1).
\end{align}
These three vectors are clearly linearly independent of one another. Due to the symmetries of the geometry, we would obtain this result for any bipartition consisting of two nodes that are next to each other on the outer pentagon and the centre node. There are obviously more possible bipartitions into two sets of three but these can all be reduced to one of the following due to the symmetry; $K=\{1,2,3\}$ and $L=\{4,5,6\}$, $K=\{1,3,6\}$ and $L=\{2,4,5\}$ or $K=\{1,2,4\} $ and $L=\{3,5,6\}$. The vectors obtained from these cases are given in Table \ref{amedivisions},
where we have grouped the three cases into the three columns of equations. Note that each column's vectors are linearly independent and so all bipartitions that split the graph into two sets of three give the correct result.

\begin{table}[h!]
\begin{center}
\begin{tabular}{ |c|c|c| } 
\hline
Case 1: $K=\{1,2,3\}$ & Case 2: $K=\{1,3,6\}$  & Case 3:  $K=\{1,2,4\} $\\
\hline
$A_1 \backslash\{1,2,3\} = (0,1,1)$ & $A_1 \backslash\{1,3,6\} = (1,0,1)$ & $A_1 \backslash\{1,2,4\} = (0,1,1)$ \\
$A_2  \backslash\{1,2,3\} = (0,0,1)$ & $A_3 \backslash\{1,3,6\} = (1,1,0)$ & $A_2 \backslash\{1,2,4\} = (1,0,1)$ \\
$A_3  \backslash\{1,2,3\} = (1,0,1)$ & $A_6 \backslash\{1,3,6\} = (1,1,1)$ & $A_4 \backslash\{1,2,4\} = (1,1,1)$ \\
\hline
\end{tabular}
\end{center}
\caption{Remaining bipartitions of AME(6,2) where $K$ and $L$ both contain 3 elements.}
\label{amedivisions}
\end{table}
 
 Now we consider the case of bipartitions into one set of two elements and one set of four elements. All possible cases can once again be reduced due to symmetry, in this case, to three simple bipartitions;  $K=\{1,2\}$ and $L=\{3,4,5,6\}$, $K=\{1,3\}$ and $L=\{2,4,5,6\}$ or $K=\{1,6\} $ and $L=\{2,3,4,5\}$. The vectors are presented in Table \ref{amedivisions2}, where they have once again been placed in columns for each bipartition, and in each column we trivially have linearly independent vectors for each possible bipartition. Hence for all possible bipartitions of the AME(6,2) graph state given in Figure \ref{AMEpentagon}, each $A_{k_i} \backslash K$ is linearly independent for all choices of $K=\{k_1,\dots, k_m\}$ with $m=\lfloor \frac{n}{2} \rfloor$. Therefore the state is indeed absolutely maximally entangled. 

\begin{table}[h!]
\begin{center}
\begin{tabular}{ |c|c|c| } 
\hline
Case 1: $K=\{1,2\}$ & Case 2: $K=\{1,3\}$  & Case 3:  $K=\{1,6\} $\\
\hline
$A_1 \backslash\{1,2\} = (0,0,1,1)$ & $A_1 \backslash\{1,3\} = (1,0,1,1)$ & $A_1 \backslash\{1,6\} = (1,0,0,1)$ \\
$A_2 \backslash\{1,2\} = (1,0,0,1)$ & $A_3 \backslash\{1,3\} = (1,1,0,1)$ & $A_6 \backslash\{1,6\} = (1,1,1,1)$ \\
\hline
\end{tabular}
\end{center}
\caption{Remaining bipartitions of AME(6,2) where $K$  contains 2 elements and $L$ contains 3 elements.}
\label{amedivisions2}
\end{table}

Given the graph state formulation of the AME(6,2) state, one may illustrate the importance of using AME states and connecting them with quantum error correcting codes. Firstly, it is well known that any AME state AME$(N,D)$ is equivalent to a pure $[[N, 0, \lfloor N/2 \rfloor +1]]_D$ quantum error correcting code \cite{HuberGrassl2019}. Further, every pure $[[N, k, d]]$ code with $N, d \ge 2$ gives rise to a family of pure codes $[[N - i, k+i, d-i]]$ where $i = \{0,\dots d-1\}$ \cite{RainsQWE}. From the graph in Figure \ref{AMEpentagon}, one may describe the full stabiliser group by a column vector of stabiliser generators found using \ref{graphstab};
\be
\label{ame62}
\mathcal{S} = \begin{bmatrix}G_1\\G_2\\G_3\\G_4\\G_5\\G_6 \end{bmatrix} = \begin{bmatrix} X&Z&I&I&Z&Z\\Z&X&Z&I&I&Z\\I&Z&X&Z&I&Z\\I&I&Z&X&Z&Z\\Z&I&I&Z&X&Z\\Z&Z&Z&Z&Z&X \end{bmatrix} \simeq \begin{bmatrix} X&Z&Z&X&I&I\\I&X&Z&Z&X&I\\X&I&X&Z&Z&I\\Z&X&I&X&Z&I\\X&X&X&X&X&X\\Z&Z&Z&Z&Z&Z \end{bmatrix} = \mathcal{S}'
\ee
where in the second step one has used basic matrix manipulations including Gaussian elimination as well as performing a transformation under the action of the local Clifford group on the final qubit. This particular transformation came using the Hadamard gate which has the following effect on the Pauli matrices; $X\rightarrow HXH^\dagger =Z$, $Y\rightarrow HYH^\dagger =-Y$, $Z\rightarrow HZH^\dagger =X$. The form given by $\mathcal{S}'$ in \ref{ame62} is the most common presentation of the stabiliser generators for AME(6,2) in the literature and is the one presented in \cite{Pastawski2015HolographicCorrespondence}. Thus, for clarity, the graph state in Figure \ref{AMEpentagon} is locally Clifford equivalent to AME(6,2) as indicated by the use of the symbol `$\simeq$'. To make the connection with the perfect tensor construction in HaPPY, we rewrite $\mathcal{S'}$ in the following way;
\be
\label{ame62iscode}
\mathcal{S}' = \begin{bmatrix} -Y&Z&Y&I&I&Z\\-Z&X&Z&I&I&X\\Y&Y&Z&I&Z&I\\Z&Z&X&I&X&I\\-Z&Y&Y&Z&I&I\\X&Z&Z&X&I&I \end{bmatrix}, \quad \begin{matrix} (3\cdot 6) \\ (2\cdot3\cdot4\cdot5) \\ (1\cdot4) \\ (1\cdot2\cdot3\cdot4) \\ (1\cdot3\cdot4) \\ (1) \end{matrix}
\ee
where we have used row multiplication, indicated by the bracketed expression to the right of each stabiliser generator where each number denotes the original stabiliser generator multiplied. Note that due to the fact operators are implicitly connected via tensor products we have simply pulled any overall negative sign of each generator to the front. 

\bigskip

The stabiliser matrix in the form (\ref{ame62iscode}) can be shown to correspond to a quantum error correcting code by removing the last qubit, as detailed in \cite{Gottesman:1997zz}, converting an $[[n,k,d]]$ code into an $[[n-1, k+1, d-1]]$. In the specific example above this translates a $[[6,0,4]]$ state into a $[[5,1,3]]$ code. To summarise the construction, one chooses $n-k$ generators such that $G_1$ ends in $Z$ and $G_2$ ends in $X$ with the remaining generators $G_3, \dots, G_{n-k}$ ending with the identity $I$. Dropping $G_1$ and $G_2$ allows a new stabiliser to form from the final $n-k-2$ generators. Now, restricting $G_1$ and $G_2$ to act only on the first $n-1$ qubits one can show these now become the logical operators $\bar{Z}$ and $\bar{X}$. The resulting $[[5,1,3]]$ code is expressed in Table \ref{ametable:1}.

\begin{table}[h!]
\begin{center}
\begin{tabular}{ |c|c|c| } 
\hline
$G'_1$ & $YYZIZ$ \\
$G'_2$ & $ZZXIX$ \\
$G'_3$ & $-ZYYZI$ \\
$G'_4$ & $XZZXI$ \\
\hline
$\bar{X}_1$ & $-ZXZII$ \\ 
$\bar{Z}_1$ & $-YZYII$ \\ 
\hline
\end{tabular}
\end{center}
\caption{The $[[5,1,3]]_2$ code derived from the AME$(6,2)$ state.}
\label{ametable:1}
\end{table}

Trivially, from the form presented in (\ref{ame62iscode}), one could continue this process converting the $[[5,1,3]]$ code into a $[[4,2,2]]$ code (given in Table \ref{ametable:2}) and then even into a $[[3,3,1]]$ code (given in Table \ref{ametable:3}). Note that after the case of the $[[3,3,1]]$ code, one can not reduce this further since no stabiliser generators remain.

\begin{table}[h!]
\begin{center}
\begin{tabular}{ |c|c|c| } 
\hline
$G''_1$ & $-ZYYZ$ \\
$G''_2$ & $XZZX$ \\
\hline
$\bar{X}'_1$ & $-ZXZI$ \\ 
$\bar{X}'_2$ & $ZZXI$ \\ 
$\bar{Z}'_1$ & $-YZYI$ \\ 
$\bar{Z}'_2$ & $YYZI$ \\ 
\hline
\end{tabular}
\end{center}
\caption{The $[[4,2,2]]_2$ code derived from the $[[5,1,3]]_2$ code.}
\label{ametable:2}
\end{table}

\begin{table}[h!]
\begin{center}
\begin{tabular}{ |c|c|c| } 
\hline
$\bar{X}''_1$ & $-ZXZ$ \\ 
$\bar{X}''_2$ & $ZZX$ \\ 
$\bar{X}''_3$ & $XZZ$ \\ 
$\bar{Z}''_1$ & $-YZY$ \\ 
$\bar{Z}''_2$ & $YYZ$ \\ 
$\bar{Z}''_3$ & $-ZYY$ \\ 
\hline
\end{tabular}
\end{center}
\caption{The $[[3,3,1]]_2$ code derived from the $[[4,2,2]]_2$ code.}
\label{ametable:3}
\end{table}

\bigskip

While any AME state AME$(N,D)$ is equivalent to a pure $[[N, 0, \lfloor N/2 \rfloor +1]]_D$ quantum error correcting code, the concept of the resultant quantum error correcting code producing further codes was explored in \cite{Mazurek2019QuantumStates}. The result is that, every $N$-qudit stabiliser AME state generates at least $\left \lfloor{\frac{N}{2}}\right \rfloor$ different stabiliser codes. In each of these codes, $m \in \{1,\dots, \left \lfloor{\frac{N}{2}}\right \rfloor \}$ logical qudits are encoded into $N-m$ physical qudits.

\bigskip

\subsection{Perfect Tensors}

We now explain the relationship between AME states and perfect tensors, an important relation in the construction of the holographic codes presented in \cite{Pastawski2015HolographicCorrespondence}. This provides the method used to concatenate AME states across the entire tessellation, forming a tensor network. This construction provides an isometric mapping from the uncontracted legs in the bulk to the uncontracted legs at the boundary of the geometric manifold and will be explained further in Section \ref{concat}. To begin we define a perfect tensor.

\bigskip

Perfect tensors naturally arise when considering AME states. One can always decompose an AME state into two subsystems $\mathcal{H}_A$ and $\mathcal{H}_B$ as
\be
\ket{\Phi} = \sum_{a,b} T_{ab} \ket{a} \ket{b},
\ee
such that $T:\mathcal{H}_A \rightarrow \mathcal{H}_B$ is a unique linear map between the two Hilbert spaces represented by a two-index tensor with 
\be 
T: \ket{a} \rightarrow \sum_{b} \ket{b} T_{ba}.
\ee
Here one denotes the complete orthonormal basis $\{\ket{a}\}$ for $\mathcal{H}_A$ and similarly, $\{\ket{b}\}$ for $\mathcal{H}_B$. This map also preserves all inner products; i.e.
\be
\sum_{b} T^\dagger_{a'b} T_{ba} = \delta_{a'a}.
\ee
Hence, each of these linear transformations applied to an AME state may be thought of as an isometry. Thus one defines the tensors associated with the isometry, isometric tensors. Similarly, the converse is true. Provided one has a perfect tensor, then one similarly has a corresponding AME state.

\bigskip 

\textbf{Definition:} \textit{Suppose one bipartitions the indices of a $2n$-index tensor $T_{a_1 a_2 \dots a_{2n}}$ into one set $A$ and its complementary set $A^c$ with $|A| \leq |A^c|$. Then, $T$ is a perfect tensor if it is proportional to an isometric tensor from $A$ and $A^c$, i.e.
\be
\sum_{a_{n+1}\dots a_{2n}} T^\dagger_{a_1 \dots a_n a_{n+1} \dots a_{2n}} T_{a_{n+1} \dots a_{2n} b_1 \dots b_n} = \delta_{a_1 b_1} \dots \delta_{a_n b_n},
\ee
where one may choose any of the $2n$ legs of the tensor to be $a_{n+1} \dots a_{2n}$.}

\bigskip
Since one can show that for every possible bipartition, one may use isometric tensors to define the transformations, then every AME state defines a perfect tensor. Each tensor can then be placed in each polytope of the tessellation, connected by the outgoing physical legs (across the facets) which represent the contraction of the tensor indices.

\subsection{Concatenation and building a tensor network}
\label{concat}

Here we explain the concatenation process of AME states which results in an isometric mapping from the uncontracted legs in the bulk to the uncontracted legs at the boundary through a tensor network built from perfect tensors. We summarise here the case for two AME states but  concatenations of one AME state and one not-necessarily AME state were explored in \cite{Mazurek2019QuantumStates}. First we note that any AME state can always be expressed such that for each of the final $m$ columns, one only has a single $X$ operator, a single $Z$ operator and everywhere else is the identity \cite{Mazurek2019QuantumStates}; relating to the statement at the end of Section \ref{AMEstates}. The two stabiliser generators that have the $X$ and $Z$ operators acting on a particular qudit, where all other generators act trivially are the logical operators encoding one qudit.

Now, consider two AME stabiliser states, denoted as left ($L$) and right ($R$) states defined on $N_L$ and $N_R$ qudits respectively. One can represent an $L$ state as a tensor product of the form $L_i \otimes \sigma_i$, for $i = 1,\dots, N_L$. Here $L_i$ denotes a tensor product of Pauli operators on qudits $l_1$ to $l_{N_L -1}$ and $\sigma_i$ is the Pauli operator individual qudit $l_{N_L}$. One can similarly define the $R$ state using the tensor product $\sigma_j \otimes R_j$ where $j=1,\dots,N_R$, with $\sigma_j$ acting on qudit $r_1$ and $R_j$ acting on qudits $r_2 \dots r_{N_R}$. Then, concatenating these two states, the joint state $N_L + N_R$ can be expressed as the first matrix in \ref{concatmeasure}.

\be
\label{concatmeasure}
\begin{pmatrix} 
{} & L_1 & I & I & \dots & I \\ {} &  L_2 & I & I & \dots & I \\ {}  &\vdots & \vdots & \vdots & \ddots & \vdots \\  {} & L_{N_L -2} & I & I & \dots & I \\ {} & L_{N_L -1} & X & I & \dots & I \\ {} & L_{N_L} & Z & I & \dots & I \\ I & \dots & I & Z & R_1 & {} \\ I & \dots & I & X & R_2 & {} \\ I & \dots & I & I & R_3 & {} \\ \vdots & \ddots & \vdots & \vdots & \vdots &{} \\  I & \dots & I & I & R_{N_R -1}  & {}\\   I & \dots & I & I & R_{N_R} & {} \\   
\end{pmatrix}
\xrightarrow{XX}
\begin{pmatrix} 
{} & L_1 & I & I & \dots & I \\ {} &  L_2 & I & I & \dots & I \\ {}  &\vdots & \vdots & \vdots & \ddots & \vdots \\ {} & L_{N_L -2} & I & I & \dots & I \\ {} & L_{N_L -1} & X & I & \dots & I \\ {} & L_{N_L} & Z & Z & R_1 & {} \\ I & \dots & X & X & \dots & I \\ I & \dots & I & X & R_2 & {} \\ I & \dots & I & I & R_3 & {} \\ \vdots & \ddots & \vdots & \vdots & \vdots &{} \\  I & \dots & I & I & R_{N_R -1}  & {}\\   I & \dots & I & I & R_{N_R} & {} \\   
\end{pmatrix}
\xrightarrow{ZZ}
\begin{pmatrix} 
{} & L_1 & I & I & \dots & I \\ {} &  L_2 & I & I & \dots & I \\ {}  &\vdots & \vdots & \vdots & \ddots & \vdots \\ {} & L_{N_L -2} & I & I & \dots & I \\ {} & L_{N_L -1} & X & X & R_2 & {} \\ {} & L_{N_L} & Z & Z & R_1 & {} \\ I & \dots & X & X & \dots & I \\ I & \dots & Z & Z & \dots & I\\ I & \dots & I & I & R_3 & {} \\  \vdots & \ddots & \vdots & \vdots & \vdots &{} \\  I & \dots & I & I & R_{N_R -1}  & {}\\   I & \dots & I & I & R_{N_R} & {} \\   
\end{pmatrix}.
\ee

\bigskip

One then performs two different measurements to obtain the second and third matrices in \eqref{concatmeasure}. Recall that the process one undergoes after performing a measurement to update the list of generators of a stabiliser state is the following \cite{Nielsen}:
\begin{itemize}
\item There is no need to update the list if the measured observable may be constructed from a product of the stabiliser generators.
\item If the measured observable commutes with all the generators of the stabiliser, however cannot be constructed from these generators, one adds this observable to the list of generators (with a phase factor determined from the outcome of the measurement).
\item If the measured observable does not commute with at least one stabiliser generator and cannot be constructed from these generators, one replaces one non-commuting  generator with the measured observable (with a phase factor determined from the outcome of the measurement). One then multiplies all other generators that do not commute with the measured observable and multiply them with the generator that was removed from the list. 
\end{itemize}
Now in \eqref{concatmeasure} one first performs a measurement of the observable $XX$ on qudits $l_{N_L}$ and $r_1$ (assuming the outcome is $+1$). Clearly there are two generators that do not commute with this observable, those being the generators that have $Z$ operators acting on either qudit  $l_{N_L}$ or $r_1$. Hence one updates the list using the above rules, replacing one of the non-commuting stabiliser generators by the measured observable before multiplying the other non-commuting generator by the one that was removed.

One then performs a similar measurement, this time measuring the observable $ZZ$, on the same pair of qudits. The result is indicated on the right hand side of \eqref{concatmeasure}. One can now see maximal entanglement between qudits $l_{N_L}$ and $r_1$ with no relation to any of the other qudits present in the system. One therefore may trace these qudits out, removing the corresponding rows and columns from the matrix:

\be
\label{entswap}
\rightarrow
\begin{pmatrix} 
{} & L_1 & I\vpmark{a} & I\vpmark{b} & \dots & I \\ {} &  L_2 & I & I & \dots & I \\ {}  &\vdots & \vdots & \vdots & \ddots & \vdots \\ {} & L_{N_L -2} & I & I & \dots & I \\ {} & L_{N_L -1} & X & X & R_2 & {} \\ {} & L_{N_L} & Z & Z & R_1 & {} \\ I\pmark{c} & \dots & X & X & \dots & I\smark{c} \\ I\pmark{d} & \dots & Z & Z & \dots & I\smark{d}\\ I & \dots & I & I & R_3 & {} \\  \vdots & \ddots & \vdots & \vdots & \vdots &{} \\  I & \dots & I & I & R_{N_R -1}  & {}\\   I & \dots & I\vsmark{a} & I\vsmark{b} & R_{N_R} & {} \\   
\end{pmatrix}
\rightarrow
\begin{pmatrix}
{} & {} & L_1 & I & \dots & I \\  {} & {} & L_2 & I & \dots & I \\ {} & {} & \vdots & \vdots & \ddots & \vdots  \\ {} & {} & L_{N_L - 2} & I & \dots & I \\ {} & {} & L_{N_L -1} & R_2 & {} & {} \\ {} & {} & L_{N_L } & R_1 & {} & {} \\ I & \dots & I & R_3 & {} & {} \\  \vdots & \ddots & \vdots & \vdots & {} & {} \\  I & \dots & I & R_{N_{R}-1} & {} & {} \\  I & \dots & I & R_{N_R} & {} & {}
\end{pmatrix}.
\ee

\bigskip

In the case where the local dimension of the qudit is $D=2$ (i.e. one is considering qubits), as in the HaPPY construction, these maximally entangled pairs correspond to EPR pairs. The final matrix on the right hand side of \eqref{entswap} is therefore the list of stabiliser generators created as a result of performing entanglement swapping on a pair of qudits shared between the two AME states $L$ and $R$. Firstly, this consists of many of the original stabiliser generators of $R$ and $L$ that act non-trivially on qudits within their original domain, but have been extended to act trivially on those outside of their original domain. Secondly, one has two truncated rows, $(L_{N_L -1} R_2)$ and $(L_{N_L} R_1)$. Recall, that for the AME state $R$, a single qudit is encoded into $N_R - 1$ qudits by the logical $X$ and $Z$ operators, which are $R_1$ and $R_2$. Thus from \eqref{entswap}, one can see that entanglement swapping across AME states is equivalent to concatenating the corresponding quantum error correcting codes that arise from these states.

\bigskip

Hence if we consider the full tessellation, we may use entanglement swapping between two AME states to concatenate the quantum codes described by them. Expanding the network by adding more AME states (or equivalently perfect tensors), filling the tessellation, builds the final holographic code by concatenating all of the codes that arise from each AME state. 

\bigskip

We can also view this as concatenating perfect tensors resulting in a tensor network that builds an encoding isometry across the geometric manifold for the holographic code. In the perfect tensor language, isometries can be understood as follows. We simply choose which indices (legs) of the tensor are incoming and which are outgoing. Usually one interprets the logical qudit as an incoming leg as well as those legs that are output legs from previous tensors. 

As an example, for the AME(6,2) state, one may have 0, 1, 2 or 3 incoming legs (including the uncontracted bulk leg corresponding to the logical qudit). Now, suppose one chooses that the only input leg is the logical qudit so there are five outgoing legs. This corresponds to the $[[5,1,3]]_2$ code encoding one logical qudit to five physical qudits.  However if one considers two input legs, then one would have the stabilisers for the $[[4,2,2]]_2$ code. Thus, each encoding isometry corresponds to the number of logical qudits being encoded within each AME state.

\bigskip

Hence, by concatenating the AME states, one analogously contracts the perfect tensors building a tensor network. From the perfect tensor perspective, one is simply concatenating the isometries of each perfect tensor layer by layer. From the AME state point of view, one can see this is due to entanglement swapping between pairs of qudits, as this section has detailed. Clearly the product of isometries is simply an isometry itself. Hence the tensor network acts as a mapping from the logical qudits in the bulk to the physical qudits at the boundary. Similarly to the isometries for individual AME states, this isometry then corresponds to the encoding transformation of the holographic code.


\section{The HaPPY Construction in higher dimensions}
\label{sec:happy3d}

In section \ref{stabconhappy} we discussed how we can associate stabiliser generators to the $\{5,4\}$ tessellation of the spatial slices of AdS$_3$. Then, one concatenates these AME states using EPR pairs building a tensor network (using perfect tensors) that acts as an exactly solvable toy model for AdS$_3$/CFT$_2$. To extend this construction to higher  dimensions we need to consider in turn the following steps of the construction:

\begin{enumerate}
\item \textbf{Tessellation:} Find a tessellation for the spatial slices of AdS$_{d+1}$ to discretise the hyperbolic geometry.
\item \textbf{Associate qudits to the geometry:} One should place both the logical and physical qudits onto the tessellated hyperbolic geometry. It should be noted that in the case of AdS$_3$, there are two possible choices for the physical qudits - either placed on the vertices or edges of the polygons. In HaPPY, one chooses to place them on edges in order to concatenate via EPR pairs. While there are more choices of cellulations as the dimension is increased (as can be visualised by the corresponding Hasse diagram), in order to maintain the concatenation principles in HaPPY, one should associate the logical qudit with the tessellated $d$-dimensional polytope itself and physical qudits with the $d-1$ dimensional facet. For example, when considering the spatial slices of AdS$_4$, one tessellates H$^3$ with three-dimensional polyhedra associating the logical qudit with the polyhedra and one physical qudit with each of its two-dimensional faces.
\item \textbf{Find the corresponding graph state:} Once the qudits have been associated to the geometry, one can build a graph state as explained in section \ref{stabconhappy}. We will see that it is at this step that one is forced to drop one of the assumptions implicit in the two-dimensional HaPPY construction.
\item \textbf{Build the tensor network:} Provided the previous steps have been accomplished, one may straightforwardly build the tensor network using perfect tensors if AME states have been used. 
\end{enumerate}

In this section we explore whether and how this formulation generalises to higher dimensions and propose several possibilities of how to overcome potential obstructions. While we focus specifically on the HaPPY construction, analogous steps would need to be followed in generalising other approaches to holographic codes to higher dimensions and we will return to this point at the end of this section. 

\begin{figure}[h!]
\centering
\begin{tikzpicture}
\coordinate (O) at (0,0,0);
\coordinate (A) at (0,\Width,0);
\coordinate (B) at (0,\Width,\Height);
\coordinate (C) at (0,0,\Height);
\coordinate (D) at (\Depth,0,0);
\coordinate (E) at (\Depth,\Width,0);
\coordinate (F) at (\Depth,\Width,\Height);
\coordinate (G) at (\Depth,0,\Height);

\coordinate (H) at (\Depth/2,\Width/2,\Height);
\coordinate (I) at (\Depth/2,\Width,\Height/2);
\coordinate (J) at (\Depth, \Width/2, \Height/2);
\coordinate (K) at (\Depth/2,\Width/2, 0);
\coordinate (L) at (\Depth/2, 0, \Height/2);
\coordinate (M) at (0, \Width/2, \Height/2);
\coordinate (P) at (\Depth/2, \Width/2, \Height/2);

\draw[black,fill=blue!20,opacity=0.8] (O) -- (C) -- (G) -- (D) -- cycle;
\draw[black,fill=purple!20] (O) -- (A) -- (E) -- (D) -- cycle;
\draw[black,fill=blue!10] (O) -- (A) -- (B) -- (C) -- cycle;
\draw[black,fill=blue!20,opacity=0.8] (D) -- (E) -- (F) -- (G) -- cycle;
\draw[black,fill=blue!20,opacity=0.6] (C) -- (B) -- (F) -- (G) -- cycle;
\draw[black,fill=blue!20,opacity=0.8] (A) -- (B) -- (F) -- (E) -- cycle;

\draw[black] (H) -- (I);
\draw[black] (H) -- (J);
\draw[black] (H) -- (L);
\draw[black] (H) -- (M);
\draw[red] (H) -- (P);

\draw[black] (K) -- (I);
\draw[black] (K) -- (J);
\draw[black] (K) -- (L);
\draw[black] (K) -- (M);
\draw[red] (K) -- (P);

\draw[black] (I) -- (J);
\draw[black] (I) -- (M);
\draw[red] (I) -- (P);

\draw[black] (L) -- (J);
\draw[black] (L) -- (M);
\draw[red] (L) -- (P);

\draw[red] (J) -- (P);
\draw[red] (M) -- (P);

\node[style={draw,shape=circle,fill=black}] at (H) (p1) {};
\node[style={draw,shape=circle,fill=black}] at (I) (p2) {};
\node[style={draw,shape=circle,fill=black}] at (J) (p3) {};
\node[style={draw,shape=circle,fill=black}] at (K) (p4) {};
\node[style={draw,shape=circle,fill=black}] at (L) (p5) {};
\node[style={draw,shape=circle,fill=black}] at (M) (p6) {};
\node[style={draw,shape=circle,fill=red}] at (P) (l1) {};

\end{tikzpicture}
\caption{Qudits placed on a cube in H$^3$. Here, the one logical qudit associated to the cube is represented by being placed in it's centre and the six physical qudits are placed on faces.}
\label{cubeAdS4}
\end{figure}
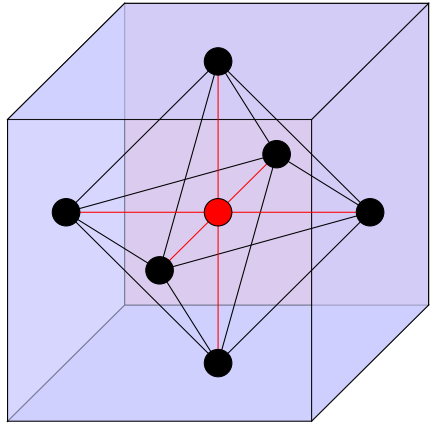

\subsection{AdS$_4$: Order-5 cubic honeycomb}

A first step towards generalising the HaPPY approach to higher spacetime dimensions is the case of AdS$_4$. We begin by considering spatial slices given by hyperbolic 3-spaces H$^3$; regular tessellations of H$^3$ were summarised in Table \ref{ads4:table}. The simplest option to visualise would clearly be the order-5 cubic honeycomb with Schl\"{a}fi symbol $\{4,3,5\}$ since the tessellated polyhedron is just the cube. We begin this section by trying to replicate the HaPPY approach of \ref{nonamecube}, before drawing the conclusion that the resultant graph state would not be AME. We then relax the assumption of preserving maximal discrete symmetry in \ref{amecube} and force the graph state to be AME to provide an alternate construction.

\subsubsection{Non-AME state}
\label{nonamecube}

\begin{figure}[h!]
\centering
\begin{tikzpicture}[mystyle/.style={draw,shape=circle,fill=black, inner sep=0pt, minimum size=4pt},scale=1.5, transform shape]
\def\ngon{6}
\node[draw, regular polygon,regular polygon sides=\ngon,minimum size=3cm] (p) {};
\foreach\x in {1,...,\ngon} \foreach\y in {1,...,\ngon} {\draw[black] (p.corner \x) -- (p.corner \y) {};
};
\foreach\x in {1,...,\ngon}{\draw[red] (p.corner \x) -- (p.center) {};
};
\foreach\x in {1,...,\ngon}{
    \node[mystyle=\x] (p\x) at (p.corner \x){};
};
\node[style={draw,shape=circle,fill=red, inner sep=0pt, minimum size=4pt}] at (p.center) {};

\end{tikzpicture}
\caption{Resultant graph state for the cube, following the HaPPY approach, that preserves maximal discrete symmetry.}
\label{7graph}
\end{figure}
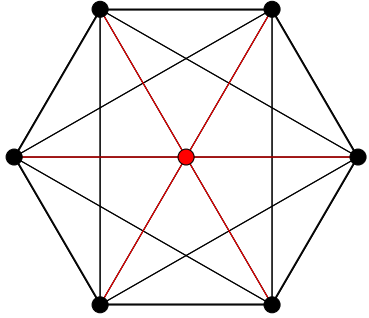

As noted in step 2 above, we can associate a single logical qudit to the centre of the cube and place physical qudits at each of the faces. To associate a graph state with this tessellation, it is natural to try to preserve maximal discrete symmetry as in the HaPPY approach. This means that if two faces are connected by an edge then the two physical qudits living on these faces should be considered as connected in the graph state. If faces are not connected by an edge, they are not connected. The logic of how the graph state is formed is illustrated diagrammatically in Figure \ref{cubeAdS4}. The graph drawn in three dimensions can then be compressed to two dimensions, as shown in Figure \ref{7graph}, so that it is in the standard representation of a graph state.

\bigskip

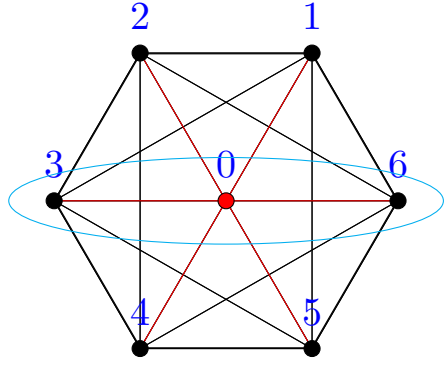
\begin{figure}[h!]
\centering
\begin{tikzpicture}[mystyle/.style={draw,shape=circle,fill=black, inner sep=0pt, minimum size=4pt},scale=1.5, transform shape,  every node/.style={draw},every newellipse node/.style={inner sep=0pt}]
\def\ngon{6}
\node[draw, regular polygon,regular polygon sides=\ngon,minimum size=3cm] (p) {};
\foreach\x in {1,...,\ngon} \foreach\y in {1,...,\ngon} {\draw[black] (p.corner \x) -- (p.corner \y) {};
};
\foreach\x in {1,...,\ngon}{\draw[red] (p.corner \x) -- (p.center) {};
};
\foreach\x in {1,...,\ngon}{
    \node[mystyle=\x,label={[font=\small,text=blue]$\x$}] (p\x) at (p.corner \x) (corner\x) {};
};
\node[style={draw,shape=circle,fill=red, inner sep=0pt, minimum size=4pt},label={[font=\small,text=blue]$0$}] at (p.center) (log) {}; 
 \path (corner6) -- (log) -- (corner3);
\node[fit=(corner6)(log)(corner3), cyan, newellipse, xscale=0.85,yscale=3.5] {};

\end{tikzpicture}
\caption{Bipartition of the graph state associated with the cube that preserves maximal discrete symmetry. Here the partition is chosen as $K=\{0,3,6\}$ and $L=\{1,2,4,5\}$. The resultant equations are not linearly independent and thus this graph state is not AME.}
\label{7graphnotame}
\end{figure}

By construction, since any graph state is by definition a stabiliser state, then Figure \ref{7graph} is a stabiliser state. Unfortunately, constructing the state in this way does not give rise to an AME state. This can be shown straightforwardly using the techniques explained in the previous section. We draw the graph state as before and choose the partition $\{0,3,6\} | \{1,2,4,5\}$ as indicated in Figure \ref{7graphnotame}. 

Following the approach described in the previous section, one then obtains
\begin{align}
&A_0  \backslash\{0,3,6\} = (1,1,1,1), \\
&A_3 \backslash\{0,3,6\} = (1,1,1,1), \\
&A_6  \backslash\{0,3,6\}= (1,1,1,1),
\end{align}
but these are not linearly independent and hence this graph state is not AME. In fact, this was to be expected since it has been shown in the literature that no AME state exists for 7 qubits or more \cite{Huber2016AbsolutelyExist}. This generic result indicates that extensions of the HaPPY construction to higher dimensions necessarily require a variation of the approach used in two dimensions. 

The classifications of  \cite{Huber2016AbsolutelyExist} show that one can have an AME state for 7 qudits or more for $D > 2$. This is possible since for $D > 2$ one can have multiple $Z$ operators between qudits. The immediate issue in exploiting such AME states for our purposes is the difficulty in maintaining maximal discrete symmetry. To maintain the discrete symmetry, if one uses multiple $Z$ operators between any pair of qudits, one has to do so everywhere. The immediate consequence of this is that none of these states can be AME since when considering $\lambda \in \{1,\dots,D-1\}$ Z operators between $D$-dimensional qudits, 
\begin{align}
&A_0  \backslash\{0,3,6\}= \lambda(1,1,1,1), \\
&A_3 \backslash\{0,3,6\} = \lambda(1,1,1,1) \\
&A_6  \backslash\{0,3,6\} = \lambda(1,1,1,1) 
\end{align}
which can never be linearly independent of one another. This of course generalises for $n\geq7$ qudits: since no 7 qubit or higher AME state can exist, no maximally symmetric construction can be formed such that the graph state is AME. Hence, in dimensions $d > 2$, one is either restricted to tessellations such that the $d$-dimensional polytope has $n<7$ $(d-1)$-dimensional facets or one lifts one of the previous assumptions. 

\bigskip

Since Figure \ref{7graph} is not AME, there is now no longer a guarantee that it represents a quantum error correcting code. In order to check that this construction does indeed correspond to a graph code, we use the following theorem developed by Schlingermann and Werner \cite{SchlingermannWerner2000}:

\bigskip

\textbf{SW-theorem:} \textit{Consider a set of input vertices $X$ and a set of output vertices $Y$. Then, given a finite abelian group $G$ and the weighted graph with adjacency (coincidence) matrix $\Xi$, a quantum error correcting code $v_\Xi$ is able to detect an error configuration $\mathcal{E} \subset Y$ iff}
\be
\label{swtheorem1}
\Xi^{I}_{X \cup \mathcal{E}} d^{X \cup \mathcal{E}} = 0
\ee
\textit{with $I= Y \backslash \mathcal{E}$ implies}
\be
\label{swtheorem2}
d^X = 0, \quad \text{\textit{and}} \quad \Xi^X_\mathcal{E} d^\mathcal{E} = 0.
\ee

\vspace{3cm}

Here, we focus on the simplest case with local dimension $D=2$. The input vertices $X$ refer to the single logical qubit in the centre of the polytope and the output vertices are the six physical qubits. The full $7 \times 7$ symmetric coincidence matrix $\Xi$ \cite{Schlingermann2001}, that describes the connectivity of the entire graph, inclusive of both input and output vertices is given by:

\be
\Xi_{[[6,1,2]]} = \begin{bmatrix} 0&1&1&1&1&1&1\\1&0&1&1&0&1&1\\1&1&0&1&1&0&1\\1&1&1&0&1&1&0\\1&0&1&1&0&1&1\\1&1&0&1&1&0&1\\1&1&1&0&1&1&0 \end{bmatrix},
\ee
and thus the $6 \times 6$ adjacency matrix $\Gamma$, describing the graph state of just the output vertices may be represented as

\be
\Gamma =  \begin{bmatrix} 0&1&1&0&1&1\\1&0&1&1&0&1\\1&1&0&1&1&0\\0&1&1&0&1&1\\1&0&1&1&0&1\\1&1&0&1&1&0 \end{bmatrix}.
\ee

We can then write down the stabiliser generators associated with the graph state of the output vertices given in Figure \ref{7graph}, following the techniques previously introduced. This gives
\begin{equation}
\label{70statestab}
\mathcal{S} = \begin{bmatrix}G_0\\G_1\\G_2\\G_3\\G_4\\G_5\\G_6 \end{bmatrix} = \begin{bmatrix} X&Z&Z&Z&Z&Z&Z\\Z&X&Z&Z&I&Z&Z\\Z&Z&X&Z&Z&I&Z\\Z&Z&Z&X&Z&Z&I\\Z&I&Z&Z&X&Z&Z\\Z&Z&I&Z&Z&X&Z\\Z&Z&Z&I&Z&Z&X \end{bmatrix}.
\end{equation}
Recall the following procedure to convert an $[[n,k,d]]$ code into an $[[n-1,k+1,d-1]]$ code: one performs standard matrix manipulations to obtain one stabiliser generator $G_X$ ending with $X$, one stabiliser generator $G_Z$ ending with $Z$ and $n-k$ generators ending with the identity $I$. Then by dropping $G_X$ and $G_Z$ and removing the final qubit, we will be left with a stabiliser code formed from the remaining $n-k-2$ generators. The resultant code has distance $d-1$ and thus encodes $(n-1) - (n-k-2) = k + 1$ qubits. Hence by performing this analysis on the stabiliser generators for the $[[7,0]]$ state (\ref{70statestab}), we can show the stabiliser generators for the $[[6,1]]$ code are
\begin{equation}
\label{newStabcorrection}
\mathcal{S} = \begin{bmatrix}G'_0\\G'_1\\G'_2\\G'_3\\G'_4 \end{bmatrix} = \begin{bmatrix} X&I&I&X&I&I\\I&X&I&I&X&I\\I&I&X&I&I&X\\Y&-Y&I&Z&Z&I\\Y&I&-Y&Z&I&Z\end{bmatrix}
\end{equation}
and the corresponding logical operators are given by $\bar{Z} = ZZZZZZ$, $\bar{X} = XZZIZZ$.

\bigskip

In order to see whether Figure \ref{7graph} realises a one-error correcting code, with one input vertex and six output vertices, we apply error detection (correction) conditions given in the SW-theorem to the ${6\choose2} = 15$ two-error correction configurations. These are denoted $\mathcal{E}_j$ with $j \in \{1,\dots,15\}$ and are defined as
\begin{align}
&\mathcal{E}_1 = \{0,1,2\}, \; \mathcal{E}_2 = \{0,1,3\}, \; \mathcal{E}_3 = \{0,1,4\}, \; \mathcal{E}_4 = \{0,1,5\}, \; \mathcal{E}_5 = \{0,1,6\}, \\ &\mathcal{E}_6 = \{0,2,3\}, \; \mathcal{E}_7 = \{0,2,4\}, \; \mathcal{E}_8 = \{0,2,5\}, \; \mathcal{E}_9 = \{0,2,6\}, \; \mathcal{E}_{10} = \{0,3,4\}, \\
&\mathcal{E}_{11} = \{0,3,5\}, \; \mathcal{E}_{12} = \{0,3,6\}, \; \mathcal{E}_{13} = \{0,4,5\}, \; \mathcal{E}_{14} = \{0,4,6\}, \; \mathcal{E}_{15} = \{0,5,6\}. \;
\end{align}

For example, consider the error configuration $\mathcal{E}_1 = \{0,1,2\}$. The resulting set of equations from \eqref{swtheorem1} are given in Table \ref{errorconfig1}. 

\begin{table}[h!]
\begin{center}
\begin{tabular}{ |c|c| } 
\hline
Vertex & Equation \\
\hline
3 & $d_0 + d_1 + d_2 = 0$\\
4 & $d_0 + d_2 = 0$\\
5 & $d_0 + d_1 = 0$\\
6 & $d_0 + d_1 + d_2 = 0$ \\
\hline
\end{tabular}
\end{center}
\caption{Set of equations for error configuration $\mathcal{E}_1$.}
\label{errorconfig1}
\end{table}

Solving this set of relations results in $d_0=d_1=d_2=d_3=0$ and thus by the SW-theorem, the error configuration $\mathcal{E}_1$ is a detectable error configuration. Following this logic, one can repeat this analysis for the remaining possible error configurations. The result of this calculation is that there are three problematic error configurations: $\mathcal{E}_3, \mathcal{E}_8$ and $\mathcal{E}_{12}$. 

As in \cite{Gottesman:1997zz}, a quantum code with stabiliser $S$ will detect all errors $\mathcal{E}$ that are either in $S$ or anticommute with at least one element in $S$ (i.e. $\mathcal{E} \in S \cup (\mathcal{P}_n - N(S)))$. For each of the problematic error configurations, three error operators exist that do not anticommute with at least one element in $S$, namely $X X$, $Y Z$ and $Z Y$. Consider the first set of these operators; $X_1 X_4, X_2 X_5$ and $X_3 X_6$, where the subscript refers to which qubit an operator acts upon. Each of these three operators is clearly an element of the stabiliser, as can be trivially read off from (\ref{newStabcorrection}). Since these error operators belong to the stabiliser of the code; $X_1 X_4, X_2 X_5, X_3 X_6 \in S$, they have no impact on the encoded state. However, the remaining six two-error operators,
\be
Y_1 Z_4, \quad Z_1 Y_4, \quad Y_2 Z_5, \quad Z_2 Y_5, \quad Y_3 Z_6, \quad Z_3 Y_6,
\ee
cannot be corrected for. Thus not all two-error configurations are either detectable or act trivially on the encoded state and so the graph state given in Figure \ref{7graph} is a quantum error correcting code with distance $d=2$, which we denote as $[[6,1,2]]_2$. Further evidence is provided by the Knill-Laflamme bound \cite{KnillLaflamme96}
\be
n \ge 2(d-1) + k,
\ee
which in this instance reduces to the statement that the code distance must satisfy $d\le 3$. We also note that for any quantum stabiliser code with distance $d$ that if $S$ contains elements of weight less than $d$, then it is a degenerate code. Since the minimum weight operator in $S$ is a weight-2 operator (e.g. $X_1 X_4 \in S$) and $d=2$, the $[[6,1,2]]_2$ code presented here is non-degenerate. 

 \begin{figure}[h!]
\begin{center}
\includegraphics[scale=0.4]{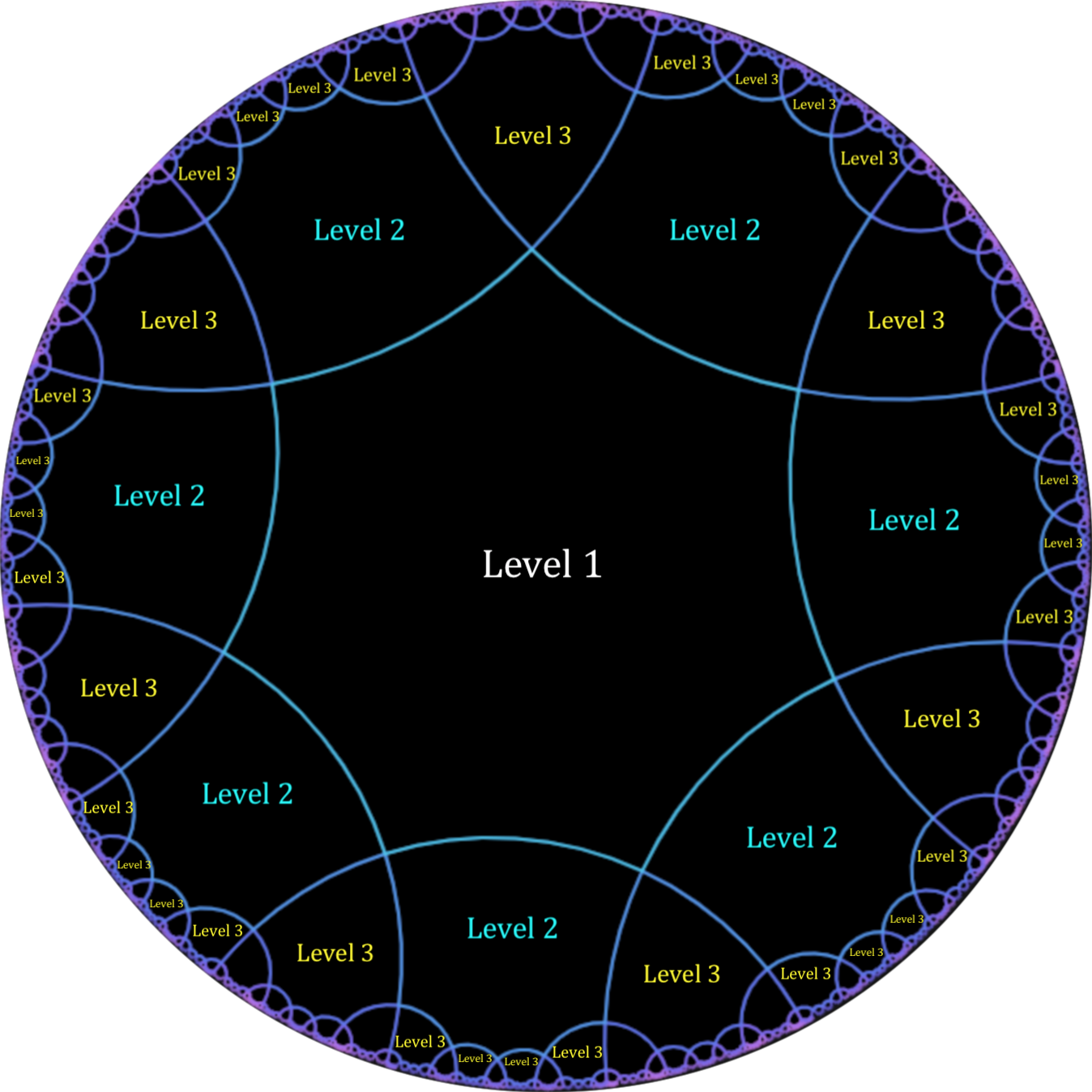}
\end{center}
\caption{\label{level74} $\{7,4\}$ tessellation of hyperbolic plane H$^2$. One shows subsequent levels beginning from some central polytope defined to be at level $L=1$ and adding a level for each polytope adjacent to a polytope at the previous level.}
\end{figure}

\bigskip

The next step is to tessellate the full spatial slices of AdS$_4$ in the form of the order-5 cubic honeycomb, tessellating cubes by associating labels to each face and consistently gluing together them together. Since our code is not AME, we need to be careful to ensure that the concatenation is consistent. In \cite{Harris2018Calderbank-Steane-ShorCodes}, the authors create a consistent construction of the $\{7,4\}$ tessellation of the spatial slice of AdS$_3$ using block-perfect tensors. Essentially the tessellation is constructed in levels, with the first level being some central heptagon, the second level being all adjacent heptagons to those in the first level and so on. We adapt this approach in the following construction of the concatenation of cells, so let us first describe how this works. 

\bigskip

 For every heptagon $h_L$ at level $L$, one assigns the set of labels $e_i$ with $i \in \{1,\dots, 7\}$ to its edges. For every $i$, the labelled edge $e_i$ at level $L$ must then be glued to some other edge of a heptagon at level $L'$. In this instance, for each $i$, $L'$ must be either level $L-1$ or $L+1$ but it cannot be both simultaneously. Thus all edges labelled $e_i$ at level $L$ are glued to the same level $L'$. 
 
 As an example, look at level $L=2$ in  \cite{Harris2018Calderbank-Steane-ShorCodes}. Each edge $e_6$ is glued to an edge in level 1 ($L-1$). Similarly, edges $e_j$ with $j=\{1,2,3,4,5,7\}$ at level $L=2$ are always glued to an edge at level 3 ($L+1$). A sketch depicting the levels in this example is shown in Figure \ref{level74}.
 
 \bigskip
 
Note, when one reaches level $L=3$, one encounters an issue. There is an inconsistency of edges connecting to other levels. Some $L=3$ heptagons are adjacent to two $L=2$ heptagons and five $L=4$ heptagons while others are only connected to one level $L=2$ heptagons and six $L=4$ heptagons. Fortunately, there is a simple fix for this. One simply treats the two cases to be distinct, as if they were there own levels (i.e. one could think of them as sub-levels $L=3.$A and $L=3.$B), that collectively make up level $L=3$. Due to the symmetry of tessellations with 4 edges meeting at each vertex ($q=4$ in the Schl\"{a}fi symbol), as one increases the level, there will always be two sub-levels for each level $L$ from level $L=3$ onwards.

\bigskip

\begin{figure}[h!]
\begin{center}
\includegraphics[scale=0.4]{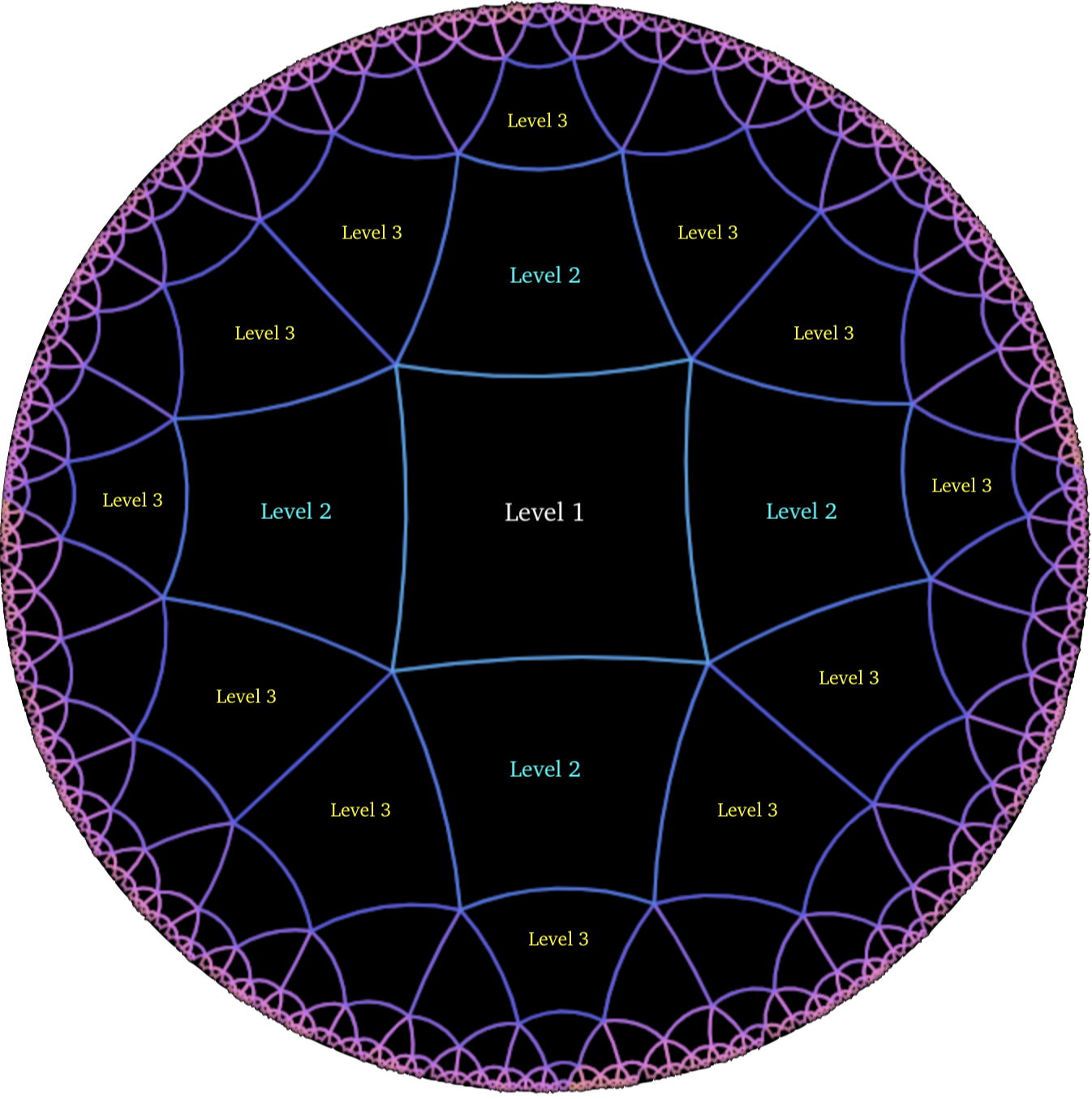}
\end{center}
\caption{\label{level45} $\{4,5\}$ tessellation of hyperbolic plane H$^2$. One shows subsequent levels beginning from some central polytope defined to be at level $L=1$ and adding a level for each polytope adjacent to a polytope at the previous level.}
\end{figure}

The ideas behind this construction can be generalised to concatenate the non-AME code across the cubic honeycomb on H$^3$. The construction is however more subtle than the two-dimensional case above. In order to visualise the 3d tessellation, we note that the structure is analogous to that of the $\{4,5\}$ tiling in H$^2$. We may proceed by labelling edges (corresponding to faces) and building the tessellation in levels as before. The result after three levels is shown in Figure \ref{level45}. 

\bigskip

The immediate issue arising is that one has inconsistencies, similar to that of the $\{7,4\}$ tiling when level $L=3$ is reached. However, these are now more intricate than before due to the difference in symmetries between the two tilings.

At $L=3$, on one hand there are some polytopes where one edge will be glued to a polytope at level $L=2$ and three edges will be glued to polytopes at level $L=4$. On the other hand, there are some polytopes where one edge will be glued to level $L=2$, one edge glued to another edge at level $L=3$ and two edges that will be glued to a polytope at level $L=4$. Hence, while one can choose one labelled edge to be glued to level $L=2$ and two to be glued to level $L=4$, one edge will sometimes be glued to level $L=3$ and sometimes to level $L=4$. Thus, one seems to have an inconsistent construction. Similarly, for the order-5 cubic honeycomb, one will analogous issues, where one face may be connected to level $L=2$, and two faces to level $L=4$ while one face will sometimes be glued to level $L=3$ and sometimes to level $L=4$. 

\bigskip

The key issue arising is that one now has polytopes at level $L$ adjacent to other polytopes at level $L$. In the $\{4,5\}$ tiling, this issue arises from the fact that the tessellation has an odd number of polytopes around each vertex (i.e. for Schl\"{a}fi symbol $\{p,q\}$, one has $q$ odd). Similarly, for the $\{4,3,5\}$ tiling this originates from the odd number of polytopes around each edge (i.e. for Schl\"{a}fi symbol $\{p,q,r\}$, one has $r$ odd).

\begin{table}[h!]
\begin{center}
\begin{tabular}{ |c|c| } 
\hline
Level $L$  & Number of sub-levels \\
\hline
$L=1$ & 1\\
$L=2$ & 1\\
$L=3$ & 2\\
$L=4$ & 2\\
$L=5$ & 3\\
$L \ge 6$ & 3\\
\hline
\end{tabular}
\end{center}
\caption{Number of sub-levels required at each level $L$ in the $\{4,3,5\}$ order-5 cubic honeycomb to form a consistent construction.}
\label{sublevels}
\end{table}

\bigskip

It is possible to resolve these subtleties and obtain consistent concatenations of codes between cells. By dividing levels into sub-levels and treating each as if it were its own level, as previously explained, consistent constructions can be obtained. Table \ref{sublevels} shows the required number of sub-levels at each level $L$ in order to maintain consistency, where we note that one sub-level simply means only the level $L$ itself.

\bigskip

If we are using a tessellation with an even structure ($q$ even in 2d or $r$ even in 3d), we can obtain a consistent construction using a similar approach to that in \cite{Harris2018Calderbank-Steane-ShorCodes}. We illustrate this for the order-4 dodecahedral honeycomb tessellation of H$^3$ with Schl\"{a}fi symbol $\{5,3,4\}$ (noting here $r$ is even) in section \ref{othertess}.

\subsubsection{AME state}
\label{amecube}

\begin{figure}[h!]
\centering
\begin{tikzpicture}[mystyle/.style={draw,shape=circle,fill=black, inner sep=0pt, minimum size=4pt}, scale=1.5, transform shape]
\def\ngon{6}
\node[regular polygon,regular polygon sides=\ngon,minimum size=3cm] (p) {};
\foreach\x in {1,...,\ngon}{\draw[black] (p.corner \x) -- (p.center) {};
};
\node[style={draw,shape=circle,fill=black, inner sep=0pt, minimum size=4pt},label={[font=\small,text=blue]$0$}] at (p.center) {}; 
\foreach\x in {1,...,\ngon}{
    \node[mystyle=\x,label={[font=\small,text=blue]$\x$}] (p\x) at (p.corner \x) (corner\x) {};
};
\draw (corner1) -- (corner6);
\draw (corner2) -- (corner3);
\draw (corner3) -- (corner4);
\draw (corner4) -- (corner5);

\draw[transform canvas={yshift=-1pt}] (corner1) -- (corner2);
\draw[transform canvas={yshift=1pt}] (corner1) -- (corner2);

\draw[transform canvas={yshift=0.35pt, xshift=-0.85pt}] (corner5) -- (corner6);
\draw[transform canvas={yshift=-0.65pt, xshift=1pt}] (corner5) -- (corner6);
\foreach\x in {1,...,\ngon}{
    \node[mystyle=\x] (p\x) at (p.corner \x){};
};

\end{tikzpicture}
\caption{Graph state corresponding to the 7 qutrit AME state.}
\label{ame7qutrit}
\end{figure}
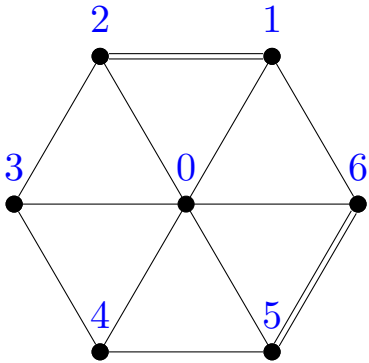

Now let us turn to an alternative approach to associating codes to H$^3$ spatial slices of AdS$_4$. If one lifts the assumption that the cells preserve maximal discrete symmetry, then graph states that are AME can be obtained. For example, for the case where the cube is the polyhedron tessellated (such as the order-5 cubic honeycomb tessellation of H$^3$), one requires a graph state consisting of six physical qudits and a single central logical qudit. The graph state given in Figure \ref{ame7qutrit} satisfies this requirement while also being absolutely maximally entangled. Here the unit of quantum information used is the qutrit $(D=3)$, allowing for up to two $Z$ operators between pairs of qutrits.

While this graph state does not possess maximal symmetry, since it is AME, many of the arguments presented in \cite{Pastawski2015HolographicCorrespondence} will still hold for this graph state. Importantly, this means this graph state still corresponds to a perfect tensor, and so when concatenated into a full tensor network, the bulk can be reconstructed using operator pushing. Now, suppose this suggestion can be implemented, representing the connectivity of the 6 physical qutrits that sit on each of the cube's faces and the single logical qutrit in its centre. 

\begin{figure}[h!]
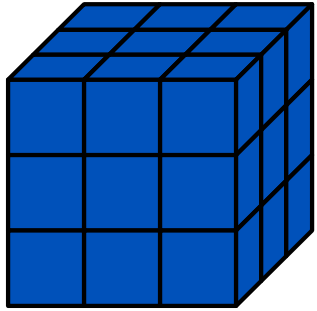

\centering
 \RubikFaceUp
        {B}{B}{B}
        {B}{B}{B}
        {B}{B}{B}
    \RubikFaceRight
        {B}{B}{B}
        {B}{B}{B}
        {B}{B}{B}
    \RubikFaceFront
        {B}{B}{B}
        {B}{B}{B}
        {B}{B}{B}
   \ShowCube{7cm}{1}{\DrawRubikCube}
\caption{Visualisation of the cubic honeycomb with Schl\"{a}fi symbol $\{4,3,4\}$ that tessellates $\mathbbm{R}^3$.}
\label{cubichoney}
\end{figure}

One can now try to consistently glue these cubes together when forming the full tessellation of $H^3$. For similar reasons to the logic presented in section \ref{nonamecube}, finding a consistent construction for the $\{4,3,5\}$ honeycomb is subtle due to there being an odd number of cubes around each edge (i.e $r$ is odd). To begin this discussion, we will consider first the construction of the cubic honeycomb with Schl\"{a}fi symbol $\{4,3,4\}$ that tessellates Euclidean 3-space $\mathbbm{R}^3$, depicted in Figure \ref{cubichoney}, noting for this tiling that $r$ is even. To our knowledge this construction has not been presented in any existing literature, and it will prove useful when considering the order-4 dodecahedral honeycomb in section \ref{othertess}. This tessellation belongs to a family of hypercube honeycombs with Schl\"{a}fi symbols $\{4,3,\dots,3,4\}$ and so the following arguments should extend straightforwardly to higher dimensions.

\bigskip

\bigskip

When considering this tessellation, the first step is to find a representation of the graph state within the cube that preserves as much symmetry as possible. Clearly the logical qutrit (labelled `0') is maximally connected to all other qutrits so it is omitted from the following discussion. One can produce a construction that preserves as much symmetry as possible in the following way:

\begin{figure}[h!]
\centering
\begin{tikzpicture}
\coordinate (O) at (0,0,0);
\coordinate (A) at (0,\Width,0);
\coordinate (B) at (0,\Width,\Height);
\coordinate (C) at (0,0,\Height);
\coordinate (D) at (\Depth,0,0);
\coordinate (E) at (\Depth,\Width,0);
\coordinate (F) at (\Depth,\Width,\Height);
\coordinate (G) at (\Depth,0,\Height);

\coordinate (H) at (\Depth/2,\Width/2,\Height);
\coordinate (I) at (\Depth/2,\Width,\Height/2);
\coordinate (J) at (\Depth, \Width/2, \Height/2);
\coordinate (K) at (\Depth/2,\Width/2, 0);
\coordinate (L) at (\Depth/2, 0, \Height/2);
\coordinate (M) at (0, \Width/2, \Height/2);
\coordinate (P) at (\Depth/2, \Width/2, \Height/2);

\draw[black,fill=blue!20,opacity=0.8] (O) -- (C) -- (G) -- (D) -- cycle;
\draw[black,fill=purple!20] (O) -- (A) -- (E) -- (D) -- cycle;
\draw[black,fill=blue!10] (O) -- (A) -- (B) -- (C) -- cycle;
\draw[black,fill=blue!20,opacity=0.8] (D) -- (E) -- (F) -- (G) -- cycle;
\draw[black,fill=blue!20,opacity=0.6] (C) -- (B) -- (F) -- (G) -- cycle;
\draw[black,fill=blue!20,opacity=0.8] (A) -- (B) -- (F) -- (E) -- cycle;

\node[style={draw,shape=circle,fill=black}, label=$Q_5$] at (H) (p1) {};
\node[style={draw,shape=circle,fill=black}, label=$Q_3$] at (I) (p2) {};
\node[style={draw,shape=circle,fill=black}, label=right:{$Q_4$}] at (J) (p3) {};
\node[style={draw,shape=circle,fill=black}, label=below:{$Q_2$}] at (K) (p4) {};
\node[style={draw,shape=circle,fill=black}, label=below:{$Q_6$}] at (L) (p5) {};
\node[style={draw,shape=circle,fill=black}, label=left:{$Q_1$}] at (M) (p6) {};
\node[style={draw,shape=circle,fill=red}] at (P) (l1) {};

\node[style={draw,shape=star,fill=yellow,star points =5, star point ratio=0.5}, label=$v_2$] at (A) (a) {};
\node[style={draw,shape=star,fill=yellow,star points =5, star point ratio=0.5}, label=below:{$v_6$}] at (C) (c) {};
\node[style={draw,shape=star,fill=yellow,star points =5, star point ratio=0.5}, label=$v_3$] at (E) (e) {};
\node[style={draw,shape=star,fill=yellow,star points =5, star point ratio=0.5}, label=$v_4$] at (F) (f) {};
\node[style={draw,shape=star,fill=yellow,star points =5, star point ratio=0.5}, label=below:{$v_5$}] at (G) (g) {};
\node[style={draw,shape=star,fill=yellow,star points =5, star point ratio=0.5}, label=below:{$v_1$}] at (O) (o) {};

\end{tikzpicture}
\caption{Assignment of qutrits to preserve the most symmetry when embedding the AME(7,3) state into a cube.}
\label{faceslabel}
\end{figure}
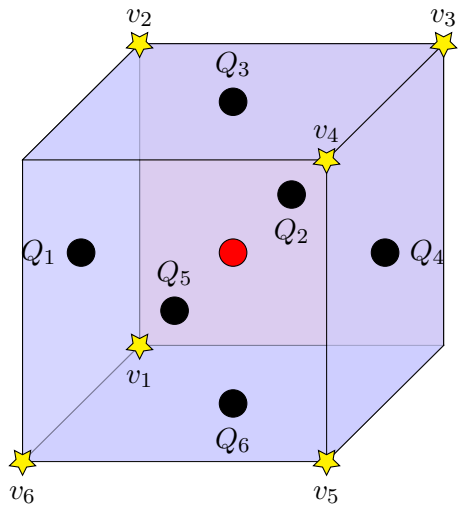

\begin{figure}[h!]
\centering
\begin{tikzpicture}[transform shape]
\coordinate (O) at (0,0,0);
\coordinate (A) at (0,\Width,0);
\coordinate (B) at (0,\Width,\Height);
\coordinate (C) at (0,0,\Height);
\coordinate (D) at (\Depth,0,0);
\coordinate (E) at (\Depth,\Width,0);
\coordinate (F) at (\Depth,\Width,\Height);
\coordinate (G) at (\Depth,0,\Height);

\coordinate (H) at (\Depth/2,\Width/2,\Height);
\coordinate (I) at (\Depth/2,\Width,\Height/2);
\coordinate (J) at (\Depth, \Width/2, \Height/2);
\coordinate (K) at (\Depth/2,\Width/2, 0);
\coordinate (L) at (\Depth/2, 0, \Height/2);
\coordinate (M) at (0, \Width/2, \Height/2);
\coordinate (P) at (\Depth/2, \Width/2, \Height/2);

\draw[black,fill=blue!20,opacity=0.8] (O) -- (C) -- (G) -- (D) -- cycle;
\draw[black,fill=purple!20] (O) -- (A) -- (E) -- (D) -- cycle;
\draw[black,fill=blue!10] (O) -- (A) -- (B) -- (C) -- cycle;
\draw[black,fill=blue!20,opacity=0.8] (D) -- (E) -- (F) -- (G) -- cycle;
\draw[black,fill=blue!20,opacity=0.6] (C) -- (B) -- (F) -- (G) -- cycle;
\draw[black,fill=blue!20,opacity=0.8] (A) -- (B) -- (F) -- (E) -- cycle;

\draw[black] (J) -- (H);
\draw[black] (I) -- (J);
\draw[black] (K) -- (I);
\draw[transform canvas={yshift=2pt, xshift=4pt}] (K) -- (M);
\draw[transform canvas={yshift=0pt, xshift=5pt}] (K) -- (M);

\draw[black] (L) -- (M);

\draw[transform canvas={yshift=0pt, xshift=-1pt}] (H) -- (L);
\draw[transform canvas={yshift=0pt, xshift=1pt}] (H) -- (L);

\draw[red] (P) -- (H);
\draw[red] (P) -- (I);
\draw[red] (P) -- (J);
\draw[red] (P) -- (K);
\draw[red] (P) -- (L);
\draw[red] (P) -- (M);

\node[style={draw,shape=circle,fill=black}, label=$Q_5$] at (H) (p1) {};
\node[style={draw,shape=circle,fill=black}, label=$Q_3$] at (I) (p2) {};
\node[style={draw,shape=circle,fill=black}, label=right:{$Q_4$}] at (J) (p3) {};
\node[style={draw,shape=circle,fill=black}, label=below:{$Q_2$}] at (K) (p4) {};
\node[style={draw,shape=circle,fill=black}, label=below:{$Q_6$}] at (L) (p5) {};
\node[style={draw,shape=circle,fill=black}, label=left:{$Q_1$}] at (M) (p6) {};
\node[style={draw,shape=circle,fill=red}] at (P) (l1) {};

\node[style={draw,shape=star,fill=yellow,star points =5, star point ratio=0.5}, label=$v_2$] at (A) (a) {};
\node[style={draw,shape=star,fill=yellow,star points =5, star point ratio=0.5}, label=below:{$v_6$}] at (C) (c) {};
\node[style={draw,shape=star,fill=yellow,star points =5, star point ratio=0.5}, label=$v_3$] at (E) (e) {};
\node[style={draw,shape=star,fill=yellow,star points =5, star point ratio=0.5}, label=$v_4$] at (F) (f) {};
\node[style={draw,shape=star,fill=yellow,star points =5, star point ratio=0.5}, label=below:{$v_5$}] at (G) (g) {};
\node[style={draw,shape=star,fill=yellow,star points =5, star point ratio=0.5}, label=below:{$v_1$}] at (O) (o) {};

\end{tikzpicture}
\caption{AME(7,3) graph state embedded on a cube $C$.}
\label{ame73cube}
\end{figure}
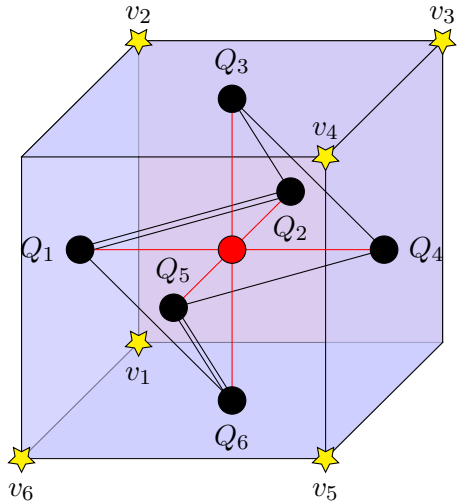

\begin{enumerate}
\item Consider the cyclically ordered set of all faces $\mathcal{F} = \{F_1, F_2, F_3, F_4, F_5, F_6\}$ and label a single vertex $v_\alpha$ for some $\alpha \in \{1,\dots,6\}$. Then label the adjacent faces $F_{\alpha -1}, F_{\alpha}, F_{\alpha+1}$. For example, label a single vertex $v_1$ and then label the adjacent faces $F_6, F_1, F_2$.
\item By construction, there will now be another vertex where both $F_\alpha$ and $F_{\alpha+1}$ meet. This vertex must then be labelled $v_{\alpha+1}$. There will also be one unlabelled face that is adjacent to both $F_\alpha$ and $F_{\alpha+1}$, which one labels $F_{\alpha+2}$. In the previous example one can hence label $v_2$ and $F_3$.
\item This process can be repeated until all faces are labelled together with the 6 vertices $v_1, \dots v_6$. One can then place a single qutrit on each of these faces labelled $Q_1, \dots, Q_6$ for faces $F_1, \dots, F_6$ respectively. The resultant diagram representing the assignment of these qutrits is shown in Figure \ref{faceslabel}.
\item One can now connect qutrits $Q_1, \dots, Q_6$ according to the graph state representation of the AME(7,3) state (Figure \ref{ame7qutrit}). The resulting graph state embedded in the cube is visualised in Figure \ref{ame73cube}.
\end{enumerate}

Now that we have a representation for the AME(7,3) state embedded on the cube, we can ask the question of how the qutrits living on the cube's faces can be concatenated. Consider two cubes $C$ and $C'$, with qutrits $Q_1, \dots, Q_6$ and $Q'_1, \dots, Q'_6$ living on each of the cube's faces respectively, assigned as depicted in Figure \ref{ame73cube}. We can concatenate $C$ and $C'$ by joining them facewise and since the graph states are AME, we can then maximally entangle the pair of qutrits (one from each face) using entanglement swapping as shown in section \ref{concat}. 

\bigskip

To maximise the symmetry of the resulting graph in the tessellation building $H^3$, one aims to consistently match pairs of qutrits. Qutrits can be categorised based on the weights of the incoming edges in the graph state, for example in Figure \ref{ame73cube}, qutrits $Q_1, Q_2, Q_5$ and $Q_6$ are associated to one weight-1 and one weight-2 operator, while qutrits $Q_3$ and $Q_4$ are associated with two weight-1 operators (in neither case including the one weight-1 operator connected with the logical qutrit). 

The simplest way to match pairs based on this, is to match qutrits $Q_i$ and $Q'_i$, $i \in \{1,\dots,6\}$ when concatenating cubes $C$ and $C'$. However, in order to successfully accomplish this, it is trivial to see that one cannot simply match the two cubes faces together without performing some operation on cube $C'$ (e.g. rotating the cube about some central axis).

\bigskip

\begin{figure}[h!]
\centering
\begin{tikzpicture}[transform shape, scale=0.6]
\coordinate (O) at (0,0,0);
\coordinate (A) at (0,\Width,0);
\coordinate (B) at (0,\Width,\Height);
\coordinate (C) at (0,0,\Height);
\coordinate (D) at (\Depth,0,0);
\coordinate (E) at (\Depth,\Width,0);
\coordinate (F) at (\Depth,\Width,\Height);
\coordinate (G) at (\Depth,0,\Height);

\coordinate (H) at (\Depth/2,\Width/2,\Height);
\coordinate (I) at (\Depth/2,\Width,\Height/2);
\coordinate (J) at (\Depth, \Width/2, \Height/2);
\coordinate (K) at (\Depth/2,\Width/2, 0);
\coordinate (L) at (\Depth/2, 0, \Height/2);
\coordinate (M) at (0, \Width/2, \Height/2);
\coordinate (P) at (\Depth/2, \Width/2, \Height/2);

\draw[black,fill=blue!20,opacity=0.8] (O) -- (C) -- (G) -- (D) -- cycle;
\draw[black,fill=purple!20] (O) -- (A) -- (E) -- (D) -- cycle;
\draw[black,fill=blue!10] (O) -- (A) -- (B) -- (C) -- cycle;
\draw[black,fill=blue!20,opacity=0.8] (D) -- (E) -- (F) -- (G) -- cycle;
\draw[black,fill=blue!20,opacity=0.6] (C) -- (B) -- (F) -- (G) -- cycle;
\draw[black,fill=blue!20,opacity=0.8] (A) -- (B) -- (F) -- (E) -- cycle;

\node[style={draw,shape=circle,fill=black}, label=$x_1$] at (A) (a) {};
\node[style={draw,shape=circle,fill=black}, label=$x_2$] at (B) (b) {};
\node[style={draw,shape=circle,fill=black}, label=below:{$x_8$}] at (C) (c) {};
\node[style={draw,shape=circle,fill=black}, label=below:{$x_6$}] at (D) (d) {};
\node[style={draw,shape=circle,fill=black}, label=$x_4$] at (E) (e) {};
\node[style={draw,shape=circle,fill=black}, label=$x_3$] at (F) (f) {};
\node[style={draw,shape=circle,fill=black}, label=below:{$x_5$}] at (G) (g) {};
\node[style={draw,shape=circle,fill=black}, label=below:{$x_7$}] at (O) (o) {};

\node[] at (H) (h) {$F_5$};
\node[] at (I) (i) {$F_3$};
\node[] at (J) (j) {$F_4$};
\node[] at (K) (k) {$F_2$};
\node[] at (L) (l) {$F_6$};
\node[] at (M) (m) {$F_1$};

\coordinate (Oo) at (14,0,0);
\coordinate (Aa) at (14,\Width,0);
\coordinate (Bb) at (14,\Width,0\Height);
\coordinate (Cc) at (14,0,\Height);
\coordinate (Dd) at (14+\Depth,0,0);
\coordinate (Ee) at (14+\Depth,\Width,0);
\coordinate (Ff) at (14+\Depth,\Width,\Height);
\coordinate (Gg) at (14+\Depth,0,\Height);

\coordinate (Hh) at (14+\Depth/2,\Width/2,\Height);
\coordinate (Ii) at (14+\Depth/2,\Width,\Height/2);
\coordinate (Jj) at (14+\Depth, \Width/2, \Height/2);
\coordinate (Kk) at (14+\Depth/2,\Width/2, 0);
\coordinate (Ll) at (14+\Depth/2, 0, \Height/2);
\coordinate (Mm) at (14, \Width/2, \Height/2);
\coordinate (Pp) at (14+\Depth/2, \Width/2, \Height/2);

\draw[black,fill=blue!20,opacity=0.8] (Oo) -- (Cc) -- (Gg) -- (Dd) -- cycle;
\draw[black,fill=purple!20] (Oo) -- (Aa) -- (Ee) -- (Dd) -- cycle;
\draw[black,fill=blue!10] (Oo) -- (Aa) -- (Bb) -- (Cc) -- cycle;
\draw[black,fill=blue!20,opacity=0.8] (Dd) -- (Ee) -- (Ff) -- (Gg) -- cycle;
\draw[black,fill=blue!20,opacity=0.6] (Cc) -- (Bb) -- (Ff) -- (Gg) -- cycle;
\draw[black,fill=blue!20,opacity=0.8] (Aa) -- (Bb) -- (Ff) -- (Ee) -- cycle;

\node[style={draw,shape=circle,fill=black}, label=$x_5$] at (Aa) (aa) {};
\node[style={draw,shape=circle,fill=black}, label=$x_6$] at (Bb) (bb) {};
\node[style={draw,shape=circle,fill=black}, label=below:{$x_4$}] at (Cc) (cc) {};
\node[style={draw,shape=circle,fill=black}, label=below:{$x_2$}] at (Dd) (dd) {};
\node[style={draw,shape=circle,fill=black}, label=$x_8$] at (Ee) (ee) {};
\node[style={draw,shape=circle,fill=black}, label=$x_7$] at (Ff) (ff) {};
\node[style={draw,shape=circle,fill=black}, label=below:{$x_1$}] at (Gg) (gg) {};
\node[style={draw,shape=circle,fill=black}, label=below:{$x_3$}] at (Oo) (oo) {};

\node[] at (Hh) (hh) {$F_2$};
\node[] at (Ii) (ii) {$F_6$};
\node[] at (Jj) (jj) {$F_1$};
\node[] at (Kk) (kk) {$F_5$};
\node[] at (Ll) (ll) {$F_3$};
\node[] at (Mm) (mm) {$F_4$};

\draw[->, thick] (7,2,2) -- (11,2,2) node[midway, label=$I$] {};
\end{tikzpicture}
\caption{Inversion of $C$ mapping to $C'$.}
\label{facesofC}
\end{figure}
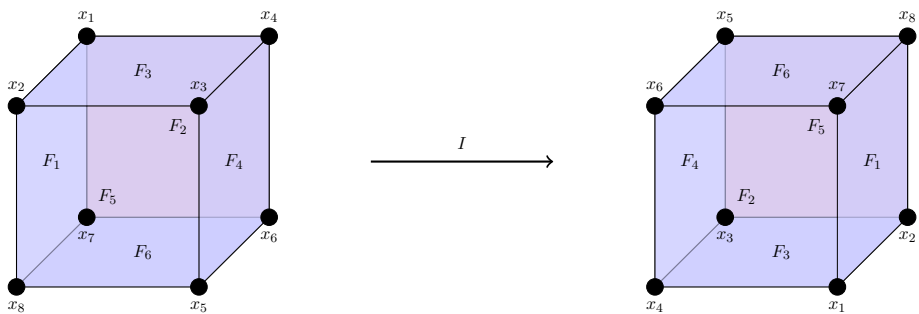

In order to preserve the internal structure of the graph state within the cube, one can attempt to choose a particular transformation mapping $C$ to $C'$ that exploits the innate symmetries of the cube. This transformation can be chosen to be an inversion, denoted $I: C \rightarrow C'$. Inversions in this context indicate mapping each vertex of the cube to the opposing vertex, and similarly with mapping each face to its opposing face, such that the centre of the cube lies at the origin. 

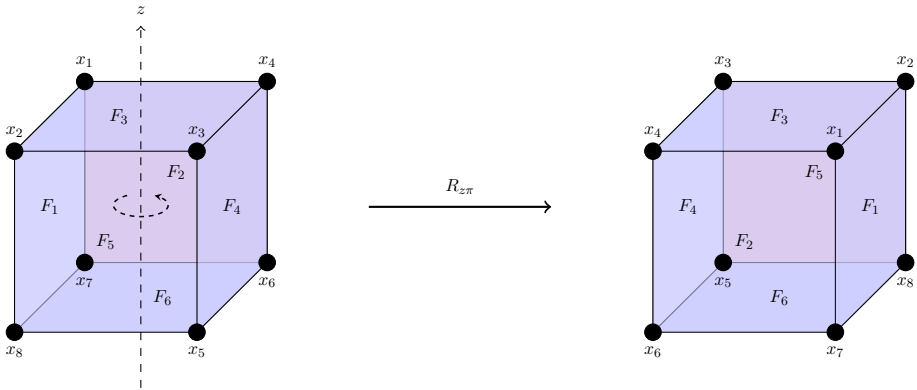
\begin{figure}[h!]
\centering
\begin{tikzpicture}[transform shape, scale=0.6]
\coordinate (O) at (0,0,0);
\coordinate (A) at (0,\Width,0);
\coordinate (B) at (0,\Width,\Height);
\coordinate (C) at (0,0,\Height);
\coordinate (D) at (\Depth,0,0);
\coordinate (E) at (\Depth,\Width,0);
\coordinate (F) at (\Depth,\Width,\Height);
\coordinate (G) at (\Depth,0,\Height);

\coordinate (H) at (\Depth/2,\Width/2,\Height);
\coordinate (I) at (\Depth/2,\Width,\Height/2);
\coordinate (J) at (\Depth, \Width/2, \Height/2);
\coordinate (K) at (\Depth/2,\Width/2, 0);
\coordinate (L) at (\Depth/2, 0, \Height/2);
\coordinate (M) at (0, \Width/2, \Height/2);
\coordinate (P) at (\Depth/2, \Width/2, \Height/2);

\draw[black,fill=blue!20,opacity=0.8] (O) -- (C) -- (G) -- (D) -- cycle;
\draw[black,fill=purple!20] (O) -- (A) -- (E) -- (D) -- cycle;
\draw[black,fill=blue!10] (O) -- (A) -- (B) -- (C) -- cycle;
\draw[black,fill=blue!20,opacity=0.8] (D) -- (E) -- (F) -- (G) -- cycle;
\draw[black,fill=blue!20,opacity=0.6] (C) -- (B) -- (F) -- (G) -- cycle;
\draw[black,fill=blue!20,opacity=0.8] (A) -- (B) -- (F) -- (E) -- cycle;

\node[style={draw,shape=circle,fill=black}, label=$x_1$] at (A) (a) {};
\node[style={draw,shape=circle,fill=black}, label=$x_2$] at (B) (b) {};
\node[style={draw,shape=circle,fill=black}, label=below:{$x_8$}] at (C) (c) {};
\node[style={draw,shape=circle,fill=black}, label=below:{$x_6$}] at (D) (d) {};
\node[style={draw,shape=circle,fill=black}, label=$x_4$] at (E) (e) {};
\node[style={draw,shape=circle,fill=black}, label=$x_3$] at (F) (f) {};
\node[style={draw,shape=circle,fill=black}, label=below:{$x_5$}] at (G) (g) {};
\node[style={draw,shape=circle,fill=black}, label=below:{$x_7$}] at (O) (o) {};

\node[] at (H) (h) {$F_5$};
\node[label=left:{$F_3$}] at (I) (i) {};
\node[] at (J) (j) {$F_4$};
\node[] at (K) (k) {$F_2$};
\node[label=right:{$F_6$}] at (L) (l) {};
\node[] at (M) (m) {$F_1$};

\draw[dashed,->] (\Depth/2,-\Width/2,\Height/2) -- (\Depth/2,1.5*\Width,\Height/2) node [midway] {\AxisRotator[rotate=-90]};
\node[label = above:{$z$}] at (\Depth/2,1.5*\Width,\Height/2) (z) {};

\coordinate (Oo) at (14,0,0);
\coordinate (Aa) at (14,\Width,0);
\coordinate (Bb) at (14,\Width,0\Height);
\coordinate (Cc) at (14,0,\Height);
\coordinate (Dd) at (14+\Depth,0,0);
\coordinate (Ee) at (14+\Depth,\Width,0);
\coordinate (Ff) at (14+\Depth,\Width,\Height);
\coordinate (Gg) at (14+\Depth,0,\Height);

\coordinate (Hh) at (14+\Depth/2,\Width/2,\Height);
\coordinate (Ii) at (14+\Depth/2,\Width,\Height/2);
\coordinate (Jj) at (14+\Depth, \Width/2, \Height/2);
\coordinate (Kk) at (14+\Depth/2,\Width/2, 0);
\coordinate (Ll) at (14+\Depth/2, 0, \Height/2);
\coordinate (Mm) at (14, \Width/2, \Height/2);
\coordinate (Pp) at (14+\Depth/2, \Width/2, \Height/2);

\draw[black,fill=blue!20,opacity=0.8] (Oo) -- (Cc) -- (Gg) -- (Dd) -- cycle;
\draw[black,fill=purple!20] (Oo) -- (Aa) -- (Ee) -- (Dd) -- cycle;
\draw[black,fill=blue!10] (Oo) -- (Aa) -- (Bb) -- (Cc) -- cycle;
\draw[black,fill=blue!20,opacity=0.8] (Dd) -- (Ee) -- (Ff) -- (Gg) -- cycle;
\draw[black,fill=blue!20,opacity=0.6] (Cc) -- (Bb) -- (Ff) -- (Gg) -- cycle;
\draw[black,fill=blue!20,opacity=0.8] (Aa) -- (Bb) -- (Ff) -- (Ee) -- cycle;

\node[style={draw,shape=circle,fill=black}, label=$x_3$] at (Aa) (aa) {};
\node[style={draw,shape=circle,fill=black}, label=$x_4$] at (Bb) (bb) {};
\node[style={draw,shape=circle,fill=black}, label=below:{$x_6$}] at (Cc) (cc) {};
\node[style={draw,shape=circle,fill=black}, label=below:{$x_8$}] at (Dd) (dd) {};
\node[style={draw,shape=circle,fill=black}, label=$x_2$] at (Ee) (ee) {};
\node[style={draw,shape=circle,fill=black}, label=$x_1$] at (Ff) (ff) {};
\node[style={draw,shape=circle,fill=black}, label=below:{$x_7$}] at (Gg) (gg) {};
\node[style={draw,shape=circle,fill=black}, label=below:{$x_5$}] at (Oo) (oo) {};

\node[] at (Hh) (hh) {$F_2$};
\node[] at (Ii) (ii) {$F_3$};
\node[] at (Jj) (jj) {$F_1$};
\node[] at (Kk) (kk) {$F_5$};
\node[] at (Ll) (ll) {$F_6$};
\node[] at (Mm) (mm) {$F_4$};

\draw[->, thick] (7,2,2) -- (11,2,2) node[midway, label={$R_{z\pi}$}] {};
\end{tikzpicture}
\caption{Rotation of vertices around the $z$-axis mapping $C$ to $\tilde{C}$.}
\label{rotation}
\end{figure}

This is implemented by considering two cubes such that, the first, $C$, has the structure shown in Figure \ref{ame73cube} and the second, $C'$, is the inversion of $C$. This transformation can be visualised as shown in Figure \ref{facesofC}. The left side represents the vertices of the original cube $C$, while the right side gives the cube's inversion $C'$. Using the particular choice of the labelling of vertices in $C$, as indicated on the left hand side of Figure \ref{facesofC}, one may define the inversion operator as
\be
I: C \rightarrow C' := \{I(x_i) = x_{i+4} \; | \; \forall x_i \in C\}
\ee
given $x_i$ are the vertices of $C$, with $i \in \{1,\dots, 8\}$ and such that the arithmetic is modulo 8 (i.e. the element $i+4 \in \mathbbm{Z}_8$). Clearly the faces bound by four vertices are also switched with their opposing face as a result. Written in terms of the permutations of vertices, the inversion can be expressed as;
\be
I(C) = (x_1 x_5) (x_2 x_6) (x_3 x_7) (x_4 x_8).
\ee
Note that an inversion can be completed from the combination of a rotation by an angle of $180^\circ$ ($\pi$ rad) about an axis defined to pass through the centre of a face and the opposing face, together with a reflection in the plane perpendicular to that axis. For example, define a set of axes such that the z-axis passes through the centre of faces $F_3$ and $F_6$ in $C$, as shown on the left hand side of Figure \ref{rotation}. 

We define the operator $R_{z\pi}: C \rightarrow \tilde{C}$ as the rotation of each vertex in $C$, about the $z$-axis by $\pi$ radians. Applying the rotation $R_{z\pi}$ to the cube $C$ corresponds to the following permutations of vertices;
\be
R_{z\pi}(C) = (x_1 x_3) (x_2 x_4) (x_5 x_7) (x_6 x_8),
\ee
mapping $C$ to $\tilde{C}$. This mapping can be seen in Figure \ref{rotation}.

\bigskip

\begin{figure}[h!]
\centering
\begin{tikzpicture}[transform shape, scale=0.6]
\coordinate (O) at (0,0,0);
\coordinate (A) at (0,\Width,0);
\coordinate (B) at (0,\Width,\Height);
\coordinate (C) at (0,0,\Height);
\coordinate (D) at (\Depth,0,0);
\coordinate (E) at (\Depth,\Width,0);
\coordinate (F) at (\Depth,\Width,\Height);
\coordinate (G) at (\Depth,0,\Height);

\coordinate (H) at (\Depth/2,\Width/2,\Height);
\coordinate (I) at (\Depth/2,\Width,\Height/2);
\coordinate (J) at (\Depth, \Width/2, \Height/2);
\coordinate (K) at (\Depth/2,\Width/2, 0);
\coordinate (L) at (\Depth/2, 0, \Height/2);
\coordinate (M) at (0, \Width/2, \Height/2);
\coordinate (P) at (\Depth/2, \Width/2, \Height/2);

\draw[black,fill=blue!20,opacity=0.8] (O) -- (C) -- (G) -- (D) -- cycle;
\draw[black,fill=purple!20] (O) -- (A) -- (E) -- (D) -- cycle;
\draw[black,fill=blue!10] (O) -- (A) -- (B) -- (C) -- cycle;
\draw[black,fill=blue!20,opacity=0.8] (D) -- (E) -- (F) -- (G) -- cycle;
\draw[black,fill=blue!20,opacity=0.6] (C) -- (B) -- (F) -- (G) -- cycle;
\draw[black,fill=blue!20,opacity=0.8] (A) -- (B) -- (F) -- (E) -- cycle;

\node[style={draw,shape=circle,fill=black}, label=$x_3$] at (A) (a) {};
\node[style={draw,shape=circle,fill=black}, label=$x_4$] at (B) (b) {};
\node[style={draw,shape=circle,fill=black}, label=below:{$x_6$}] at (C) (c) {};
\node[style={draw,shape=circle,fill=black}, label=below:{$x_8$}] at (D) (d) {};
\node[style={draw,shape=circle,fill=black}, label=$x_2$] at (E) (e) {};
\node[style={draw,shape=circle,fill=black}, label=$x_1$] at (F) (f) {};
\node[style={draw,shape=circle,fill=black}, label=below:{$x_7$}] at (G) (g) {};
\node[style={draw,shape=circle,fill=black}, label=below:{$x_5$}] at (O) (o) {};

\draw[pattern=north west lines] (0, \Width/2, 0) -- (\Depth, \Width/2, 0) --  (\Depth, \Width/2, \Height) -- (0, \Width/2, \Height) -- cycle;

\coordinate (Oo) at (14,0,0);
\coordinate (Aa) at (14,\Width,0);
\coordinate (Bb) at (14,\Width,0\Height);
\coordinate (Cc) at (14,0,\Height);
\coordinate (Dd) at (14+\Depth,0,0);
\coordinate (Ee) at (14+\Depth,\Width,0);
\coordinate (Ff) at (14+\Depth,\Width,\Height);
\coordinate (Gg) at (14+\Depth,0,\Height);

\coordinate (Hh) at (14+\Depth/2,\Width/2,\Height);
\coordinate (Ii) at (14+\Depth/2,\Width,\Height/2);
\coordinate (Jj) at (14+\Depth, \Width/2, \Height/2);
\coordinate (Kk) at (14+\Depth/2,\Width/2, 0);
\coordinate (Ll) at (14+\Depth/2, 0, \Height/2);
\coordinate (Mm) at (14, \Width/2, \Height/2);
\coordinate (Pp) at (14+\Depth/2, \Width/2, \Height/2);

\draw[black,fill=blue!20,opacity=0.8] (Oo) -- (Cc) -- (Gg) -- (Dd) -- cycle;
\draw[black,fill=purple!20] (Oo) -- (Aa) -- (Ee) -- (Dd) -- cycle;
\draw[black,fill=blue!10] (Oo) -- (Aa) -- (Bb) -- (Cc) -- cycle;
\draw[black,fill=blue!20,opacity=0.8] (Dd) -- (Ee) -- (Ff) -- (Gg) -- cycle;
\draw[black,fill=blue!20,opacity=0.6] (Cc) -- (Bb) -- (Ff) -- (Gg) -- cycle;
\draw[black,fill=blue!20,opacity=0.8] (Aa) -- (Bb) -- (Ff) -- (Ee) -- cycle;

\node[style={draw,shape=circle,fill=black}, label=$x_5$] at (Aa) (aa) {};
\node[style={draw,shape=circle,fill=black}, label=$x_6$] at (Bb) (bb) {};
\node[style={draw,shape=circle,fill=black}, label=below:{$x_4$}] at (Cc) (cc) {};
\node[style={draw,shape=circle,fill=black}, label=below:{$x_2$}] at (Dd) (dd) {};
\node[style={draw,shape=circle,fill=black}, label=$x_8$] at (Ee) (ee) {};
\node[style={draw,shape=circle,fill=black}, label=$x_7$] at (Ff) (ff) {};
\node[style={draw,shape=circle,fill=black}, label=below:{$x_1$}] at (Gg) (gg) {};
\node[style={draw,shape=circle,fill=black}, label=below:{$x_3$}] at (Oo) (oo) {};

\node[] at (Hh) (hh) {$F_2$};
\node[] at (Ii) (ii) {$F_6$};
\node[] at (Jj) (jj) {$F_1$};
\node[] at (Kk) (kk) {$F_5$};
\node[] at (Ll) (ll) {$F_3$};
\node[] at (Mm) (mm) {$F_4$};

\draw[->, thick] (7,2,2) -- (11,2,2) node[midway, label={$\gamma_z$}] {};
\end{tikzpicture}
\caption{Reflection of vertices in the plane $z=0$ mapping $\tilde{C}$ to $C'$.}
\label{mirrorz}
\end{figure}
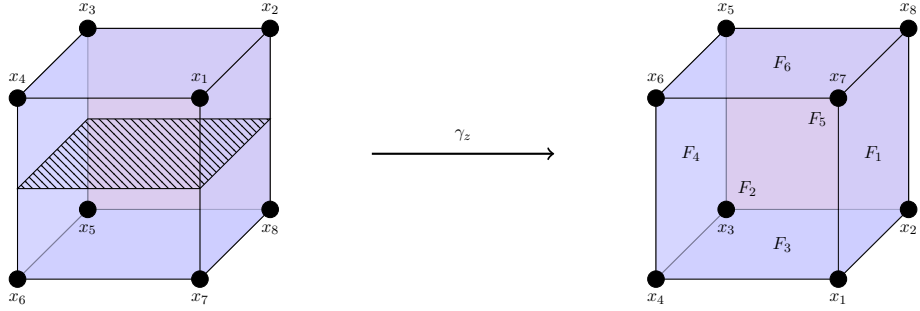

We can subsequently apply the reflection through the plane $z=0$, denoted by $\gamma_z$, to the new cube $\tilde{C}$, with permutations
\be
\gamma_z (\tilde{C}) = (x_1 x_7) (x_2 x_8) (x_3 x_5) (x_4 x_6),
\ee
as shown in Figure \ref{mirrorz}. This results in the cube $C'$, thus showing this set of transformations is equivalent to the inversion shown in Figure \ref{facesofC}.

\bigskip

We should note that performing an inversion on a cube does not change the structure of the graph state within $C$, but simply provides a `mirror' of it. Since faces of $C$ and $C'$ are opposite by construction, when building a tessellation and thus concatenating states, one may join $C$ and $C'$ in any direction. For example, in $C$, $F_4$ is on the `right' face whereas in $C'$, $F_4$ is on the `left' face (as indicated in Figure \ref{facesofC}). Thus these can be joined, concatenating qubits $Q_4$ in $C$ and $Q'_4$ in $C'$. 

\bigskip

We could do this with each face of cube $C$. This central cube would thus be joined to six cubes at its faces such that each of these adjacent cubes are the inversions of $C$, given by $C'$. However, one could also provide a construction so each cube $C'$ may be attached to six cubes such that all of the adjacent cubes are given by $C$. Since applying the inversion operator twice results in the identity $I^2 = e$, and hence $I(C') = I^2(C) = C$, these two constructions are the same and can be consistently built. 

\begin{figure}[h!]
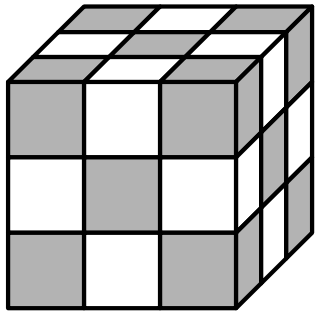

\centering
 \RubikFaceUp
        {X}{W}{X}
        {W}{X}{W}
        {X}{W}{X}
    \RubikFaceRight
        {X}{W}{X}
        {W}{X}{W}
        {X}{W}{X}
    \RubikFaceFront
        {X}{W}{X}
        {W}{X}{W}
        {X}{W}{X}
   \ShowCube{7cm}{1}{\DrawRubikCube}
\caption{Tessellation of cubes $C$ and $C'$ forming a checkerboard cubic honeycomb in $\mathbbm{R}^3$.}
\label{checkerboard}
\end{figure}

Thus one can tessellate the entirety of $R^3$ with cubes $C$ and $C'$ such that qubits $Q_i$ are only ever concatenated with qubits $Q'_i$ and $C'$ is the inversion of $C$. This construction may be thought of as a 3-dimensional checkerboard where one draws all cubes $C$ to be white cubes and all cubes $C'$ to be grey cubes, in turn filling $H^3$, as depicted in Figure \ref{checkerboard}. Clearly, this construction has no preferred direction and while not maximising symmetry on a single polytope (which is achieved by the construction in Figure \ref{cubeAdS4}), there is a clear structure with remarkably high symmetry, when considering the full three-dimensional space.

\bigskip

\bigskip

Now let us return to the cubic tessellation of $H^3$. In $H^3$ five cubes meet at each edge and accordingly the construction described above does not work and one cannot consistently concatenate the AME codes associated with each cube in hyperbolic space. It is possible that a similar construction may exist however it is extremely non-trivial, thus one presents the simpler case of the $\{5,3,4\}$ honeycomb in the following section.

\subsection{AdS$_4$: Other regular tessellations}
\label{othertess}

Turning to the other regular tessellations of H$^3$ (the icosahedral honeycomb $\{3,5,3\}$, the order-4 dodecahedral honeycomb $\{5,3,4\}$ and the order-5 dodecahedral honeycomb $\{5,3,5\}$), the main conclusions are unchanged: one cannot realise AME states with qubits. The polytope tessellated in the icosahedral honeycomb is the icosahedron $\{3,5\}$ and both other cases are constructed from the dodecahedron $\{5,3\}$. Just as for the cubic case one can construct two types of graph states corresponding to each polytope. 

Interestingly, in this construction, the graph state is a graph corresponding to the connectivity of the faces of the polytope. However, when representing Platonic solids as graphs, one usually considers the graph corresponding to the connectivity of the vertices and this is how many graphs have been stated in the literature. Hence, when sketching a graph state for a polytope $p$, one can effectively think of drawing the graph $G$ for the dual polytope $\tilde{p}$, corresponding to the connectivity of the vertices. 

\bigskip

\begin{figure}[h!]
\centering
\begin{tikzpicture}[mystyle/.style={draw,shape=circle,fill=black, inner sep=0pt, minimum size=7pt}]
	\GraphInit[vstyle=Simple]
	\SetUpEdge[lw  = 0.5pt]     
	\SetVertexNoLabel
	\tikzset{EdgeStyle/.append style={black}}
	\tikzset{VertexStyle/.append style={mystyle}}
	\grCirculant[RA=3]{12}{1}
	\draw (a3) -- (a0);
	\draw (a3) -- (a11);	
	\draw (a3) -- (a7);
	\draw (a1) -- (a10);
	\draw (a1) -- (a9);	
	\draw (a1) -- (a5);
	\draw (a11) -- (a8);
	\draw (a11) -- (a7);	
	\draw (a9) -- (a6);
	\draw (a9) -- (a5);	
	\draw (a7) -- (a4);
	\draw (a5) -- (a2);
	\draw (a2) -- (a0);
	\draw (a4) -- (a2);	
	\draw (a6) -- (a4);
	\draw (a8) -- (a6);
	\draw (a10) -- (a8);	
	\draw (a0) -- (a10);
	\tikzset{VertexStyle/.append style={draw,shape=circle,fill=red, inner sep=0pt, minimum size=7pt}}
	\Vertex{d1}
	\SetUpEdge[lw  = 0.5pt]     
	\tikzset{EdgeStyle/.append style={red}}
	\EdgeFromOneToAll{d}{a}{1}{12}
\end{tikzpicture}
\caption{Graph state preserving maximal discrete symmetry for regular, uniform tessellations of $H^3$ consisting of dodecahedrons, commonly referred to as an `icosahedral graph' in the mathematics literature.}
\label{dodecgs}
\end{figure}
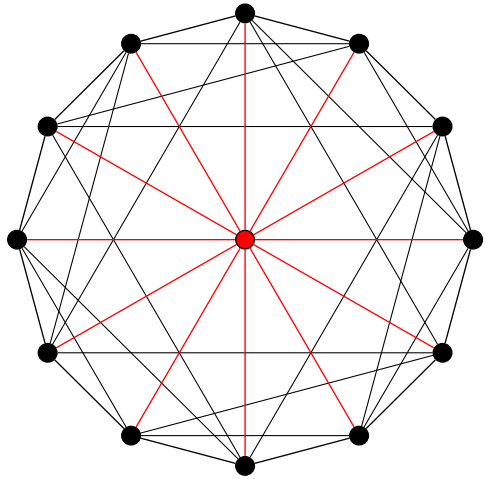

\vspace{1cm}

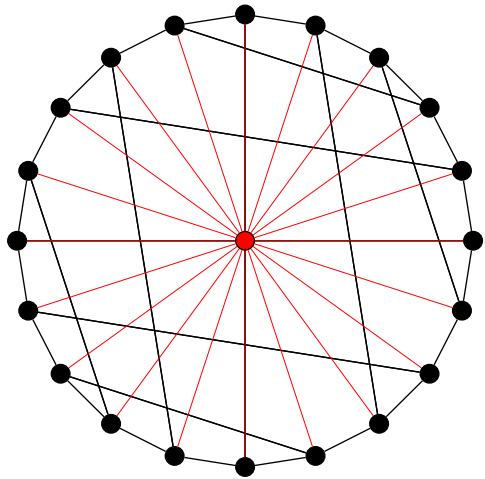
\begin{figure}[h!]
\centering
\begin{tikzpicture}[mystyle/.style={draw,shape=circle,fill=black, inner sep=0pt, minimum size=7pt}]
	\GraphInit[vstyle=Simple]
	\SetUpEdge[lw  = 0.5pt]     
	\SetVertexNoLabel
	\tikzset{EdgeStyle/.append style={black}}
	\tikzset{VertexStyle/.append style={mystyle}}
	\grDodecahedral[form=4, RA=3]
	\tikzset{VertexStyle/.append style={draw,shape=circle,fill=red, inner sep=0pt, minimum size=7pt}}
	\Vertex{d1}
	\SetUpEdge[lw  = 0.3pt]     
	\tikzset{EdgeStyle/.append style={red}}
	\EdgeFromOneToAll{d}{a}{1}{20}
\end{tikzpicture}
\caption{Graph state preserving maximal discrete symmetry for regular, uniform tessellations of $H^3$ consisting of icosahedrons, commonly referred to as a `dodecahedral graph' in the mathematics literature.}
\label{iscosgs}
\end{figure}

For example, drawing the graph state for the icosahedron, gives rise to the graph that is often named the `dodecahedral' graph. Similarly, the graph state for the dodecahedron corresponds to what is commonly referred to as the `icosahedral' graph. Note that for the cube this issue was not discussed since the cube's dual is simply the cube itself. The two graph states corresponding to tessellations of $H^3$ consisting of dodecahedrons and icosahedrons which preserve maximal discrete symmetry are depicted in Figure \ref{dodecgs} and Figure \ref{iscosgs} respectively. 

\bigskip

As for the previous discussion for the order-5 cubic honeycomb using a cubic tessellation, since one cannot have graph states with $n\geq7$ qudits that is both AME and preserves maximal discrete symmetry, we must choose to relax one or both of these assumptions. In this section, we present both a construction using non-AME graph states that preserves maximal discrete symmetry and a construction using AME states that eases the restriction on maximal discrete symmetry.

\subsubsection{Order-4 dodecahedral honeycomb non-AME state}

Similarly to the order-5 cubic honeycomb, we first try to build a maximally symmetric model for the order-4 dodecahedral honeycomb. We begin by embedding the graph state (Figure \ref{dodecgs}) into a single dodecahedron. Here, each of the qudits is placed such that one lives on each face of the dodecahedron with a singular logical qudit represented in the centre of the dodecahedron. Thus, edges connecting qudits in the graph state correspond to two faces being adjacent. 

Since we already know this graph state cannot be AME, we do not need to explicitly check this. However, it is required to check that this graph does in fact represent a quantum error correcting code. Here we choose the local dimension to be $D=2$ for simplicity to demonstrate that the graph is a code. The resulting stabiliser matrix for the graph state in Figure \ref{dodecgs} can therefore be expressed as
\begin{equation}
\mathcal{S}_{13} = \begin{pmatrix} X & Z & Z & Z & Z & Z & Z & Z & Z & Z & Z & Z & Z \\ Z & X & Z & I & Z & Z & I & I & I & Z & I & I & Z \\ Z & Z & X & Z & Z & I & I & I & I & I & I & Z & Z \\ Z & I & Z & X & Z & I & Z & Z & I & I & I & Z & I \\ Z & Z & Z & Z & X & Z & Z & I & I & I & I & I & I \\ Z & Z & I & I & Z & X & Z & I & Z & Z & I & I & I \\ Z & I & I &  Z & Z & Z & X & Z & Z & I & I & I & I \\ Z & I & I & Z & I & I & Z & X & Z & I & Z & Z & I \\ Z & I & I & I & I &  Z & Z & Z & X & Z & Z & I & I \\ Z & Z & I &  I & I & Z & I & I & Z & X & Z & I & Z \\ Z & I & I & I & I & I & I &  Z & Z & Z & X & Z & Z\\ Z & I & Z & Z & I &  I & I & Z & I & I & Z & X & Z \\ Z & Z & Z & I & I & I & I & I & I &  Z & Z & Z & X \end{pmatrix}.
\end{equation}

We can then repeat the analysis used for the maximally symmetric graph state for a cube (Figure \ref{7graph}) however this time using the dodecahedron graph state (Figure \ref{dodecgs}). We note that it is easier to use the restated version of the SW-theorem presented in \cite{Cafaro2014}:

\bigskip

\textbf{Restated SW-theorem:}  \textit{Consider a set of input vertices $X$ and a set of output vertices $Y$. Then, given a finite abelian group $G$ and the weighted graph with adjacency (coincidence) matrix $\Xi$, a quantum error correcting code $v_\Xi$ is able to detect an error configuration $\mathcal{E} \subset Y$ iff given}
\be
\label{restate1}
d^X = 0, \quad \text{\textit{and}} \quad \Xi^X_\mathcal{E} d^\mathcal{E} = 0
\ee
\textit{then}
\be
\label{restate2}
\Xi^{I}_{X \cup \mathcal{E}} d^{X \cup \mathcal{E}} = 0 \Rightarrow d^{X \cup \mathcal{E}} = 0, \qquad \textit{with $I= Y \backslash \mathcal{E}$.}
\ee

\bigskip

We need to apply this theorem to the ${12\choose2} = 66$ possible two-error configurations. Here we omit the full details of this lengthy calculation but after performing this analysis it turns out that the graph state shown in Figure \ref{dodecgs} is indeed a valid quantum error correcting code. While further analysis detailing some of the properties of this code would be very interesting, we do not deeply investigate them here as it is sufficient for our construction to show that one is indeed working with a quantum error correcting code.

\bigskip

We now construct the tessellation of the spatial slices of AdS$_4$ in the form of the order-4 dodecahedral honeycomb by using a consistent method to tessellate the dodecahedrons together, associating labels to each face and gluing them together. For simplicity, we present the arguments for the two-dimensional analogue of the order-4 dodecahedral tessellation, which corresponds to the familiar pentagonal tiling of H$^2$, with Schl\"{a}fi symbol $\{5,4\}$. 

\bigskip

\begin{figure}[h]
\begin{center}
\includegraphics [scale =0.4] {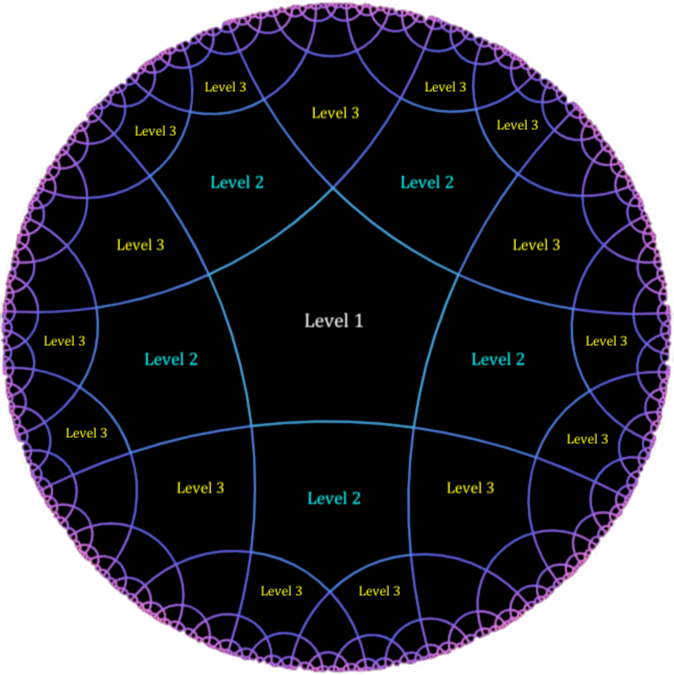}
\end{center}
\caption{\label{pent54} $\{5,4\}$ tessellation of hyperbolic plane H$^2$. We show subsequent levels beginning from some central polytope defined to be at level $L=1$ and adding a level for each polytope adjacent to a polytope at the previous level.}
\end{figure}

We may then simply use the method presented in section \ref{nonamecube}, where one associates a distinct set of labels $e_i$ to each pentagon (where now $i \in \{1,\dots, 5\}$) at level $L$ and glues pentagons by edges accordingly to level $L'$. The relevant diagram showing the $\{5,4\}$ tessellation and subsequent levels beginning from some central pentagon at level $L=1$ is shown in Figure \ref{pent54}. 

\bigskip

\bigskip

Similarly to the $\{7,4\}$ tessellation of H$^2$, we must divide levels into sub-levels from level $L=3$ onwards for the $\{5,4\}$ tessellation. However, as explained in section \ref{nonamecube}, since we are now working with a case in which $q=4$, there are only two sub-levels for all levels $L>3$. Hence, one can continuously build this consistent tessellation for any number of discrete levels approaching the spatial boundary of the spacetime.

The construction works in the exact same manner for the full three-dimensional order-4 dodecahedral honeycomb although it is more difficult to visualise. Here, polytopes are matched face-wise rather than edge-wise but otherwise the construction trivially generalises. The important feature is that one now has Sch\"{a}fi symbol $\{5,3,4\}$ and thus there are four dodecahedra around each edge (since $r=4$), analogous to the four pentagons around each edge in the $\{5,4\}$ tessellation. Therefore, one will similarly have only two sub-levels for all levels $L>3$.

\subsubsection{Order-4 dodecahedral honeycomb AME state}

When discussing the order-5 cubic honeycomb in section \ref{amecube}, we began by considering the graph state corresponding to the 7 qutrit AME state denoted AME(7,3). If we wish to begin with an analogous construction for the order-4 dodecahedral honeycomb, we require a 13 qudit AME state since a dodecahedron possesses 12 faces, on each of which one of these qudits lives, with one qudit associated with the dodecahedron itself. 

\bigskip

Clearly the AME(13,2) state does not exist \cite{Huber2016AbsolutelyExist} due to the theorems we have previously mentioned. It turns out that in local dimension $D=3$, no AME state containing 13 parties exists \cite{HuberShadow} (i.e. AME(13,3) does not exist) due to constraints imposed by the so-called shadow inequalities. In fact, there is very little literature about the existence of any AME(13,$D$) state. The AME(13,7) and AME(13,8) states are thought to exist due to arguments presented in \cite{MDScodes} (e.g. the existence of the $[[14,0,8]]_7$ QMDS code), though little is known about their properties.

\bigskip

Since not much is known about AME(13,$D$) states, to our knowledge no corresponding graph states have been produced in the literature. However, one would expect that some states do exist provided the AME state is a stabiliser state. In what follows, we assume that for a certain local dimension $D$ there does exist a stabiliser state AME(13,$D$) and that one can draw a corresponding graph state. 

\bigskip

The advantage of tessellating H$^3$ using the order-4 dodecahedral honeycomb over the order-5 cubic honeycomb is that when considering their Schl\"{a}fi symbols $\{p,q,r\}$, the order-4 dodecahedral honeycomb has $r$ even, while the order-5 cubic honeycomb has $r$ odd. Further, since $r=4$, one can draw many parallels with the order-4 cubic honeycomb that tessellates $\mathbbm{R}^3$. Recall the procedure we used for this Euclidean tessellation. We began by attempting to maximise the symmetry of the embedded graph state within the polytope (cube). Then we explicitly defined an inversion mapping $I$ as a product of a rotation $R_{z\pi}$ and a reflection $\gamma_z$, such that it possessed the property $I^2 = e$, where $e$ is the identity. We then had two possible cubes, $C$ and its inversion $C'$. When tessellating $\mathbbm{R}^3$, each face of $C$ associated with a qutrit $Q_i$ would be glued to a face of $C'$ associated with a qutrit $Q'_i$ and vice versa (see Figure \ref{checkerboard}).

\bigskip

\begin{figure}[h]
\begin{center}
\includegraphics [scale =0.3] {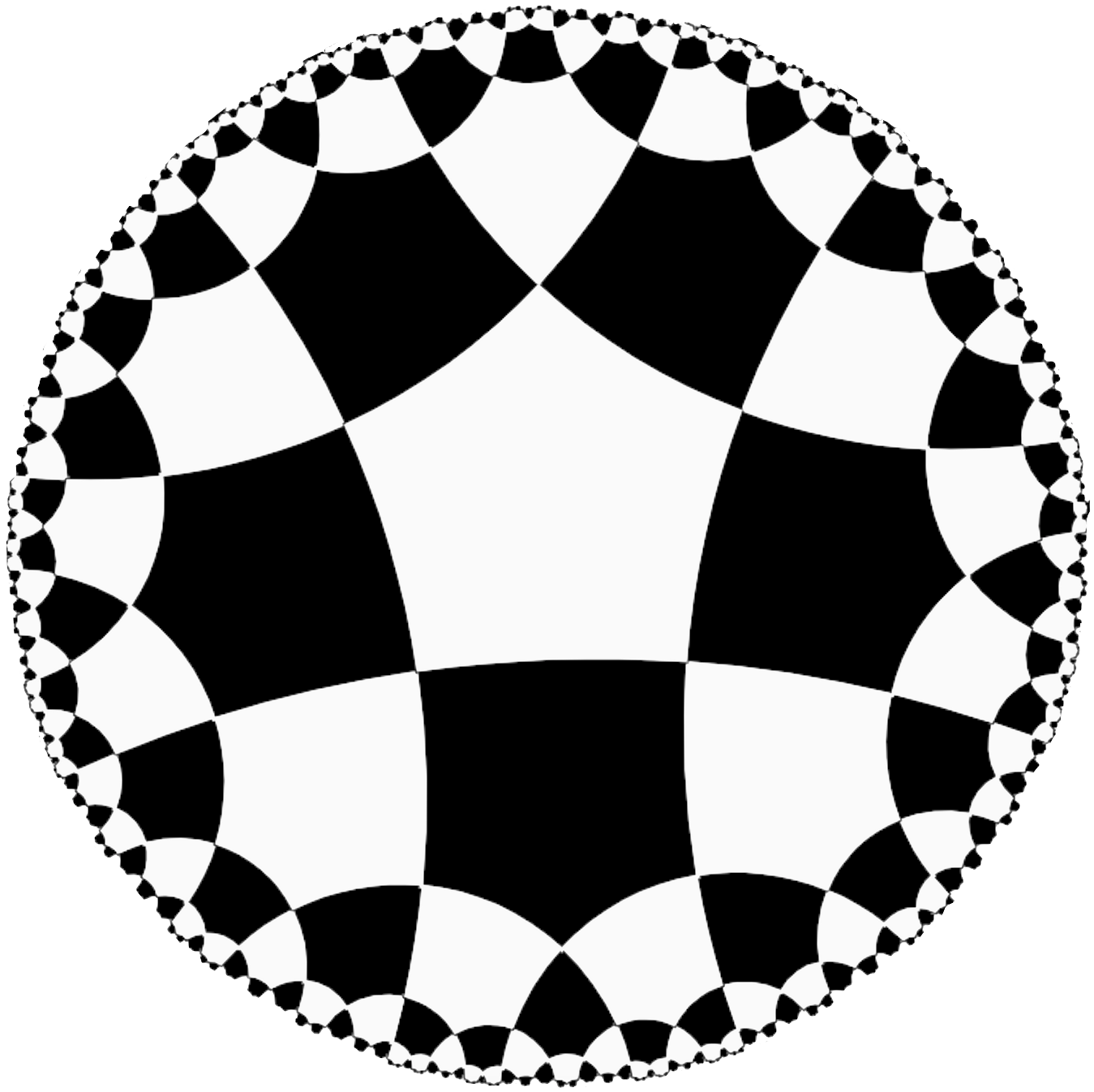}
\end{center}
\caption{\label{pent54bw} Checkerboard of the $\{5,4\}$ tessellation of H$^2$, analogous to the checkerboard construction of the order-4 dodecahedral honeycomb in H$^3$.}
\end{figure}

In order to produce a similar construction for the order-5 dodecahedral honeycomb, we would begin as for the order-4 cubic honeycomb in $\mathbbm{R}^3$ case by attempting to maximise the symmetry of the embedded graph state within the polytope. Since dodecahedrons have a more complex structure than cubes, it is likely this would be less straightforward. One would then define an inversion for the dodecahedron analagous to that of the cube consisting of rotations and reflections, so that after the inversion, all faces and vertices are `opposite' to those in the original dodecahedron. One would once again then have two types of polytope, the original and it's inversion. One can think of this tessellation therefore to be alternating between black dodecahedra and white dodecahedra. Since visualising this tessellation in hyperbolic geometry is challenging, we present its analogous counterpart in Figure \ref{pent54bw}, alternating between black and white pentagons forming a checkerboard in H$^2$.

\subsection{AdS$_5$ and higher dimensions}

One may wonder whether the obstructions encountered in applying the HaPPY construction to AdS$_4$ apply generically in higher dimensions. It is straightforward to show that they indeed do, exemplifying the discussion with AdS$_5$. Spatial slices of AdS$_5$ correspond to the hyperbolic 4-space $H^4$ and its possible regular uniform tessellations are given by the Schl\"{a}fi symbols given in (\ref{regular4d}). Since four dimensional objects are hard to visualise, it is useful to introduce the {\it configuration matrix} $C$, an object that stores all of the relevant information about a polytope within a single matrix. 

\textbf{Definition:} For any regular $n$-dimensional polytope consisting of $N_i$ $i$-faces with $0 \le i < n$, one can write its configuration matrix such that for each $i$-face element $N_i$, the number of $j$-faces incident is denoted $N_{ij}$, where $i\neq j$. Trivially, 
\be
N_i N_{ij} = N_j N_{ji}.
\ee
One can construct a matrix from these configurational numbers with $N_i = N_{ii}$ as the diagonal elements and $N_{ij}$ as the non-diagonal elements;
\be 
\begin{bmatrix} 
N_{0,0} & N_{0,1} & N_{0,2} & \dots & N_{0,n-1} \\
N_{1,0} & N_{1,1} & N_{1,2} & \dots & N_{1,n-1} \\
\vdots & \vdots & \vdots & \ddots & \vdots \\
N_{n-1,0} & N_{n-1,1} & N_{n-1,2} & \dots & N_{n-1,n-1} \\
\end{bmatrix}.
\ee

\bigskip

From (\ref{regular4d}), we note that the first possible hyperbolic tessellation of $H^4$ is the order-5 5-cell honeycomb with Schl\"{a}fi symbol $\{3,3,3,5\}$. The convex, regular 4-dimensional polytope associated with this tessellation is the 5-cell (also referred to as the 4-simplex), with Schl\"{a}fi symbol $\{3,3,3\}$ and it is bounded by five regular tetrahedra. Hence, provided one follows the HaPPY approach, for each 5-cell there will be five physical qudits (one for each tetrahedra) and one logical qudit associated with the 5-cell, represented by being placed in it's centre. 

In order to draw the maximal discrete symmetry-preserving graph state, we need to know which tetrahedra are connected via faces to one another. In this instance we can show each tetrahedron is connected to four other tetrahedra - one for each of its faces. The configuration matrix is rather useful to summarise all of this information
\be
C_5 = \begin{bmatrix} 
5 & 4 & 6 & 4 \\
2 & 10 & 3 & 3 \\
3 & 3 & 10 & 2 \\
4 & 6 & 4 & 5 \\
\end{bmatrix}.
\ee

\begin{figure}[h!]
\centering
\begin{tikzpicture}[mystyle/.style={draw,shape=circle,fill=black, inner sep=0pt, minimum size=7pt}]
	\GraphInit[vstyle=Simple]
	\SetUpEdge[lw  = 0.5pt]     
	\SetVertexNoLabel
	\tikzset{EdgeStyle/.append style={black}}
	\tikzset{VertexStyle/.append style={mystyle}}
	\grComplete[RA=3] {5}
	\tikzset{VertexStyle/.append style={draw,shape=circle,fill=red, inner sep=0pt, minimum size=7pt}}
	\Vertex{d1}
	\SetUpEdge[lw  = 0.5pt]     
	\tikzset{EdgeStyle/.append style={red}}
	\EdgeFromOneToAll{d}{a}{1}{5}
\end{tikzpicture}
\caption{Graph state preserving maximal discrete symmetry for regular, uniform tessellations of $H^4$ consisting of 5-cells, a 4-dimensional polytope bounded by five tetrahedra.}
\label{5cell}
\end{figure}
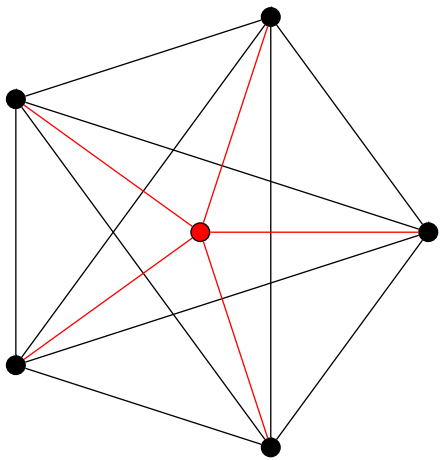

The diagonal shows that the 5-cell has a total of 5 vertices, 10 edges, 10 faces and 5 cells. Importantly, the element in the fourth row and third column indicates that each cell has 4 faces showing these are  comprised of tetrahedra. Since each physical qudit lives on its own tetrahedron, each physical qudit must therefore be connected to 4 other physical qudits in the graph state. However, since there are only 5 physical qudits in the system, the graph state must be maximally connected, as shown in Figure \ref{5cell}, also imposing the central logical qudit is connected to each physical qudit.

Trivially, since this graph is maximally connected, then for any bipartition, one cannot obtain linearly independent vectors when performing the standard check and so this state is not AME. This result can similarly be obtained using the fact that Figure \ref{5cell} is not locally Clifford equivalent to the AME(6,2) state. Hence as before one must either ease the maximal discrete symmetry assumption allowing one to work with an AME state (where, in this case one could use the AME(6,2) state) or alternatively, one can work with states that are not AME but preserve maximal discrete symmetry.

\bigskip

Other possible regular compact tessellations of $H^4$ are given in (\ref{regular4d}), but these can be reduced to considering the tessellations of the 120-cell (120-cell honeycomb $\{5,3,3,3\}$, order-4 120-cell honeycomb $\{5,3,3,4\}$ and order-5 120-cell honeycomb $\{5,3,3,5\}$) and that of the tesseract/four-dimensional hypercube (order-5 tesseractic honeycomb $\{4,3,3,5\}$). The 120-cell has Schl\"{a}fi symbol $\{5,3,3\}$ and configuration matrix $C_{120}$ while the tesseract has Schl\"{a}fi symbol $\{4,3,3\}$ and configuration matrix $C_{8}$ where
\be
C_{120} = \begin{bmatrix} 
600 & 4 & 6 & 4 \\
2 & 1200 & 3 & 3 \\
5 & 5 & 720 & 2 \\
20 & 30 & 12 & 120 \\
\end{bmatrix}, \qquad
C_8 = \begin{bmatrix} 
16 & 4 & 6 & 4 \\
2 & 32 & 3 & 3 \\
4 & 4 & 24 & 2 \\
8 & 12 & 6 & 8 \\
\end{bmatrix}.
\ee

\bigskip

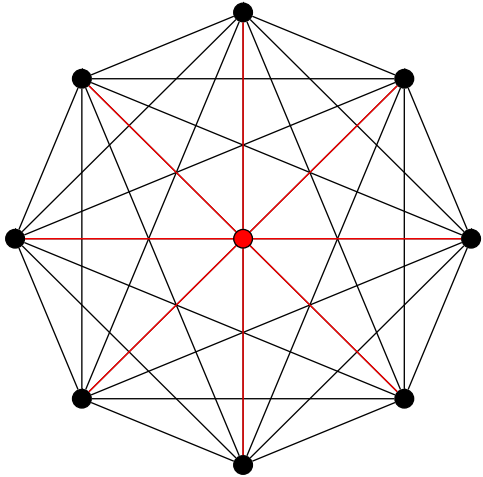
\begin{figure}[h!]
\centering
\begin{tikzpicture}[mystyle/.style={draw,shape=circle,fill=black, inner sep=0pt, minimum size=7pt}]
	\GraphInit[vstyle=Simple]
	\SetUpEdge[lw  = 0.5pt]     
	\SetVertexNoLabel
	\tikzset{EdgeStyle/.append style={black}}
	\tikzset{VertexStyle/.append style={mystyle}}
	\grComplete[RA=3] {8}
	\tikzset{VertexStyle/.append style={draw,shape=circle,fill=red, inner sep=0pt, minimum size=7pt}}
	\Vertex{d1}
	\SetUpEdge[lw  = 0.5pt]     
	\tikzset{EdgeStyle/.append style={red}}
	\EdgeFromOneToAll{d}{a}{1}{8}
\end{tikzpicture}
\caption{Graph state preserving maximal discrete symmetry for regular, uniform tessellations of $H^4$ consisting of tesseracts, a 4-dimensional polytope bounded by eight cubes.}
\label{tesseract}
\end{figure}

Following the previous arguments, since both of these polytopes would provide $n>7$ physical qudits (120 for the 120-cell and 8 for the tesseract), neither polytope can be used to create an AME graph state that also preserves maximal discrete symmetry. For completeness, Figure \ref{tesseract} is the corresponding maximally symmetric graph state for the tesseract. We do not show the graph state for the 120-cell since it is very complex and does not add anything new to the discussion.

\bigskip

\bigskip

Since there are no compact, regular tessellations of $H^5$ or higher dimensional space, these constructions cannot be explored using the construction that has been presented. Hence, if one wanted to consider AdS$_6$ or higher dimensions using a HaPPY type approach, then spatial slices would need to be discretised in a different manner. One could consider paracompact, regular tilings for $H^5$ but none of these exist for hyperbolic spaces of dimension 6 or higher. In order to be able to extend to general $(d+1)$-dimensional spacetime AdS$_{d+1}$, it seems that one would need to consider non-compact regular tessellations of the spatial slices $H^d$ or use a qualitatively different approach based on irregular tessellations. We will return to this point in the following subsection as well as in section \ref{discussion}.

\subsection{Summary}

In this section we have shown that the HaPPY construction can be generalised to uniform regular tessellations of hyperbolic space in higher dimensions, but with important differences relative to two dimensions. Firstly, one either needs to relax the assumption of maximal discrete symmetry of the graph within the cell to get an AME code, or one needs to work with non AME states, with corresponding subtleties in concatenating cells. Secondly, only a sparse number of such codes can exist in $d > 2$, due to the limited
number of uniform regular tessellations. 

In previous literature it has been noted that the perfect tensor/absolutely maximally entangled properties are constraining and various alternatives have been proposed in the context of two spatial dimensions.  One class of approaches is based around relaxing the perfect tensor condition, for example to block perfect tensors as in \cite{Harris2018Calderbank-Steane-ShorCodes,Harris2020DecodingDecoder}. A block perfect tensor is one which is isometric for partitions into adjacent sets of indices and with this approach one can associate codes to 2d hyperbolic tilings that are not compatible with perfect tensors. There are many other approaches that are similarly related to generalisation to different types of tensors \cite{Hayden:2016cfa,Qi:2018shh}. These constructions are very much analogous in spirit to the non-AME constructions above. 

In two dimensions, more complicated generalizations of HaPPY have been considered. For example, one can consider hyperbolic tilings that are not regular but alternate different polygons \cite{Cao:2020ksw,Cao:2021ibt}; these can represent Bacon-Shor type codes which include gauge degrees of freedom.  Generalizing this approach to higher dimensions would be interesting and would rely on classifications of hyperbolic tessellations involving more than one type of polytope. For example, one could envisage using rectified honeycombs, which alternate different polytopes, as well as runcinated honeycombs which alternate polytopes and have irregular vertex figures. In higher dimensions where no compact regular tessellations exist one would need to use semi-regular honeycombs as a basis for constructing codes.

Other generalizations of HaPPY include adding degrees of freedom on a tensor network connected to copies of HaPPY by an isometry, to give models for bulk gauge fields and gravitons \cite{Donnelly:2016qqt}. Clearly for this construction to be lifted to higher dimensions one would need to first develop an AME code before connecting this code to an auxiliary tensor network.

\section{CSS stabilizer codes and tessellations} \label{css-tes}

In this section we will consider a qualitatively different class of quantum error correcting codes that can be associated with hyperbolic tessellations. This class of codes has the advantage that the generalisation from two to higher dimensions is straightforward as the structure of the code follows directly from the properties of the tessellation. We will compare and contrast this construction to the AME/perfect tensor approach used in the previous section. 

\bigskip

The codes discussed in this section are CSS codes, a specific class of stabilizer codes, the best known example of which is perhaps the Steane seven qubit code that is reviewed in the appendix \ref{key-codes}. As in earlier sections, the basic principle of stabilizer codes is to encode $k$ logical qubits in $n$ physical qubits\footnote{From here onwards we restrict to qubits, although the constructions can straightforwardly be extended to qudits.} using a code space that is the eigenspace of a set of commuting stabilizers \cite{Gottesman:1997zz}. To make the link with tessellations and cellulations the relevant feature of a CSS code is that the parity check matrix, which characterises the error upon corruption of the message, can be expressed in terms of $\frac{1}{2} (n-k)$ generators that consist of (only) $X$ operations on qubits and $\frac{1}{2} (n-k)$ generators that consist of (only) $Z$ operations on qubits. For example, we can express one of the three $X$ (only) generators on the Steane seven qubit code as 
$X_1 X_5 X_6 X_7$ where the subscripts denote the physical qubits on which the operators act. The remaining five independent generators are given in \eqref{steane-checks}. 

\bigskip

\begin{figure}[h!]
\begin{center}
\begin{tikzpicture}
	\checkpauli (u3) at (-1,1) {}; 
	\checkpauli (y3) at (0,1) {};
	\checkpauli (w3) at (1,1) {};
	\vertex (a2) at (-3,0) {};
	\vertex (b2) at (-2,0) {};
	\vertex (c2) at (-1,0) {};
	\vertex (d2) at (0,0) {};
	\vertex (e2) at (1,0) {};
	\vertex (f2) at (2,0) {};
	\vertex (g2) at (3,0) {};
	\checkpauli (u1) at (-1,-1) {}; 
	\checkpauli (y1) at (0,-1) {};
	\checkpauli (w1) at (1,-1) {};
	
	\path
		(u1) edge (a2)
		(u1) edge (e2)
		(u1) edge (f2)
		(u1) edge (g2)
		(w1) edge (b2)
		(w1) edge (d2)
		(w1) edge (f2)
		(w1) edge (g2)
		(y1) edge (c2)
		(y1) edge (d2)
		(y1) edge (e2)
		(y1) edge (f2)
		(w3) edge (a2)
		(w3) edge (c2)
		(w3) edge (d2)
		(w3) edge (g2)
		(y3) edge (b2)
		(y3) edge (c2)
		(y3) edge (e2)
		(y3) edge (g2)
		(u3) edge (a2)
		(u3) edge (b2)
		(u3) edge (c2)
		(u3) edge (f2)

	;
\end{tikzpicture}
\end{center}
\caption{\label{fig:tanner} Tanner graph for the  [[7,1,3]] Steane code. This graph is tripartite with the middle row of nodes representing the physical qubits and the top and bottom rows representing $Z$ stabiliser checks and $X$ stabiliser checks respectively.}
\end{figure}
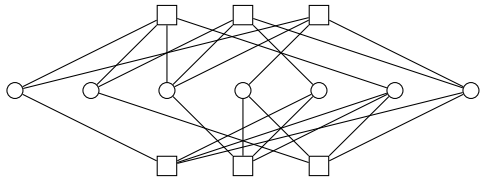

\bigskip

The Tanner graph for a CSS code provides a visualisation for the parity check matrix. The middle row of a Tanner graph shows the physical qubits as nodes. Stabilizer checks are shown as boxes in the top (Z) and bottom (X) row, with there being an edge between the stabilizer and a qubit if the operator acts on that qubit. Tanner graphs are tripartite, with the partitions being the qubits, the X-checks and the Z-checks. From the check matrix for the Steane code given in \eqref{steane-checks} we can draw the associated Tanner graph, shown in Figure~\ref{fig:tanner}. 

The main fact that we will use in this section is that any three layer Hasse diagram for a cellulation may be reinterpreted in terms of a Tanner graph for a CSS code. In the case of the tetrahedron shown in Figure~\ref{fig:hasse}, the top row of faces represents the Z checks; the middle row of edges represents the physical qubits and the bottom row of vertices represents the X checks. However, one needs to take into account that not all of the checks are linearly independent. For example, the four Z checks are:
\begin{equation}
Z_1 Z_2 Z_3 \qquad
Z_1 Z_4 Z_5 \qquad
Z_2 Z_5 Z_6 \qquad
Z_3 Z_4 Z_6,
\end{equation}
where as above we use subscripts to denote the qubit on which the check acts. Clearly the product first three gives the fourth, and therefore only three of the Z checks are linearly independent. Similarly, only three out of four of the X checks are linearly independent. Accordingly this code is trivial, as the number of checks is equal to the number of physical qubits, and therefore no logical information can be encoded. 

However, even though the code associated with the individual cell encodes no logical qubits, a cellulation constructed from tetrahedrons may encode logical qubits, as explained below. In other words, this approach to constructing codes thus relies inherently on the global properties of the tessellation, rather than on the properties of the individual cells, as the codes in the previous section did. 

\bigskip

To exemplify cellulations associated with non-trivial codes, let us consider the case of $\{r, s\}$ regular tessellations of the hyperbolic plane. We can identify $\{r, s\}$ tessellations with CSS codes as follows, see \cite{Bravyi1998QuantumCO}. Each stabilizer X check corresponds to a vertex of the tessellation, acting on all edges incident to the vertex, so has weight $s$. Each stabilizer Z check corresponds to a face of the tessellation, acting on all associated edges, so has weight $r$. Starting from a single $r$-gon, one can reflect in its edges to generate further $r$-gons, and then keep repeating the process. The polygons of each iteration are labelled by their level, starting from the original polygon (level 1) and continuing to level $k$. This process clearly generates a tessellation with a boundary. 

However, the code associated with such a tessellation encodes no qubits. Associated with each vertex is an X check, but there is one linear dependency between these checks, so the total number is $V-1$, with $V$ the number of vertices. Associated with each face is a Z check and these are linearly independent, so the total number is $F$, with $F$ the number of faces. The graph is planar and therefore the standard relationship between edges $E$, $F$ and $V$ holds:
\begin{equation}
E = V - 1 + F.
\end{equation} 
The number of edges is identified with the number of physical qubits, and therefore the total number of physical qubits is equal to the total number of stabilizers, so the code is trivial. More generally, the properties of a code are determined by homology, see Appendix \ref{homology}. The number of encoded qubits is zero since the first homology class is trivial. 

\bigskip

To obtain non-trivial codes one clearly needs constructions with non-trivial first homology class. There are several distinct approaches to obtaining non-trivial codes in two dimensions and there is considerable literature on the subject. One approach to consider codes associated with closed two-dimensional surfaces; the homology of a closed surface is generically non-trivial, and can be understood in terms of the identifications made to close the surface. A second approach was initiated in \cite{Bravyi1998QuantumCO}, which gave a systematic procedure for removing certain X and Z checks at the boundary of an open surface so that the code encodes $(k-1)$ qubits; we will discuss this further below. Adding holes and defects in the interior also gives non-trivial homological codes \cite{Freedman2001ProjectivePA}. 

\bigskip

One of the main topics of exploration is how to maximise the distance of the code, for a given ratio of logical to physical qubits $(k/n)$, see for example \cite{10.1007/978-3-642-01877-0_21, PhysRevLett.104.050503, Delfosse2013TradeoffsFR}. While much of the focus has been on planar and toric codes, recent literature has also explored codes associated with hyperbolic surfaces, see for example \cite{Breuckmann_2017} which in particular studies codes associated with closed hyperbolic surfaces i.e. quotients $H^2/\Gamma$. Here it is assumed that the symmetry group $\Gamma$ has no fixed points although it could be interesting to generalise these constructions to quotients associated with the spatial sections of wormholes \cite{Krasnov:2000zq,Skenderis:2009ju}. 

\bigskip

Much of the literature has focused on codes associated with tessellations of two-dimensional manifolds. However, the procedure for associating a code with a cellulation works in any dimension. For physical applications, one clearly needs to be able to implement the code in a system with three spatial dimensions but various properties of higher dimensional codes have nevertheless been explored in the literature, see \cite{Dennis2002TopologicalQM,Arakawa:2004qf,Alicki2010OnTS,Guth2014QuantumEC,Bombin2014SingleShotFQ, Breuckmann2017LocalDF, Breuckmann_2017,Londe2017GoldenCQ,breuckmann2020singleshot}. For example, codes associated with tori have been constructed not just for three dimensional tori, but also four dimensional tori \cite{Dennis2002TopologicalQM}. 
To make contact with holography, we will again focus on codes associated with hyperbolic space and these have been studied in a number of recent works \cite{Breuckmann2017LocalDF,Breuckmann_2017,Londe2017GoldenCQ,breuckmann2020singleshot}. 

\bigskip

In a three-dimensional cellulation one identifies qubits with faces (2-cells), and stabilizer checks with edges (1-cells) and 3-cells. The interpretation of the vertices is that their connections with the edges define linear codes acting on the X-checks. By contrast there is no such linear code acting on the Z checks, and Z checks are linearly independent. 

\begin{figure}[h!]
\begin{center}
\begin{tikzpicture}
	\checkpauli at (-1.5,1) (u1) {}; 
	\checkpauli at (-0.5,1) (y1) {};
	\checkpauli at (0.5,1) (w1) {}; 
	\checkpauli at (1.5,1) (x1) {};
		\vertex at (-2.5,0) (a2) {}; 
	\vertex at (-1.5,0) (b2) {};	
		\vertex at (-0.5,0) (c2) {}; 
	\vertex at (0.5,0) (d2) {};
	\vertex at (1.5,0) (e2) {}; 
	\vertex at (2.5,0) (f2) {};	
	
	\checkpauli at (-1.5,-1) (u3) {}; 
	\checkpauli at (-0.5,-1) (y3) {};
	\checkpauli at (0.5,-1) (w3) {}; 
	\checkpauli at (1.5,-1) (x3) {};
	
	\node[draw,diamond, scale=0.5] at (-1.5,-2) (u4) {}; 
	\node[draw,diamond, scale=0.5]  at (-0.5,-2) (y4) {};
	\node[draw,diamond, scale=0.5]at (0.5,-2) (w4) {}; 
	\node[draw,diamond, scale=0.5] at (1.5,-2) (x4) {};
		
	\path
		(u1) edge (a2)
		(u1) edge (b2)
		(u1) edge (c2)
		(y1) edge (a2)
		(y1) edge (d2)
		(y1) edge (e2)
		(w1) edge (b2)
		(w1) edge (e2)
		(w1) edge (f2)
		(x1) edge (c2)
		(x1) edge (d2)
		(x1) edge (f2)	
		(u3) edge (a2)
		(u3) edge (b2)
		(u3) edge (e2)
		 
		(y3) edge (a2)
		(y3) edge (c2)
		(y3) edge (d2)
		(w3) edge (b2)
		(w3) edge (c2)
		(w3) edge (f2)
		(x3) edge (d2)
		(x3) edge (e2)
		(x3) edge (f2)
		
		(u4) edge (u3)
		(u4) edge (y3)
		(y4) edge (y3)
		(y4) edge (x3)
		(w4) edge (y3)
		(w4) edge (w3)
		(w4) edge (x3)
		(x4) edge (w3)
		(x4) edge (x3)

	;
\end{tikzpicture}
\end{center}
\caption{\label{fig:3code} Tanner graph for toy example of a 3d code.}
\end{figure}
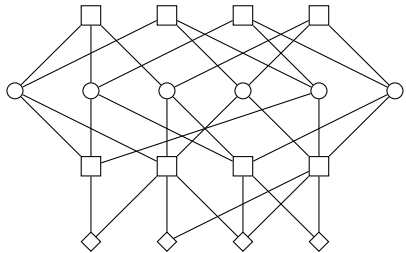

A toy example of a 3d code is illustrated in Figure~\ref{fig:3code}. The top row shows the Z stabilizers, the second row shows the qubits and the third row shows the X stabilizers. The bottom row is associated with the vertices, and shows the action of the linear code on the X checks. Note that this is a toy example, as in realistic 3d cellulations there would be many more nodes, and many more connections between nodes: the graph captures all cells within the cellulation and the code properties relate to the global structure of the cellulation. 

\bigskip

\bigskip

For four dimensional cellulations, codes have been constructed by again identifying qubits with faces, and stabilizers with edges and 3-cells. In four dimensions there are linear codes acting on both the X checks and the Z checks, associated with the vertices and 4-cells, respectively. The Tanner graph in this case would have one additional row relative to Figure~\ref{fig:3code}, above the Z checks, capturing the action of the linear code on these stabilizers. While in principle the construction of codes could be extended to arbitrary dimensions, there does not appear to be previous literature discussing codes in dimensions higher than four. 

Most of the four-dimensional hyperbolic codes discussed in the literature are associated with closed manifolds. 
Properties of codes associated with tessellations of closed 4d hyperbolic manifolds were explored in \cite{Guth2014QuantumEC}. Geometric arguments were used to show that such codes would have constant rate ($k/n$ constant) while the distance scales as $n^{\epsilon}$ with $0 < \epsilon < 0.3$. Explicit examples of codes associated with closed 4d hyperbolic cellulations have been constructed recently in \cite{Breuckmann2017LocalDF,Breuckmann_2017,Londe2017GoldenCQ,breuckmann2020singleshot}. These are based on the local regular $H^4$ cellulations \eqref{regular4d}, identifying 3-cells to give a closed manifold. A well-known example of such a closed 4d hyperbolic manifold is the Davis manifold, obtained by identifying boundary dodecahedra in the $\{5,3,3,5\}$ (120-cell) tessellation. The properties of the associated code can be determined from homology, see appendix \ref{homology}, and the homology of the David manifold was derived in \cite{RATCLIFFE2001327}. The examples of closed 4d hyperbolic codes given in \cite{Breuckmann2017LocalDF,Breuckmann_2017,Londe2017GoldenCQ,breuckmann2020singleshot} are based on analogous constructions.

\bigskip

\subsection{CSS codes for the hyperbolic plane}

In this section we will describe the explicit construction of non-trivial CSS codes associated with hyperbolic planes with boundary.  From the discussions above, the code associated with a hyperbolic tessellation generated by iterative reflections of a cell is trivial, due to the trivial homology. For holographic applications the most natural way to generate a non-trivial code is following a construction analogous to that of \cite{Bravyi1998QuantumCO}, i.e. modifying the qubits and checks at the boundary.  We first review the two-dimensional construction of \cite{Bravyi1998QuantumCO} and then explain how this approach can be generalised to higher dimensions. We will illustrate the two-dimensional construction with hyperbolic tessellations following the discussions in \cite{Breuckmann_2017}. 

\bigskip

For an $\{r, s \}$ tessellation our starting graph is obtained by iterative reflection of polygons, beginning from a single $r$-gon. The starting graph encodes no qubits and all of the boundaries are smooth boundaries: the edges of the polygons, at which a string of X-errors can start and end. The reason why X-errors can start and end on smooth boundaries is because the X stabilisers act on all qubits on edges adjacent to vertices; the weights of the X checks at the boundary are two or $s$ (for even $s$) or $(s-1)$ (for odd $s$) while in the interior the X checks have weight $s$. 

Suppose the starting graph is generated by $k$ reflections of the original cell. The boundary then consists of the edges that the level $k$ polygons would be reflected in to generate the graph of $(k+1)$ reflections. The idea is to divide this boundary into $2k$ equally sized regions. Half of these are smooth boundaries while the other half are so-called rough boundaries at which Z-errors can start or end. 

The process for creating a rough boundary used in \cite{Bravyi1998QuantumCO,Breuckmann_2017} is:
\begin{itemize}
\item Remove all boundary X-checks of weight two that have two edges within the boundary region considered. 
\item Remove all qubits on which only a single Z-check acts. 
\item Add certain weight two ZZ checks to the stabilizer (which ensures that Z-errors can only run between rough regions). 
\end{itemize}

\bigskip

 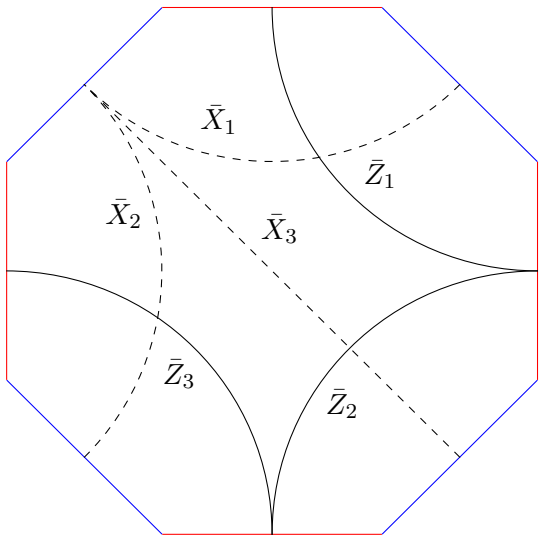
\begin{figure}[h]
        \centering
        \begin{tikzpicture}[scale=1.4]
        \def\lstcolors{"red","blue"}
        \foreach \X [remember=\X as \LastX] in {0,...,8}
        {\ifnum\X=0
            \node [scale=0.01](X\X) at (67.5:2.7){};
            \else
            \node [scale=0.01] (X\X) at
            ({67.5-\X*360/8}:2.7){};
            \pgfmathsetmacro{\mycolor}{{\lstcolors}[Mod(\X,2)]}             
            \draw[color=\mycolor]  (X\LastX) --  (X\X);
            \fi};
         
         \foreach \Y [remember=\Y as \LastY] in {0,...,8}
        {\ifnum\Y=0
            \node [scale=0.01](Y\Y) at (90:2.5){};
            \else
            \node [scale=0.01] (Y\Y) at
            ({90-\Y*360/8}:2.5){};
            \fi};
            \draw[] (Y0) to[out=-90,in=-180] node[label={[xshift=0.4cm, yshift=-0.2cm]$\bar{Z}_1$}]  {}  (Y2);
            \draw[dashed] (Y1) to[out=-135,in=-45] node[label={[xshift=-0.7cm, yshift=0.1cm]$\bar{X}_1$}]  {} (Y7);
            \draw[] (Y2) to[out= -180,in=90] node[label={[xshift=-0.1cm, yshift=-1.2cm]$\bar{Z}_2$}]  {}  (Y4);
            \draw[] (Y4) to[out= 90,in=0]  node[label={[xshift=-0.2cm, yshift=-0.8cm]$\bar{Z}_3$}]  {} (Y6);
            \draw[dashed] (Y5) to[out=45,in=-45] node[label={[xshift=-0.5cm, yshift=0.3cm]$\bar{X}_2$}] {}  (Y7);
            \draw[dashed]  (Y3) -- node[label={[xshift=0.1cm, yshift=0.1cm]$\bar{X}_3$}]  {}  (Y7);

        \end{tikzpicture}
\caption{\label{fig:area2} Example with eight boundary segments. Logical string operators running between rough and smooth boundaries, respectively, are shown.}
    \end{figure}  
    
\bigskip

With such a construction the code can encode $(k-1)$ logical qubits \cite{Bravyi1998QuantumCO}. Note that one can interpret the smooth and rough boundaries in terms of the lattice and its dual, see discussions in \cite{Bravyi1998QuantumCO,Breuckmann_2017}. Here the logical $\bar{Z}_i$ operators run from the $i$th to the $(i+1)$th rough boundary with $i = 1,\cdots, (k-1)$. Similarly the logical $\bar{X}_i$ operators run from the $i$th to the $(i+1)$th smooth boundary. (The definition of logical operators for stabiliser codes is reviewed in the Appendix.) An example with eight boundary segments is illustrated in Figure~\ref{fig:area2}. 

\bigskip

While the details of the construction can be adjusted e.g. to remove boundary X-checks of higher weight within the rough regions, the number of logical qubits was argued in \cite{Bravyi1998QuantumCO} to be optimised at around $k$ for a tessellation based on $k$ symmetry operations acting on an initial cell. Conceptually this limit follows from demanding that the distance between rough regions is minimised by going through the bulk, rather than along the boundary. If the rough regions can be connected via a shorter distance along the boundary, there are no logical qubits encoded, and this leads to the optimal division of the boundary into around $2k$ (equal sized) regions.  

\subsection{CSS codes for hyperbolic manifolds in $d > 2$}

Here we outline the construction of CSS codes for higher dimensional hyperbolic manifolds, emphasising the differences relative to two dimensions. In two dimensions
the boundary of the starting graph generated by $k$ reflections of the original cell is topologically a circle, and the boundary is tessellated by segments. The construction above is based on alternating rough and smooth segments. 

Now let us consider three dimensional hyperbolic space as the simplest prototype for higher dimensions. Our starting configuration is obtained 
by iteratively reflecting polyhedra beginning from a single cell. This starting configuration as above encodes no qubits: the boundary is topologically a sphere and using the homology as described in Appendix \ref{homology} one can show that no logical qubits are encoded. 

While the CSS code approach does not inherently rely on uniformity or regularity, let us first consider the case of uniform regular tessellations of hyperbolic space. The two dimensional boundary of the starting configuration will by construction be a regular monogonal tessellation i.e. all of the cells of the tessellation are one type of regular polygons. However, the boundary tessellation will not be vertex transitive (isogonal). 

One can see the latter immediately from considering cubic tessellations for which the boundary sphere is tessellated by squares. The only uniform regular isogonal tessellation of a sphere by squares is by six squares, with three edges meeting at each vertex; this tessellation is of course associated with the cube itself. Now consider a tessellation of Euclidean space by cubes, as shown in Figure~\ref{checkerboard}. Clearly the boundary is topologically a sphere, tessellated by squares, but there are two types of vertices, of order four and three respectively, and it is not isogonal. The boundary of the hyperbolic cubic honeycomb is also not isogonal, but has vertices of order four, three and six. 
As one reduces the symmetry of the bulk tessellation, the symmetry of the boundary correspondingly decreases, with the number of different vertices and the size of the fundamental region increasing. 

\bigskip

\begin{figure}[h!]
\begin{center}

\begin{tikzpicture}
\draw[ red] (-0.5,1) to (1,1);
 \draw[{|[right]}-{|[left]}, blue] (0,1) to (1,1);  
    \draw[{|[right]}-{|[left]}, red] (1,1) to (2,1);  
\draw[{|[right]}-{|[left]}, blue] (2,1) to (3,1);
  \draw[{|[right]}-{|[left]}, red] (3,1) to (4,1);  
   \draw[blue] (4,1) to (4.5,1);
  \end{tikzpicture}

\end{center}
\caption{\label{fig:segment} Rough and smooth segments of one-dimensional boundary.}
\end{figure}
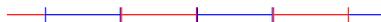

For a one-dimensional boundary one can subdivide the boundary in a binary way into rough and smooth segments, with rough regions separated from each other as shown in Figure~\ref{fig:segment}. For a higher dimensional boundary the vertices in the boundary tessellation are of order greater than two and such a binary division is not possible. 

\bigskip

\begin{figure}[h!]
\begin{center}

\begin{tikzpicture}
    [
        box/.style={rectangle,draw=black,thick, minimum size=1cm},
    ]

\foreach \x in {0,1,...,3}{
    \foreach \y in {0,1,...,3}
        \node[box] at (\x,\y){};
}

\node[box,fill=cyan] at (3,3){};  
\node[box,fill=red  ] at (2,3){};  
\node[box,fill=blue ] at (2,2){};  
\node[box,fill= magenta] at (3,2){};  

\node[box,fill=cyan] at (1,3){};  
\node[box,fill=red  ] at (0,3){};  
\node[box,fill=blue ] at (0,2){};  
\node[box,fill= magenta] at (1,2){};  

\node[box,fill=cyan] at (3,1){};  
\node[box,fill=red  ] at (2,1){};  
\node[box,fill=blue ] at (2,0){};  
\node[box,fill= magenta] at (3,0){};  

\node[box,fill=cyan] at (1,1){};  
\node[box,fill=red  ] at (0,1){};  
\node[box,fill=blue ] at (0,0){};  
\node[box,fill= magenta] at (1,0){};

\end{tikzpicture}
\end{center}
\caption{\label{fig:area} Square Euclidean tessellation. Rough area elements are shown in blue and cyan while smooth elements are shown in red and magenta.}
\end{figure}
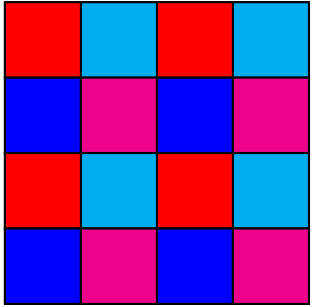

Let us consider a uniform square tessellation, illustrated in Figure~\ref{fig:area}, as a simple prototype. The simplest generalization of the lower-dimensional construction is to define distinct operations on each of the four cells in the fundamental region, and repeat these throughout the tessellation. Just as above, we can define roughing of boundary area elements through removing qubits and X checks. However, we will need to define two distinct types of roughing $R_1$ and $R_2$ for diagonally opposite cells, shown in blue and cyan. The corresponding adjoining smooth regions are also of two distinct types, reflecting the removal of different X checks on adjoining edges etc. These smooth regions $S_1$ and $S_2$ are shown in red and magenta. For a square tessellation with $4k^2$ cells, one would accordingly obtain $\approx k^2$ qubits encoded, associated with logical string operators extending in the bulk between neighbouring regions. 

\bigskip

\begin{figure}[h!]
\begin{center}
\includegraphics[scale=0.2]{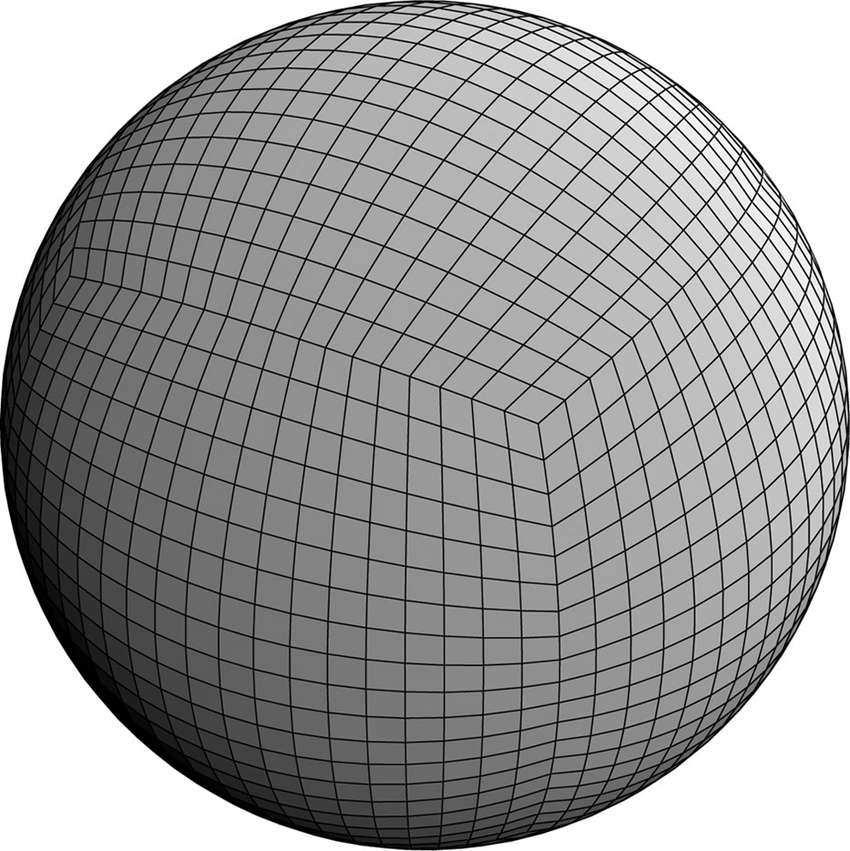}
\end{center}
\caption{\label{fig:cubic} Cubic tessellation and bounding sphere.}
\end{figure}

Now let us consider how such an encoding would work over the entire spherical boundary, illustrating with the case of a cubic tessellation of Euclidean space, as shown in  Figure~\ref{fig:cubic}. One can use the construction outlined above to encode logical qubits associated with vertices of order four. However, there are also vertices of order three (the corners of the cube projected onto the sphere) and one would need to adjust the roughing and smoothing at these vertices. As discussed above, cubic tessellations of hyperbolic space are associated with boundary tessellations in which the vertices are of order three, four and six. Accordingly one would need to define alternating roughing and smoothing over the entire fundamental region of the tessellation to construct a consistent code. We leave the detailed construction for future work, but based on the arguments above one would expect to be able to encode around $k^2$ qubits starting from a level $k$ tessellation. Note that constructions of Euclidean surface codes in three dimensions can be found in works such as \cite{Vasmer2018Three-dimensionalArchitectures, Kubica2018TheBy}. 
 
\subsection{Summary and holographic interpretation}

In this section we have discussed CSS codes associated with hyperbolic tessellations. While logical information is encoded in every cell of a HaPPY type construction, the logical information in these CSS code constructions is inherently associated with the boundary of the tessellation. As in the HaPPY approach, we again encounter subtleties in extending from the two dimensional constructions in the literature to higher dimensional constructions. In two dimensions one can divide the boundary into rough and smooth segments, associated with logical $\bar{Z}$ and $\bar{X}$  operators respectively. Since the vertices in the boundary tessellation are of order higher than two in higher dimensions, one will need to use more types of area elements associating them with distinct logical operators. The boundary tessellation is not isogonal and the rough/smooth construction will need to be defined over the fundamental region of the tessellation. 

\bigskip

These code constructions are interesting in their own right for quantum information theory. From the perspective of holography, one could envisage two distinct applications of these constructions. The first is to take the limit of large $k$ so that the code is associated with the entire bulk holographic space. The logical string operators running through the bulk should then have an interpretation in terms of the dual gauge theory, perhaps in terms of correlations between operators inserted at the boundary locations.  

The second application could be in the context of local holography. A level $k$ code would encode around $k^2$ qubits. One could envisage gluing together different finite $k$ constructions to cover the full holographic space i.e. using each CSS building block code as an analogue of the unit cell in the HaPPY construction. Consistent concatenation 
of these CSS codes to cover the bulk would require approaches of the type discussed in the previous section. This approach could be viewed as local holography, in that 
the boundary of each block is associated with the encoding of logical information within the block. Using local building blocks to cover the full space is a natural first step towards dynamics: one could envisage that dynamical evolution could change the CSS codes in each block, reflecting the local changes in curvature through dynamics.



\section{Conclusions and outlook}
\label{discussion}

The main goal of this paper has been to generalise quantum error correction codes to new classes of holographic geometries. Much of our discussion
focuses on hyperbolic manifolds of dimensions greater than two. We noted at the beginning of the paper that general classes of Einstein/dilaton theories
can be understood in terms of toric reductions of negative curvature Einstein spaces so our constructions provide the starting point for analysing these holographic 
dualities also. 

To associate a code with (the spatial section of) a holographic geometry, we discretise the geometry; associate a graph state to each cell; analyse the properties of 
the code corresponding to the graph state and then explore the concatenation between cells. We begin by exploring the generalisation of 
the HaPPY construction to dimensions greater than two, i.e. working with uniform regular tessellations and associated perfect tensors/AME states. While it is well-known that HaPPY does not give rise to the required behaviour of entanglement and correlation functions for a 2d CFT \cite{Hayden:2016cfa}, many code constructions are based on adaptations of HaPPY e.g. modifications of the tensors to be nearly perfect \cite{Bhattacharyya:2016hbx,Bhattacharyya:2016zyk}. In two dimensions quasiperiodic boundaries of regular hyperbolic tilings can on average give correlation function decays in line with those of CFTs, see \cite{Jahn2019MajoranaCodes,Jahn2019CentralModels,JahnHolographicReview}. Beginning with HaPPY type constructions therefore seems like a natural starting point for higher dimensional codes. 

\bigskip

We have shown that non-trivial codes based on uniform regular hyperbolic tessellations can be constructed. However, there are key differences relative to the two-dimensional construction. Firstly, there are only a finite number of uniform regular tessellations in $H^3$ and $H^4$, with no uniform regular compact 
tessellations at all for $H^{d}$ with $d > 5$. Accordingly the number of codes is much more sparse than in two dimensions. 

Secondly, we are forced to break the maximal discrete symmetry associated with the polytope of the tessellation. If we work with AME codes/perfect tensors, the breaking of the symmetry automatically arises in the mapping between the AME state and the cell, and follows immediately from the properties of AME states. In the codes constructed from non-AME states, the graph state apparently respects the discrete symmetry of the polytope cell, but in concatenating cells one does not preserve transitivity i.e. different faces/edges are not equivalent. Our constructions very much mirror the constructions of 2d codes based on block perfect tensors \cite{Harris2018Calderbank-Steane-ShorCodes}, for which there is also no vertex and edge transitivity.

It would be interesting to establish in future work the performance properties of these codes, together with the behaviour for entanglement and correlation functions in the dual theory which they capture. The relative sparsity of AME codes in dimensions higher than two may be linked with the fact that AME codes do not capture CFT behaviour. It would be intriguing to explore whether AME behaviour in dimensions higher than two is parametrically more distant from CFT behaviour than it is in two dimensions i.e. one may not be able to obtain CFT behaviour from small corrections to perfect tensors as in \cite{Bhattacharyya:2016hbx}. 

\bigskip

In section \ref{sec:tess} we noted that general classes of non-conformal holographic dualities can be obtained from reducing hyperbolic spaces on tori and we illustrated this discussion in section \ref{sec:tor} with the example of the reduction of the hyperbolic plane on a circle, which is relevant in the context of AdS$_3$ and JT gravity. We noted that there are two distinct ways of dealing with tessellations of toroidally reduced spaces: one can either choose the tessellation to respect the toroidal symmetry (at the price of imposing by hand equal area of cells) or one can exploit the discrete symmetry of the hyperbolic tessellation to make identifications in the toroidal directions. Either way one has, as expected, a regulated boundary excising the conical deficit in the interior of the space. It would be interesting to extend the circle reduction discussed in section \ref{sec:tor} to more general toroidal reductions of higher dimensional hyperbolic tessellations, and to explore the properties of the codes obtained by discrete identifications.  

\bigskip

In section \ref{css-tes} we have discussed an alternative approach to associating codes to hyperbolic tessellations, based on the relation between Hasse diagrams of cellulations and Tanner graphs of CSS codes. In this approach the logical qubits are encoded through global properties, rather than being associated with each cell. There are several distinct ways to obtain non-trivial encoding, from including defects in the bulk lattice to topologically non-trivial identifications. 

For codes associated with manifolds with boundary, it is natural to use the approach of adjusting the boundary tessellation, adding and removing checks and qubits. Here we discussed the encoding of logical qubits through dividing the boundary into different sections, with logical operators running between disconnected sections of the boundary. In two-dimensional cellulations the boundary is one-dimensional and can be divided in a binary way between rough and smooth regions. In higher dimensions one would need more complicated constructions, with the number of different types of regions increasing with the order of the vertices in the boundary tessellation. Leaving aside applications to holography, our hyperbolic CSS codes would be interesting in their own right and their properties will be explored further in future work. 
 
From a holographic perspective, one could envisage two distinct applications of the hyperbolic CSS codes. In the case that one takes the number of levels of the tessellation $k$ to be very large, the tessellation could be viewed as being associated with the entire regulated hyperbolic spatial slice. The logical operators would then be viewed as logical strings running through the bulk between disconnected regions on the boundary. This seems somewhat analogous to capturing the correlation between local operators in the boundary by geodesics running through the bulk. It would be interesting to explore potential interpretations of the logical strings from the perspective of discrete versions of the AdS/CFT correspondence \cite{Boyle2020ConformalHolography,JahnTensorAdS/qCFT}, as well as connections to the codes in lattice CFTs explored in \cite{Dymarsky:2020qom}.  

One could also consider gluing together finite $k$ hyperbolic CSS codes, to obtain a discretisation of hyperbolic space in which each CSS block captures a certain amount of logical information. This could provide a way to think about local holography with each extended block capturing logical information, cf each individual cell within the HaPPY type codes. The concatenation of the blocks would determine the entanglement and correlation functions structure in the dual field theory.     

\bigskip

Throughout this paper we have focused on the spatial slice of (static) holographic geometries and the code construction uses explicitly the constant negative 
curvature. AdS gravity in three dimensions is not dynamical and all gravitational solutions have constant negative curvature. For AdS gravity in higher dimensions gravity is dynamical and Einstein solutions generically have non-trivial Riemann curvature. Accordingly, even without including matter fields, one would expect that the lattice should be such that dynamics can change it in spatial dimensions higher than two i.e. dynamics should be able to change from exactly hyperbolic lattice to a lattice with local curvature variations. This issue would seem to link with the long running attempts to discretise dynamical gravity in dimensions higher than two, and all the associated challenges that are encountered. It could also potentially relate to dynamical hyperbolic networks described by simplicial complexes, see for example \cite{Bianconi2016EmergentGeometry}. 

Finally, we note that graphs associated with the hyperbolic plane are used in certain deep learning algorithms \cite{NIPS2017_59dfa2df,DBLP:journals/corr/abs-1904-02239,DBLP:journals/corr/abs-2101-04562}. Networks associated with hyperbolic geometry have various advantages, including reduction in model parameters and over-fitting of data. The constructions of graphs associated with hyperbolic geometry in this work could be used to generalise these deep learning algorithms with potential applications in natural language processing and image classification \cite{DBLP:journals/corr/abs-2101-04562}. 

\section*{Acknowledgments}
MT is supported in part by the Science and Technology Facilities Council (Consolidated Grant “Exploring the Limits of the Standard Model and Beyond”). CW is supported by the STFC DISCnet CDT and the School of Mathematical Sciences at the University of Southampton.

\appendix

\renewcommand{\theequation}{\Alph{section}.\arabic{equation}}

\setcounter{section}{0}

\section*{Appendix}
\setcounter{section}{0}

\section{Review of key properties of codes} \label{key-codes}

\subsection{Classical linear codes}

Quantum CSS codes are constructed from classical linear codes, and it is useful to summarise here the defining properties of the latter. More details may be found in \cite{Nielsen}. A linear code which encodes $k$ bits of information within an $n$ bit code space, i.e. an $[ n,k ]$ code, may be specified by an $(n \times k)$ generator matrix $G$ whose elements are zeroes and ones. This matrix maps the message to the coded equivalent i.e. a message $x$ is encoded as $y = G \cdot x$. The set of possible codewords for the code is the space spanned by the columns of the generator matrix $G$. 

Error correction for linear codes may be captured by the parity check matrix, an $(n-k) \times n$ matrix $H$. An $[n,k]$ code consists of all $n$ element vectors $y$ such that $h \cdot y = 0$; the code is the kernel of $H$, which is $k$ dimensional. Using matrix manipulations one can bring the parity check matrix into the standard form $[A_{(n-k) \times k}  |  I _{(n-k) \times (n-k)} ]$ where $A$ is an $(n-k) \times k$ matrix. 

Suppose that the code $y$ is corrupted, so that the received message is $y'$, then
\begin{equation}
y' = y + e
\end{equation}
where $e$ is the error. Then by construction $H \cdot y' = H \cdot e$ captures the error (the error syndrome). 

The (Hamming) {\it distance} between code words counts the number of places at which the $n$ bit code words differ. The distance of a code $d$ is defined to be the minimum distance between any two (distinct) codewords, and classical codes are often referred to as $[n,k,d]$, i.e. specifying this distance. If $d \ge 2 t + 1$ for some integer $t$ then one can correct errors on up to $t$ bits. The distance is bounded from above according to the Singleton bound
\begin{equation}
d \le (n-k) + 1. 
\end{equation}

\subsection{Stabilizer codes}
We next review the basic properties of stabilizer quantum codes. The number of physical qubits is $n$ and the Pauli group acting on $n$ qubits is 
\begin{equation}
{\cal P}_n := \langle i,  X_j, Z_j \rangle = \{ \phi \otimes_{j=1}^n P_j \}
\end{equation}
where $j = \{ 1, \cdots, n \}$, $\phi \in \{ \pm 1, \pm i \}$ and $P_j \in \{I,X,Y,Z \}$. $X$, $Y$ and $Z$ are the Pauli matrices with $Y = i X Z$. The Pauli group has $(2n+1)$ generators and its order is $4^{n+1}$. 

The stabilizer approach uses properties of the Pauli group to define subspaces of the Hilbert space. A stabilizer group $S$ is a subgroup of the Pauli group which is abelian and does not contain $- I$. Elements of $S$ are called stabilizers and usually the stabilizer group has a distinguished set of generators, whose properties will be defined below. All stabilizers have the eigenvalue $1$ and stabilizers are independent if the group they generate becomes smaller if any of them are omitted. 

A {\it stabilizer code} $C$ is the eigenspace of all elements of a stabilizer group $S$
\begin{equation}
C = \{ | \psi \rangle | \; s | \psi \rangle = | \psi \rangle \; \forall s \in S \}. 
\end{equation}
The dimension of the code space is $2^{n-r}$ where there are $r$ independent generators of $S$. Accordingly $C$ encodes $k$ logical qubits 
where 
\begin{equation}
k = (n-r).
\end{equation}
$S$ has a minimal representation in terms of the $r$ independent generators $G_1, G_2, ...$, each of which functions in the same way as a parity check does on a classical linear code. 

The {\it logical operators} are those elements of the Pauli group that act non-trivially on the code space but leave the code space as a whole invariant. These operators form the normalizer of the stabilizer group in the Pauli group:
\begin{equation}
N(S) = \{ g \in P_n | \; g s g^{\dagger} \in S \; \; \forall s \in S \}.
\end{equation}
Clearly $S \in N(S)$ and, since all elements of $S$ have trivial action on the code space, the elements of $N(S) \backslash S$ form the logical operators. The group of logical operators is isomorphic to the Pauli group on $k$ qubits, up to phases, and it is therefore usual to represent the generators of the logical group as $\bar{X}_1,\cdots \bar{X}_k$ and $\bar{Z}_1 \cdots \bar{Z}_k$. 

The distance $d$ of a stabilizer code is the minimum weight of a logical operator
\begin{equation}
d = {\rm min}_{g \in N(S) \backslash S} {\rm wt} (g),
\end{equation}
where the weight ${\rm wt} (g)$ of a Pauli group element is the number of qubits on which it acts non-trivially. The distance is a measure of how well the code can protect against qubit errors. Stabilizer codes are often denoted as $[[ n, k, d ]]$ where $n$ is the number of physical qubits, $k$ is the number of logical qubits and $d$ is the distance. 

\subsection{CSS construction}

In this section we summarise the main features of the CSS (Calderbank and Shor \cite{Calderbank:1995dw}, and Steane \cite{Steane:1995vv}) construction of codes using classical linear codes. The CSS construction begins with two classical linear codes $W_1$ ($[n,k_1]$) and $W_2$ ($[n,k_2]$) such that all elements of $W_1$ and $W_2$ are orthogonal i.e. $\langle a, b \rangle =0$ for all $a= (a_1, \cdots , a_n) \in W_1$ and $b = (b_1, \cdots , b_n) \in W_2$. A stabilizer group can then be constructed as 
\begin{equation}
S = \langle X^a, Z^b \; | \; a \in W_1, b \in W_2 \rangle
\end{equation}
where 
\begin{equation}
X^a = X^{a_1} \otimes \cdots \otimes X^{a_n} \qquad
Z^b = Z^{b_1} \otimes \cdots \otimes Z^{b_n} 
\end{equation}
The group $S$ is Abelian, since $X^a$ and $Z^b$ commute; the latter follows from the orthogonality of $W_1$ and $W_2$. The number of independent generators of $S$ is given by 
\begin{equation}
R = {\rm{dim}} (W_1) + { \rm{dim}} (W_2).
\end{equation}
The distance is the minimum weight Pauli operator that commutes with all elements in $S$. 

The check matrix for a CSS code may be expressed in the form
\begin{equation}
\begin{bmatrix}
H(W_1) & 0 \\ 
0 & H(W_2^{\perp}) 
\end{bmatrix}
\end{equation}
where $H(W_1)$ and $H(W_2^{\perp})$ are the check matrices of the associated classical codes. 

\bigskip

For any stabilizer code (i.e. not necessarily CSS) we can express the check matrix as an $l \times 2n$ matrix, where there are $l = (n-k)$ independent generators of the stabilizer group. The left hand side of the matrix (i.e. an $l \times n$ matrix) contains ones to indicate which generators contain $X$ matrices, and zero otherwise. Similarly, the right hand side of the matrix shows ones for generators containing $Z$ matrices and zero otherwise. If there is a one in the same position on both sides, then there is a $Y$ in the generator as the product of $X$ and $Z$ gives $Y$. 

By matrix manipulations, the check matrix for any stabilizer code can be brought into a standard form: 
\begin{equation}
\begin{bmatrix}
I_{r \times r} & A_{r \times (n-k-r)} & B_{r \times k} & | & C_{r \times r} & 0_{r \times (n-k-r)} & D_{r \times k}  \\
0_{(n-k-r) \times r} & 0_{(n-k-r) \times (n-k-r)} & 0_{(n-k-r) \times r} & |  & E_{(n-k-r) \times r} & I_{(n-k-r) \times (n-k-r)} & F_{(n-k-r) \times k} 
\end{bmatrix},
\end{equation}
where $r$ is the rank of left hand side of the check matrix. In the case of a CSS code, this standard form simplifies further, with the matrices $C$ and $D$ being zero. 

A well known example of a CSS code is the Steane $[[7,1,3]]$ code \cite{Steane:1995vv}, which is constructed from $[7,4]$ and $[7,3]$ classical codes. The standard form for the check matrix for this code is discussed in detail in \cite{Nielsen},
\begin{equation}
\begin{bmatrix}
1 & 0 & 0 & 0 & 1 & 1 & 1 & | & 0 & 0 & 0 & 0 & 0 & 0 & 0 \\
0 & 1 & 0 & 1 & 0 & 1 & 1 & | & 0 & 0 & 0 & 0 & 0 & 0 & 0 \\
0 & 0 & 1 & 1 & 1 & 1 & 0 & | & 0 & 0 & 0 & 0 & 0 & 0 & 0 \\
0 & 0 & 0 & 0 & 0 & 0 & 0 & | & 1 & 0 & 1 & 1 & 0 & 0 & 1 \\
0 & 0 & 0 & 0 & 0 & 0 & 0 & | & 0 & 1 & 1 & 0 & 1 & 0 & 1 \\
0 & 0 & 0 & 0 & 0 & 0 & 0 & | & 1 & 1 & 1 & 0 & 0 & 1 & 0 
\end{bmatrix}. \label{steane-checks}
\end{equation}
This can be represented in a Tanner graph as shown in Figure~\ref{fig:tanner}.

\subsection{Relation between CSS codes and homology of cellulations} \label{homology}

Any cellulation of a manifold can be associated with a CSS code and the properties of the code are associated with homological properties of the cellulation. Let the $i$ cells (the number of which is denoted ${\rm dim} (C_i)$) be associated with qubits i.e. $n = {\rm dim} (C_i)$. The boundaries of the $(i+1)$ cells are used to define Z checks and the coboundaries of $(i-1)$ cells define X checks. The number of each is denoted ${\rm dim} (B_i)$ and ${\rm dim} (B^i)$ respectively.  The number of encoded qubits $k$ is calculated by subtracting the number of stabilizers from the number of physical qubits:
\begin{equation}
k = {\rm dim} (C_i) - {\rm dim} (B_i) - {\rm dim} (B^i) = {\rm dim} (H_i),
\end{equation}
where $H_i$ is the $i$th homology group. 

The generating sets of stabilizers are not in general independent. For the Z checks, the number can be expressed in terms of the number of cells with dimension greater than $i$ and the dimensions of the homology groups of dimension greater than $i$:
\begin{equation}
{\rm dim} (B_i) = \sum_{j=1}^{D-i} (-1)^{j+1} \left ( {\rm dim} (C_{i+j}) - {\rm dim} (H_{i+j}) \right ),
\end{equation}
where we have used recursively the relationship 
\begin{equation}
{\rm dim} (B_i) = {\rm dim} (C_{i+1}) - {\rm dim} (H_{i+1}) + {\rm dim} (B_{i+1}). 
\end{equation}
Similarly for the X checks the relation is
\begin{equation}
{\rm dim} (B^i) = \sum_{j=1}^{i} (-1)^{j+1} \left ( {\rm dim} (C_{i-j}) - {\rm dim} (H^{i-j}) \right ).
\end{equation}
In the case of a $D=2$ cellulation that is topologically a disk, only $H^{0} = 1$ is non-trivial: the number of Z checks is equal to the number of faces, while the number of X edges is equal to the number of vertices minus one.  

The distance of the code is the same as the minimum length of an essential $i$ cycle in the cell complex, or its dual. 


\subsection{Codes and Hilbert spaces}

Quantum error correcting codes are those in which all the information within the code subspace of the Hilbert space is accessible from a subset of the physical degrees of freedom. Accordingly the full Hilbert space ${\cal H}$ can be expressed as a direct product
\be
{\cal H} = {\cal H}_{\cal R} \otimes {\cal H}_{\bar{\cal R}}
\ee
where 
\begin{itemize}
\item All logical operators may be represented on ${\cal R}$.
\item There is no correlation between ${\bar{\cal R}}$ and the encoded information.
\end{itemize}
Suppose that the Hilbert space ${\cal H}_{\cal R}$ factorises as
\be
{\cal H}_{\cal R} = {\cal H}_{{\cal R}_1} \otimes {\cal H}_{{\cal R}_2},
\ee
where the dimension of ${\cal H}_{ {\cal R}_1}$ is the same as the logical dimension; ${\cal H}_{ {\cal R}_1}$ is the logical subspace. The other 
factor is associated with the redundancy that protects from errors. In this context we can represent a state in the code subspace as
\be
| \tilde{\psi} \rangle = {\cal U}_{\cal R} \left ( | \psi \rangle_{ {\cal R}_1} | \chi \rangle_{ {\cal R}_2 {\bar{\cal R}}} \right ),
\ee
where $| \psi \rangle_{ {\cal R}_1}$ is the logical state; $| \chi \rangle_{ {\cal R}_2 {\bar{\cal R}}}$ is an entangled state and ${\cal U}_{\cal R}$ is a unitary operator on ${\cal H}_{\cal R}$. In the context of holography, ${\cal R}$ is associated with a subset of the conformal boundary where ${\bar {\cal R}}$ is its complement. 

Suppose that we want to encode $k$ logical qudits into $n$ physical qudits i.e. the dimension of the code subspace is $d^k$ and that of the full Hilbert space is $d^n$. The quantum Singleton bound says that the information encoded in the logical qudits can be recovered from $m$ qudits where
\be
m \ge \frac{1}{2} \left (n + k \right ) .
\ee
For holographic encodings associated with two dimensional geometries, the encoding map is related to a tensor structure i.e. 
\be
T_{i_1 \cdots i_n; j_1 \cdots j_k} \propto \langle i_1 \cdots i_n | \tilde{j}_1 \cdots \tilde{j}_k \rangle,
\ee
where the physical qudits are denoted by $i_i \cdots i_n$ and the logical qudits are denoted by $\tilde{j}_1 \cdots \tilde{j}_k$.


\section{Generalised Pauli Operators}
\label{Pauli}

The generalised Pauli operators \cite{Patera1998, GottesmanKitaevPreskill2000, BartlettdeGuise2001} for the D-dimensional qudits can be defined as
\be
Z \ket{k} := \omega^{k} \ket{k},
\ee
\be
X \ket{k} := \ket{k+1},
\ee
where $\omega = e^{2\pi i/D}$ and $k\in\mathbbm{Z}_D$. One can similarly generalise controlled gates to qudits, for example the controlled-Z operator between two qudits, $i$ and $j$ can be implemented as
\be
\label{contZ}
CZ_{ij} := \sum_{k=0}^{D-1} \ket{k} \bra{k}_i \otimes Z^k_j = \sum_{k,l=0}^{D-1} \omega^{kl} \ket{k} \bra{k}_i \otimes \ket{l} \bra{l}_j .
\ee
Trivially, one can check the commutation relation is
\be
ZX = \omega XZ
\ee
and each generalised operator has the property that when applied to a state $D$ times, one is simply left with the identity operator;  i.e. $Z^D = X^D = CZ^D = \mathbbm{1}$. Another useful operator is the Fourier gate,
\be
F = \frac{1}{\sqrt{D}} \sum_{k=0}^{D-1} \omega^{kl} \ket{k} \bra{l}
\ee
which allows one to transform between the $Z$-eigenbasis and the $X$-eigenbasis;
\be
\ket{\bar{k}} = F^\dagger \ket{k} = \frac{1}{\sqrt{D}} \sum_{l=0}^{D-1}\omega^{-kl} \ket{l}.
\ee
Hence, this is simply the generalisation of the Hadamard gate. 

Stabiliser states can also be written in terms of generalised operators when extending from qubits to qudits \cite{Ashikhmin2000, Ketkar2005}. The generalised Pauli group acting on $n$ qudits can be defined as
\be
{\cal P}_n :=  \{ \omega^a \otimes_{j=1}^n P_j \}
\ee
where $\omega = e^{2\pi i/D}$ and $a \in \mathbbm{Z}_D$.




\bibliographystyle{JHEP}
\bibliography{references1}

\end{document}